\documentclass[12pt,review,doublespacing]{elsarticle/elsarticle}
\usepackage[left=1in,right=1in,top=1in,bottom=1in]{geometry}
\journal{}
\emergencystretch=\maxdimen
\usepackage[scaled=0.92]{helvet}
\graphicspath{{Figures}{Figures/Figure_01}{Figures/Figure_02}{Figures/Figure_03}{Figures/Figure_04}{Figures/Figure_05}{Figures/Figure_06}{Figures/Figure_07}{Figures/Figure_08}{Figures/Figure_09}{Figures/Figure_10}}
\usepackage{sectsty}
\sectionfont{\color{blue}\MakeUppercase}
\subsectionfont{\color{blue}}
\biboptions{authoryear}
\usepackage{natbib}
\usepackage{hyperref}
 \hypersetup{
 	plainpages = false, 
	bookmarksopen = true,
	bookmarksnumbered = true,
	breaklinks = true,
	linktocpage,
	colorlinks = true,
	linkcolor = purple,
	urlcolor  = black,
	citecolor = purple,
	}
\usepackage{multicol}
\usepackage{multirow}
\usepackage{imakeidx}
\usepackage{nomencl}
\usepackage[pdf]{pstricks}
\makenomenclature
\usepackage[xindy]{glossaries}
\makeglossaries
\usepackage{graphics}
\usepackage{graphicx}
\usepackage{epsf,psfrag}
\usepackage{epstopdf}

\usepackage[para]{threeparttable}
\usepackage{subfig}
\usepackage{epsfig}
\usepackage{pdflscape}
\usepackage{rotating}
\usepackage{lscape}
\usepackage{float}
\usepackage{stfloats}
\usepackage{textgreek}
\usepackage{longtable}
\usepackage{lscape}
\usepackage{capt-of}
\usepackage{comment}
\usepackage{tablefootnote}
\usepackage{footnote}
\usepackage{caption}
\usepackage[autopunct=true]{csquotes}
\usepackage{rotating}
\usepackage{txfonts}
\usepackage[normalem]{ulem}
\usepackage{nccmath}
\usepackage{enumitem}
\usepackage[nodisplayskipstretch]{setspace} 
\usepackage{color,xcolor,soul}
\usepackage{amsfonts,amsmath,amssymb}
\usepackage{latexsym,array}%,subeqn}
\usepackage{textcomp}
\usepackage{mathtools}

\newcommand{\RomanNumeralCaps}[1]
\linenumbers
\makesavenoteenv{tabular}
\makesavenoteenv{table}
\emergencystretch=\maxdimen
\usepackage[left,displaymath, mathlines]{lineno}
\providecommand\bnabla{\boldsymbol{\nabla}}

\DeclareMathAlphabet{\mathpzc}{OT1}{pzc}{m}{it}
\def\fig{Figure~}
\def\figs{Figures~}
\def\eqn{Eq.~}
\def\eqns{Eqs.~}
\def\tab{Table~}

\def\micro{\textmu}
\providecommand\bnabla{\boldsymbol{\nabla}}

\providecommand\p{{\partial}}

\newcommand\cac{Ca_{\text{c}}}
\newcommand\qr{Q_{\text{r}}}
\newcommand\qc{Q_{\text{c}}}
\newcommand\qd{Q_{\text{d}}}

\providecommand\bnabla{\boldsymbol{\nabla}}
\providecommand\p{{\partial}}
\newcommand\wrr{w_{\text{r}}}
\newcommand\wc{w_{\text{c}}}
\newcommand\wdp{w_{\text{d}}}
\newcommand\rhor{\rho_{\text{r}}}
\newcommand\rhoc{\rho_{\text{c}}}
\newcommand\rhod{\rho_{\text{d}}}
\newcommand\mur{\mu_{\text{r}}}
\newcommand\muc{\mu_{\text{c}}}
\newcommand\mud{\mu_{\text{d}}}

\newcommand\uc{u_{\text{c}}}

\def\tsc#1{\csdef{#1}{\textsc{\lowercase{#1}}\xspace}}
\tsc{WGM}
\tsc{QE}
\tsc{EP}
\tsc{PMS}
\tsc{BEC}
\tsc{DE}
\newcommand{\rev}[1]{\textcolor{black}{#1}}     
\newcommand{\ttxt}[1]{\textcolor{blue}{#1} }     

\begin{document}
%
%\input{reviewer-1.tex}
%\newpage
%\setcounter{page}{1}
%\input{second-review2.tex}
%\newpage
%
%\begin{comment}
%---------------------
%\section*{Highlights}
%---------------------
%\input{highlights.tex}
%---------------------
%
\setcounter{page}{1}
%
%\end{comment}
%
\begin{frontmatter} % for elsevier jrnls
% 
%\openup -0.5em % 1em for double spacing
%\title{\ttxt{Effects of surface wettability on interface evolution and  droplet pinch-off mechanism in two-phase flow through T-junction microfluidic system}}
% 
%\title{\ttxt{CFD analysis of effects of surface wettability and flow rates on the interface evolution and droplet pinch-off mechanism in the cross-flow microfluidic systems}}
% 
\title{\ttxt{Effects of surface wettability and flow rates on the interface evolution and droplet pinch-off mechanism in the cross-flow microfluidic systems}}
%
%\begin{comment}
\author[labela]{Akepogu Venkateshwarlu}
%\ead{avenkateshwarlu@ch.iitr.ac.in}
\author[labela]{Ram Prakash Bharti\corref{coradd}}
\ead{rpbharti@iitr.ac.in}
\address[labela]{Complex Fluid Dynamics and Microfluidics (CFDM) Lab, Department of Chemical Engineering, Indian Institute of Technology Roorkee, Roorkee - 247667, Uttarakhand, INDIA}
%
%\address[labela]{Department of Chemical Engineering, Indian Institute of Technology Roorkee, Roorkee - 247667, Uttarakhand, INDIA}
%
\cortext[coradd]{\textit{Corresponding author. }}
%\end{comment}
%
%%%%%%%%%%%%%%%%%%%%%%%%%%%%%%%%%%%%%%%%%%%%%%%%%%%%%%%%%%%%%%%%%%%%%%%%%%%%%%%%%%%%%%
\begin{abstract}
%\linespread{1.3}
\fontsize{11}{17pt}\selectfont
%----Text of abstract
%\subsection*{Background:}    
%\subsection*{Methods:}
%\subsection*{Significant Findings:}
\noindent In this study, the effect of surface wettability on the two-phase immiscible \rev{fluids} flow and dynamics of droplet pinch-off in a T-junction microchannel has been numerically investigated using the finite element method. A conservative level set method (CLSM) has been adopted to capture the interface topology in the squeezing regime ($\cac <10^{-2}$) for wide flow rate ratio ($1/10 \leq \qr \leq 10$) and contact angle ($120^{\circ} \leq \theta \leq 180^{\circ}$). This study has revealed that the wettability is a dominant factor in determining the hydrodynamic features of the droplet. Based on the instantaneous phase \rev{flow} profiles, the droplet formation stages are classified as: initial, filling, squeezing, pinch-off and stable droplet. Wettability effects are insignificant in the filling stage. However, the hydrophobic effects are more vital in the squeezing and pinch-off stages. In general, it is shown that engineering parameters have complex dependence on the dimensionless parameters ($\cac$, $\qr$, $\theta$). Capturing the instantaneous interface evolution has revealed that the droplet shape is sensitive to the contact angle. Interface shape profiles transform from convex into concave immediately for hydrophobic conditions ($120^{\circ} \leq \theta \leq 150^{\circ}$) whereas slowly for the super hydrophobic conditions ($150^{\circ} <
 \theta \leq 180^{\circ}$). In contrast to the literature, the pressure in the dispersed phase is not constant, but it is an anti-phase with the pressure in the continuous phase. Maximum pressure in the continuous phase, \rev{and} neck width of the interface are complex function of the \rev{governing conditions ($\cac$, $\qr$, $\theta$)}. Comparison of the filling and pinch-off time based on the pressure and phase profiles has brought new insights that the droplet pinch-off mechanism can be elucidated by installing the pressure sensors even without the flow visualization and phase profiles. The interface curvature adopts a flattened to a more concave shape when the Laplace pressure varies from a smaller to higher value. The interface neck width ($2r$) shows an increasing trend up to a threshold value and then decreases linearly with the contact angle. 
\end{abstract}
%%%%%%%%%%%%%%%%%%%%%%%%%%%%%%%%%%%%%%%%%%%%%%%%%%%%%%%%%%%%%%%%%%%%%%%%%%%%%%%%%%%%%%
%%%%%%%%%%%%%%%%%%%%%%%%%%%%%%%%%%%%%%%%%%%%%%%%%%%%%%%%%%%%%%%%%%%%%%%%%%%%%%%%%%%%%%
\end{frontmatter}
%\end{document}
%%%%%%%%%%%%%%%%%%%%%%%%%%%%%%%%%%%%%%%%%%%%%%%%%%%%%%%%%%%%%%%%%%%%%%%%%%%%%%%%%%%%%%
%\fontsize{12}{14pt}\selectfont
%\linespread{1.6}
%\doublespacing
%\openup 1em % 1em for double spacing
%\setstretch{2} 
%
%\clearpage
%
%%%%%%%%%%%%%%%%%%%%%%%%%%%%%%%%%%%%%%%%%%%%%%%%%%%%%%%%%%%%%%%%%%%%%
\section{Introduction}\label{sec:intro}
%%%%%%%%%%%%%%%%%%%%%%%%%%%%%%%%%%%%%%%%%%%%%%%%%%%%%%%%%%%%%%%%%%%%%
%
\noindent
Microfluidics has become an emerging field in the contemporary research world due to its versatile applications. It has triggered technological revolutions in interdisciplinary engineering, science, biochemical and biomedical like emulsions, mixing reactions, diagnostics, protein encapsulation, DNA analysis, new material synthesis, paints, inks, and coating \citep{Barnes1994,Abate2013,Wu2014,Hou2017,Kaminski2017, Boruah2018,Weng2019,Schroen2021,BhartiBook2022,JUNG2022}. Microfluidic devices are frequently applied as a potential platform to generate droplet emulsions by using various geometric configurations and methods. Commonly used microfluidic devices to generate the droplets are co-flow, cross-flow, and flow-focusing devices \citep{Carrier2014,MOON2014}. Amongst others, T-junction cross-flow microfluidic device (\fig\ref{fig:1a}) is prevalent for producing monodispersed droplets due to its simplicity and ease of controlling the generation \citep{venkat2021,venkat2022a,venkat2022b}. The displacement of the dispersed phase and formation of the droplet through the microchannels is governed mainly by important determinants like capillary number (ratio of viscous to interfacial tension forces), and Reynolds number (\rev{ratio of}  inertial to viscous forces) \citep{MASTIANI2017a,Agarwal2020,venkat2021,venkat2022a,venkat2022b}.  

\noindent
The dynamic behavior of the droplet generation is governed by the three fundamental characteristics which express the exerting surface forces: (a) {contact angle} - described by the surface energies where they meet; (b) {pressure jump across the surfaces and interfaces} - higher pressure on the concave side of the surface, and (c) {the shear stress caused by the surface tension gradient} -  the pressure difference between each side of the interface is caused by the surface tension; the pressure essentially causes the bulge, and hence it is always higher on the concave side of the interface \citep{NCFMF1969,Bashir2014, Shi2014,venkat2022a}.

\noindent
Surface wettability, i.e., fluid-solid interaction (FSI), is known to play a vital role in many physical, chemical,  biological, and industrial processes related to agriculture (efficiency of pesticides), medicine (integration of implants with bone), food packaging (prolonged release of antimicrobial agents and increased shelf life), painting and printing (ensured suitable adhesion of liquid to solid), heat transfer and lubrication (consideration of surface energy), and oil recovery (selective absorption of oil in solid materials) \citep{Huhtamaki2018}. 
Surface wettability influences the degree of contact with the fluidic environment and is quantified in terms of the contact (or wetting) angle, inversely proportional to the surface energy. 
The contact angle  ($\theta$, degrees) is defined geometrically as the angle between the tangent to the liquid-liquid interface and the solid surface at the three-phase (liquid-vapor/gas-solid or liquid-liquid-solid) contact line (\fig\ref{fig:1c}). 
The interfacial tensions of solid-liquid ($\sigma_{\text{sl}}$), solid-vapor ($\sigma_{\text{sv}}$), and liquid-liquid ($\sigma$) form the equilibrium contact angle of wetting ($\theta$) (\fig\ref{fig:1c}). 
It represents the strength of FSI (fluid-solid interaction) and is measured conventionally from the liquid (dispersed or heavier) side. For example, high surface energy (i.e., high $\sigma_{\text{sv}}$) would exhibit a low $\theta$ and the liquid tendency to spread and adhere to the surface. In contrast, low surface energy (i.e., low $\sigma_{\text{sl}}$)  demonstrates high $\theta$  and the surface tendency to repel the liquid.
The hydrophilicity diminishes\footnotemark\ with increasing contact angle \citep{Law2014}. 
\noindent
The contact angle between the ideal solid surface (i.e., atomically smooth, non-reactive, chemically homogeneous, and perfectly rigid) and liquid is defined by the well-known Young equation ($\sigma_{\text{sv}}=\sigma_{\text{sl}}+\sigma\cos\theta$). 
\footnotetext{$\theta=0^{\circ}$: perfect wetting (or super-hydrophilic); $0^{\circ} <\theta< 90^{\circ}$: spreading (or wettable or hydrophilic); $\theta = 90^{\circ}$: neutral ($\sigma_{\text{sl}}=\sigma_{\text{sv}}$, i.e., equal cohesive and adhesive forces); $90^{\circ} < \theta < 180^{\circ}$: repelling (or non-wettable or hydrophobic); $\theta = 180^{\circ}$: completely liquid-repellent (or non-wetting or super-hydrophobic).}

\noindent
The surface wettability plays a significant role in droplet evolution. The contact angle greatly influences the droplet dynamics as it is affected by surface roughness, impurities on the solid surface, porosity, surface energy, and functional groups present on the surface \citep{Boruah2018,DEKA2022}. 
%Therefore, for a perfect wettability condition, the contact angle is $0^{\circ}$, and the non-wetting contact angle is $180^{\circ}$. 
%
The microfluidic devices are easier to control and manipulate as they do not require any secondary fluid to trigger the formation of the droplet, and the minimum contact angle of about $\theta=120^{\circ}$ is needed to form a droplet \citep{KAWAKATSU2001,MAAN2013,Eggersdorfer2018}. 
The mono-dispersed droplets are generated by operating a T-junction microchannel under a squeezing or dripping regime. The geometry confinement suppresses the capillary instabilities; hence, the produced droplets are regular and stable \citep{VanSteijn2007,Glawdel2012a,venkat2021,venkat2022a}. 

\noindent
Extensive literature is available on the dynamics of droplet generation through the cross-flow microfluidic systems \citep{Thorsen2001,Nisisako2002,Garstecki2006,Demenech2008,Christopher2008,
Gupta2009,Wong2017,venkat2021,venkat2022a,venkat2022b,PradeepChap13,ShuvamChap14,BhartiBook2022}.  The droplet formation process is divided into three stages (filling, squeezing, and instant pinch-off). The breakup (or pinch-off) step is, essentially, initiated by the pressure developed due to the reverse flow in the gutters between the channel wall and liquid-liquid interface (LLI). Further, the hydrodynamic pressure becomes equal on both sides of the interface at the pinch-off. 
%\rev{Further, it is reported that at the pinch-off, the Laplace pressure between the backside of the interface and front side becomes equal.} 
The linear relationship predicts the dependence of the droplet size on the flow rate ratio in the squeezing flow regime \citep{Volkert2009,Glawdel2012a,venkat2021}. Limited efforts \citep{VanderGraaf2006,Bashir2014,Shi2014,Boruah2018} are also made to explore the influence of surface wettability on the dynamics of droplet generation in the microchannel.
%\noindent
Although several attempts have been made to elucidate the droplet pinch-off mechanism by exploiting the interface curvature but still lacks to understand the effect of the contact angle on the interface curvature and the esoteric reason behind the shape evolution \citep{Wang2020}. 

\noindent
The main challenges in an in-depth study of the droplet pinch-off mechanisms are complexity in capturing the topological changes of the interface curvature and non-linearity of the equations. In addition, a detailed understanding of the mechanisms requires rigorous and precise post-processing and analysis of the results. Further, a detailed systematic study depicting the wettability effects on the interface evolution and droplet breakup mechanisms will be instrumental in controlling droplet-based phenomena of mixing, dispersion, paints, and coatings. 
Therefore, the present work aims to explore the influences of surface wettability on the dynamics of the interface evolution and droplet generation in cross-flow microfluidic devices.
%
%---------------------
\section{Physical and Mathematical Modelling}
%---------------------
%
%
\begin{figure}[tb]
	\centering
	\subfloat[]{\includegraphics[width=1\linewidth]{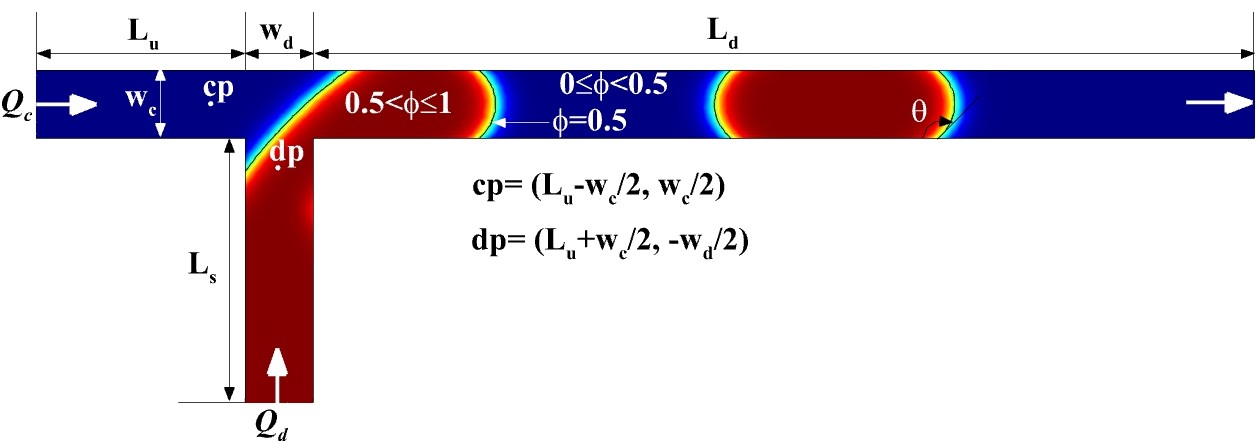}\label{fig:1a}}\\
	\subfloat[]{\includegraphics[width=0.45\linewidth]{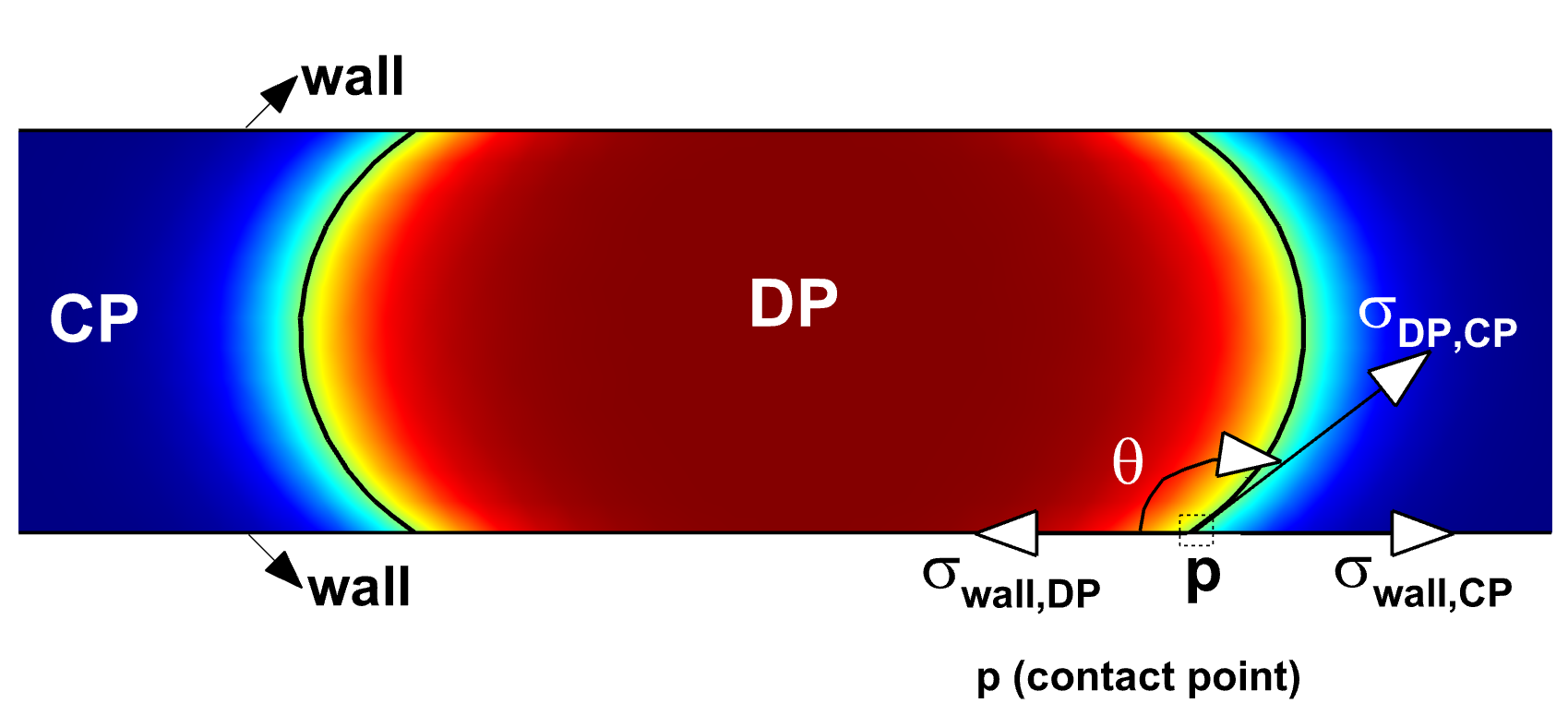}\label{fig:1c}}
	\subfloat[]{\includegraphics[width=0.45\linewidth]{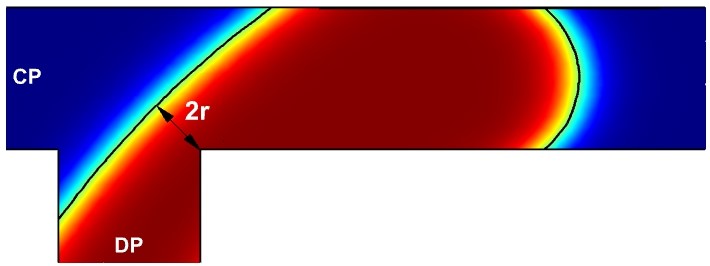}\label{fig:1b}}
	\caption{Schematic representation of  (a)  the two-phase flow  in T-junction microfluidic device, (b) contact  angle ($\theta$), and (c) neck thickness of the interface ($2r$).}
	\label{fig:1}
\end{figure}
\noindent 
Consider the two-dimensional laminar flow of two immiscible fluids through a cross-flow rectangular slit-type microfluidic system, as illustrated in \fig\ref{fig:1a}. The microfluidic system depicts a T-junction by geometrically arranging a side (or vertical) channel perpendicularly to the horizontal main (or primary) channel. The geometrical dimensions (length, width) of the primary and side channels are ($L_{\text m}$, $w_{\text c}$) and ($L_{\text s}$, $w_{\text d}$), respectively.  The upstream length (i.e., the distance between the inlet of the primary channel and the front of the side channel) and downstream length (i.e., the distance between the rear of the side channel and the outlet of the primary channel) are  $L_{\text u}$ and $L_{\text d}$, respectively. The length of the primary channel  ($L_{\text m}=L_{\text u}+w_{\text d}+L_{\text d}$) is taken to be sufficient to avoid the end effects. All the length quantities ($L$, $w$) are measured in microns (\micro m).

\noindent
Both fluids are assumed to be Newtonian (i.e., constant viscosity, $\mu$ Pa.s), isothermal (i.e., constant temperature, $T~^\circ$C), non-reactive, and incompressible (i.e., constant density, $\rho$ kg/m$^{3}$). The Marangoni and dynamic interfacial effects are ignored, and hence, the surface tension ($\sigma$ N/m) is uniform throughout the device. Further, the device walls are considered to be ideal solid surfaces (i.e., atomically smooth, non-reactive, chemically homogeneous, and perfectly rigid). 

\noindent
The continuous phase (CP, indicated by blue color) and dispersed phase (DP, indicated by red color) enter (refer \fig\ref{fig:1a}) from the inlets of the primary and side channels, with the fixed volumetric flow rates ($\mu$l/min) of $Q_{\text c}$ and $Q_{\text d}$, respectively. Both phases interact at the T-junction point, flow downstream, and exit through the primary channel open to the atmosphere (i.e., $p=0$). Further, the flow is confined within the no-slip walls.    
%
%----------------------------------
%\subsection{Governing equations} 
%----------------------------------
%

\noindent 
The above-stated physical problem is mathematically expressed by the conservation equations of mass (\eqn\ref{eq:1}), momentum (\eqn\ref{eq:2}), and phases (\eqn\ref{eq:6}), as follows.
\begin{gather}
\nabla \cdot \mathbf{u} = 0  
\label{eq:1} 
\end{gather}
\begin{gather}
\rho\left[\dfrac{\p \mathbf{u}}{\p t}+\mathbf{u} \cdot \bnabla \mathbf{u}\right]=-\bnabla p+\bnabla \cdot \boldsymbol{\tau}+\mathbf{F}_{\sigma} 
\label{eq:2} 
\end{gather}
where $\mathbf{u}$, $p$, and $t$ are the velocity vector, pressure, and time, respectively. The deviatoric stress tensor ($\boldsymbol{\tau}$) is defined as follows.
\begin{gather}
	\boldsymbol{\tau}=2\mu \mathbf{D} \label{eq:3}
\end{gather}
where $\mathbf{D}$ is the rate of deformation tensor. The physical properties ($X = \rho$ and $\mu$) of the two fluid phases (CP and DP) at any point are expressed as follows. 
\begin{gather}
	X=X_{\text c}+(X_{\text d}-X_{\text c})\phi
	%,\qquad \text{where}\qquad
%	X=\mu,~\rho
	\label{eq:4}
\end{gather}
\rev{where, the level set function ($\phi$) represents the liquid phase composition whose value varies between $\phi=0$ (for pure CP) and $\phi=1$ (for pure DP). The values of $\phi< 0.5$, $\phi = 0.5$ and $\phi > 0.5$ indicate for the CP, liquid-liquid interface (LLI) and DP, respectively.} 

\noindent The interfacial tension force (IFT, $\mathbf{F}_{\sigma}$) acting between the two immiscible liquids is given as follows by the continuum surface force (CSF) model \citep{Brackbill1992}.
\begin{gather}
	\mathbf{F}_{\sigma}=\sigma \kappa \delta(\phi) \boldsymbol{n}
	\label{eq:5}
\end{gather}	 
{where}
\begin{gather*}
	\delta(\phi)=6 |\bnabla \phi| |\phi (1-\phi)|,
	\qquad  
	\kappa=(1/{R})=-(\mathbf{\bnabla} \cdot \boldsymbol{n}) \qquad\text{and}
	\qquad \boldsymbol{n}=\frac{\mathbf{\bnabla \phi}}{|\mathbf{\bnabla \phi}|}	
\end{gather*}
where $\sigma$, $\boldsymbol{n}$, $\delta$, $\kappa$, and ${R}$ are the liquid-liquid surface tension coefficient (N/m), unit normal, the Dirac delta function approximated by a smooth function,  mean curvature, and the radius of curvature, respectively.

\noindent 
The conservative level set method (CLSM) is well-suited to capture the topological changes of moving interfaces. It is a robust scheme that is relatively easy to implement \citep{Osher1988,Olsson2005,AkhlaghiAmiri2013,Gada2009,Bashir2011,Wong2017}. The following level set equation (LSE) expresses the conservation of the phase field.
\begin{gather}
	\dfrac{\p \phi}{\p t}+\mathbf{u} \cdot {\bnabla} \phi=\gamma {\bnabla} \cdot \left[\epsilon_{\text{ls}}{\bnabla} \phi- \phi (1-\phi){\boldsymbol{n}}\right]
	\label{eq:6}
\end{gather}
The level set parameters $\gamma$, and $\epsilon_{\text{ls}}$ acts for the re-initialization and stabilization of $\phi$, and controlling interface thickness, respectively.
%
%....................
%\subsection{Boundary conditions}
%....................
%

\noindent
The physical boundary conditions for the governing equations (\eqns\ref{eq:1} - \ref{eq:6})  stated above are as follows.
\begin{enumerate}[label=(\alph*)]
\item Fixed flow rate of CP ($Q_{\text{c}}$ ${\mu}$l/min) at the inlet of the primary channel.
\item Fixed flow rate of DP ($Q_{\text{d}}$ ${\mu}$l/min) at the inlet of vertical channel.
\item Constant ambient pressure ($p=0$) and fully developed velocity and phase fields at the outlet of the primary channel.
\item 
The wetted wall boundary condition\rev{, which} allows the fluid-fluid interface to move along the wall\rev{, has been imposed on all the no-slip solid walls}. This condition requires a slip length ($\beta$ \micro m), i.e., the distance outside the solid surface at which the tangential component of the velocity extrapolated to zero \citep{Bashir2011,Yan2012,Hernandez-cid2020}. The value of $\beta$ is generally equivalent to the local mesh size element ($h$ \micro m). 

\noindent For laminar flow, the boundary condition for the wetted wall to enforce a no-penetration condition on the solid surface (wall) is $\mathbf{u} \cdot \boldsymbol{n_{\text{s}}}=0$, where $\boldsymbol{n}_{\text{s}}$ is a unit normal to the solid surface. 
\noindent
The tangential stress is defined as $\mathbf{F}_{\text {fr}}=-(\mu/\beta)\mathbf{u}$. The boundary force to enforce the contact angle is defined as $\mathbf{F}_{\theta}=\sigma \delta (\boldsymbol{n}_{\text {s}} \cdot \boldsymbol{n}-\cos\theta)\boldsymbol{n}$, where $\theta$ is the static contact angle ($\theta_{\text{s}}$) between the solid wall and the fluid interface \citep{GERBEAU2009,Bashir2014,Mirzaaghaian2020, COMSOL60}. 
\noindent
In weak form, the wetted wall boundary condition is expressed as:
\begin{equation}
	\int_{\p \Omega} \textbf{u}_{\text t} \cdot [\sigma (\boldsymbol{n}_{\text{s}}-(\boldsymbol{n} \cos \theta_{\text{s}}))\delta]dS=0
	\label{eq:weakform_wettedwall}
\end{equation}
where $\textbf{u}_{\text t}$ is the test function used in the Galerkin finite element method; it is a function of velocity ($\textbf{u}$) and domain (${\p \Omega}$). The Lagrange multipliers are used to implement this condition in COMSOL by the Galerkin finite element approximation (variational approach).
\end{enumerate}
Finally, it is appropriate to define the non-dimensional parameters used in the present work. 
\begin{align}
		Re_{\text{c}} = \frac{\rhoc \uc \wc}{\muc},\qquad
		\cac = \frac{\uc \muc}{\sigma},\qquad 
		\qr = \frac{\qd}{\qc}, \qquad
		\wrr = \frac{\wdp}{\wc}, \nonumber\\
		\rhor = \frac{\rhod}{\rhoc}, \qquad
		\mur = \frac{\mud}{\muc}, \qquad
		p={p^{*}}\left(\frac{\wc}{\uc \muc}\right), \qquad
		t=t^{*}\left(\frac{\uc}{\wc}\right),  \qquad
		2r=\frac{2r^{*}}{\wc}
	\label{eq:7}
\end{align}
where $Re$ and $Ca$ are the Reynolds, and capillary numbers. The subscripts `r', `c', and `d' denote for ratio, CP, and DP, respectively. 
The variables with asterisk ($\ast$) superscript are dimensional, but denoted without asterisk ($\ast$) before \eqn(\ref{eq:7}). Further, the neck width ($2r$) is defined \citep{venkat2022a} as the shortest distance from the receding interface to the lower-right corner of the junction (refer \fig\ref{fig:1b}). 

\noindent
The numerical solution of the mathematical model, subject to the boundary conditions, provides an instantaneous flow ($\mathbf{u}$, $p$) and phase ($\phi$) fields as a function of dimensionless parameters ($Re_{\text{c}}$, $\cac$, $\qr$, $\wrr$, $\rhor$, $\mur$). The post-processing of these fields is further performed to gain insights of the dynamics of interface development and droplet generation. The next section details the numerical modelling and governing parameters used in this work. 
%
%---------------------------------------------------
\section{Numerical modelling and parameters}
\label{sec:sanp}
%---------------------------------------------------
%
In this study, the computational fluid dynamics (CFD) simulations of the mathematical model %(\eqns\ref{eq:1} - \ref{eq:6}) subject to the boundary conditions 
have been performed using the finite element method (FEM) based COMSOL multiphysics for determining the phase ($\phi$), velocity ($\mathbf{u}$), and pressure ($p$) fields. 
Since the detailed solution approach is presented \rev{in our recent studies} \citep{venkat2021,venkat2022a}, only salient features are included here.
\noindent 
The present mathematical model is expressed in COMSOL by using the ``two-dimensional (2-D) $\rightarrow$ fluid flow (\textit{ff}) $\rightarrow$ multiphase flow (\textit{mpf}) $\rightarrow$ two-phase flow, level set (\textit{tpfls}) $\rightarrow$ laminar flow (\textit{lf})'' modules. The computational domain (\fig\ref{fig:1a}) is discretized using 2-D, linear, non-uniform, unstructured, triangular mesh elements. 
%
%The fundamental laws or flow governing equations are generally expressed in terms of partial differential equations (PDEs). In most cases, the PDEs are complicated to solve by analytical methods. Hence, these PDEs are transformed into approximated equations based on the discretization techniques \citep{Sartipzadeh2020,DEKA2022}. 
%\rev{In the present study, the finite element method has been adopted to compute the approximated discretization methods.} In spatial discretization, the order of the polynomials is considered as p$^{\text{th}}$ and q$^{\text{th}}$ (i.e., P$_{\text{p}}$ + P$_{\text{q}}$) for shape functions of velocity and pressure fields, respectively. However, in the present study, the order of both p$^{\text{th}}$ and q$^{\text{th}}$ is taken as 1 \citep{Jakub2021}. Further, the accurate implicit backward differentiation formula (BDF) is chosen for ODEs temporal discretization, and it has resulted in smooth convergence of non-linear complex dynamic problems. The obtained discretization equations are solved by using a fully coupled approach, i.e., Parallel Direct Sparse Solver (PARDISO) and Newton’s non-linear solvers \citep{Asghari2020,Sartipzadeh2020}. 
%
\noindent 
The transient PDEs (partial differential equations) are transformed into ODE (ordinary differential equations) by using the finite element method (FEM). The spatial interpolation of both pressure and velocity fields is performed using the shape functions of the first-order polynomial (i.e., P$_1$+P$_1$). Further, the temporal derivatives in ODEs have been discretized using an implicit backward differentiation formula (BDF). The accuracy of the solution is generally a trade-off with the stable convergence for the higher- to lower-order BDF approximation. The discretization process results in differential-algebraic equations (DAEs) with variable time steps ($\Delta t$). 

\noindent
The fully converged numerical solution of DAEs have been obtained using a fully coupled Newton’s non-linear and PARDISO solvers with a sufficiently {low} time step ($\Delta t=10$ \micro s) and the relative tolerance ($5 \times 10^{-3}$) for the following ranges of conditions.
\begin{itemize}
\item Geometrical parameters: $w_{\text c}=100$ \micro m; $w_{\text r}=1$; $L_{\text u}=L_{\text s}=9w_{\text c}$; $L_{\text d}=30w_{\text c}$; $L_{\text m}=40w_{\text c}$
\item Mesh parameters: $N_{\text{e}}=13766$; DoF $=53029$; $h_{\text {max}}=10$ \micro m
\item Level set parameters: $\gamma=1$ m/s; $\epsilon_{\text{ls}}=h_{\text{max}}/2$ 
\item Fluid and flow parameters: $\cac < 10^{-2}$; $Re_{\text c}=0.1$; $120^{\circ} \leq \theta \leq 180^{\circ}$; $0.1 \leq \qr \leq 10$; $Q_{\text d}=0.14$ \micro l/s;  $\rho_{\text d}=1000$ kg/m$^3$; $\mu_{\text d}=0.001$ Pa.s; $7.143\times 10^{-3}\le\mur\le 7.143\times 10^{-1}$; $1.96\times 10^{-6}\le\sigma \le 1.96\times 10^{2}$ N/m
\end{itemize}	
{The above noted geometrical, mesh and simulation parameters have been tested in our recent stud\rev{y} \citep{venkat2021} for their dependence on the numerical results.} 
%	
%The detailed mesh independence study for the geometry and validation has been done \citep{venkat2021,venkat2022a} in our recent work and found that 13766 elements (maximum size of the element, 10 µm) are sufficient to carry out the present study to reduce the computational cost and efforts. 
%
%---------------------------------
\section{Results and discussion}
%---------------------------------
%
\noindent
In this section, the effects of surface wettability ($120^{\circ} \leq \theta \leq 180^{\circ}$) and the flow rate ratio ($0.1 \leq \qr \leq 10$) on the instantaneous evolution of the interface and the droplet pinch-off mechanism have been elucidated through an instantaneous phase ($\phi$) and pressure ($p$) evolutions, neck thickness or width ($2r$), and the Laplace pressure ($p_{\text{L}}$) acting on the interface for the broad conditions of the microfluidic flow. 
\rev{It is noted here the results for the droplet pinch-off dynamics for a fixed contact angle ($\theta=135^\circ$) published \citep{venkat2022a} recently are included in this study, wherever required for comparison purposes only. Furthermore, our recent studies \citep{venkat2021,venkat2022a,venkat2022b} have established the reliability and accuracy of the present modeling and simulation approaches via a thorough comparison of results with the earlier experimental and numerical studies in terms of droplet length, effective diameter, and radius of the interface evolution for broader flow operating conditions. 	The validation of results, thus, is not reported herein to avoid repetition. Based on our earlier experience \citep{venkat2021,venkat2022a,venkat2022b}, the subsequent presented results are believed to have excellent (±1-2 \%) accuracy.}
%---------------------------------------------------------------
\subsection{Instantaneous phase flow and droplet formation profiles}
\begin{figure}[htbp]
	(a) S0\hspace{0.3in} (b) S1 \hspace{0.6in} (c) S2 \hspace{1in} (d) S3 \hspace{1in} (e) S4 \hspace{0.2in}\\
	\centering
	\subfloat[$\qr=10$]{\includegraphics[width=1\linewidth]{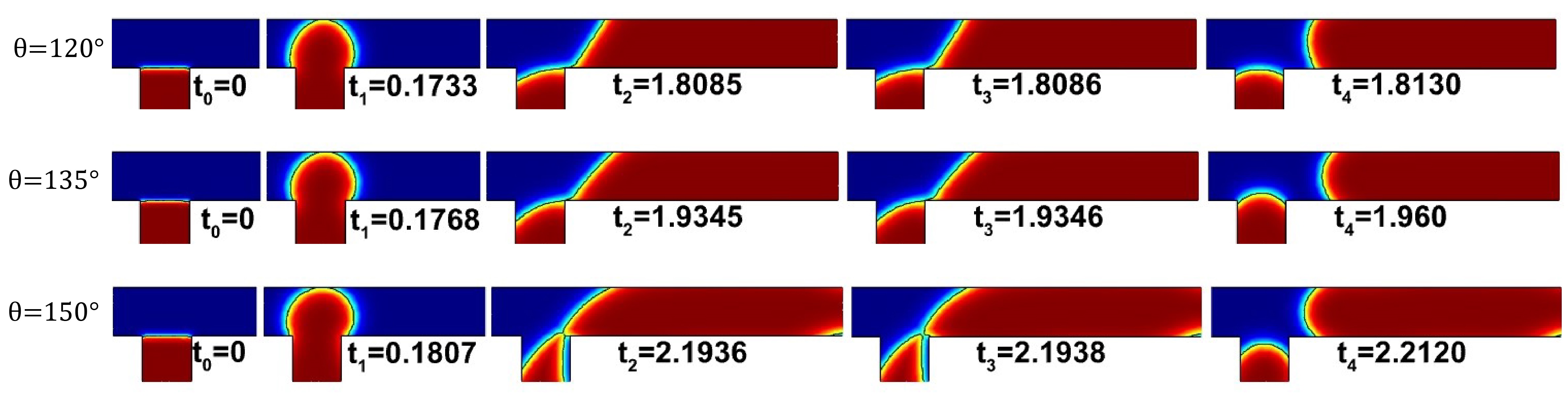}}\\
	\subfloat[{$\qr=1$}]{\includegraphics[width=1\linewidth]{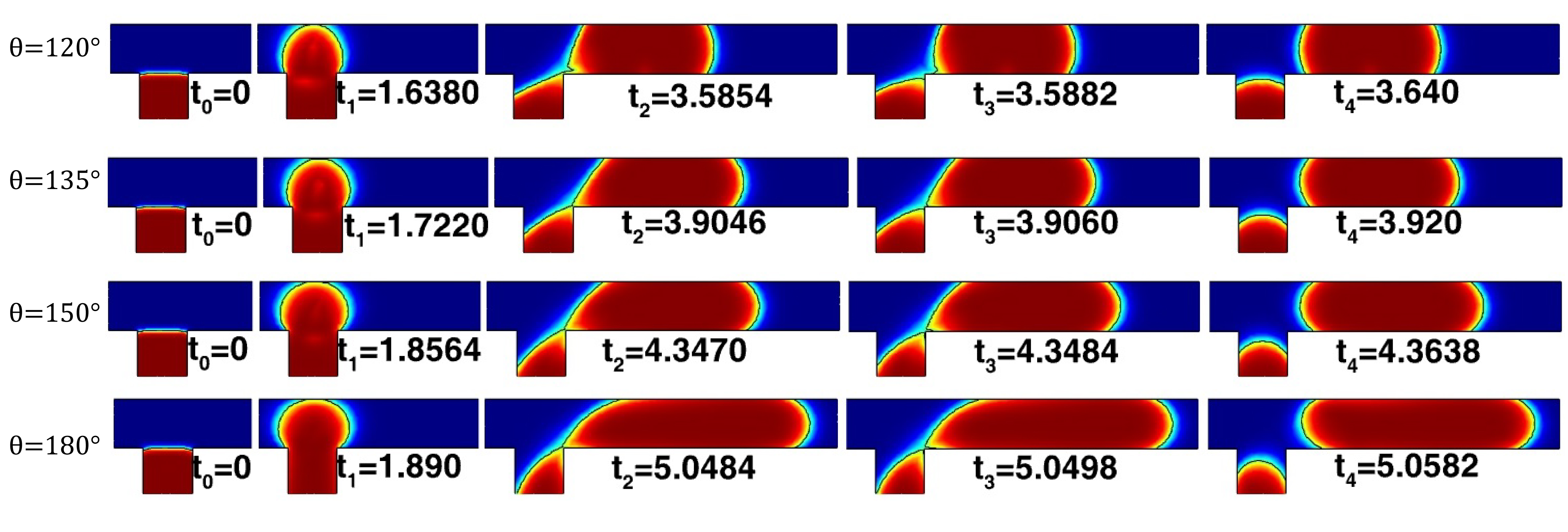}}\\
	\subfloat[\rev{$\qr=1/10$}]{\includegraphics[width=1\linewidth]{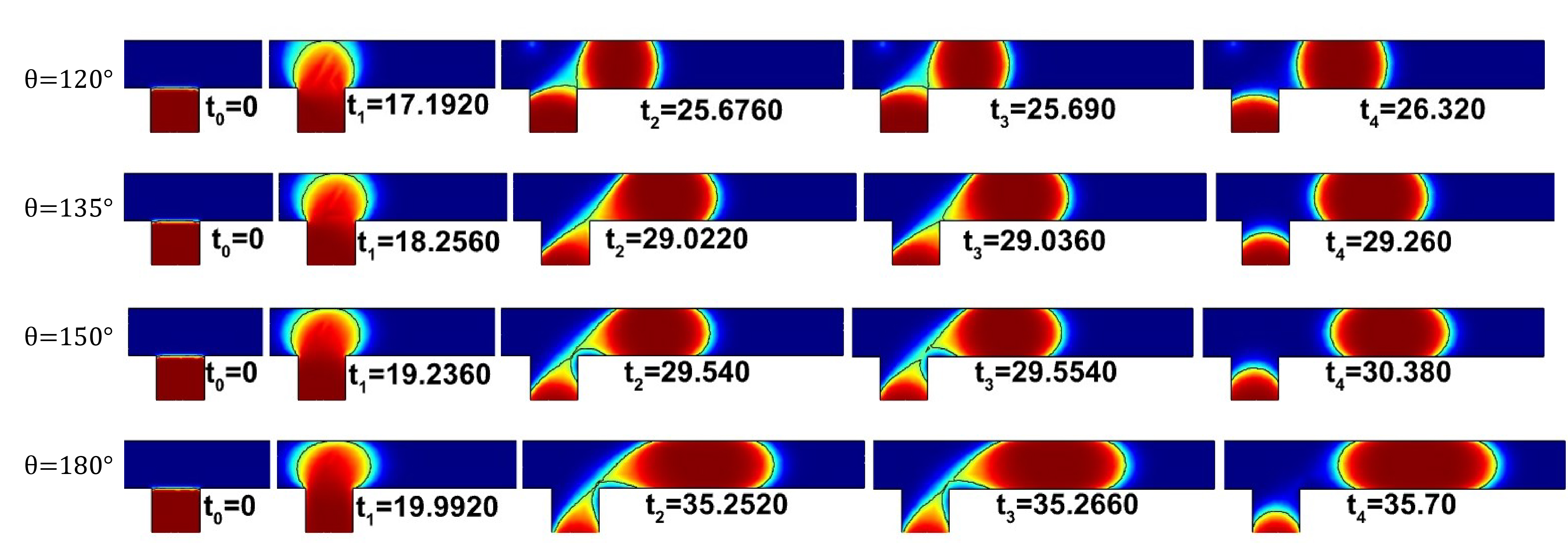}}	
	\caption{Instantaneous phase composition profiles and the droplet formation stages (S0 - initial, S1 - filling, S2 - squeezing, S3 - pinch-off, and S4 - stable droplet) as a function of flow rate ratio ($\qr$) and contact angle ($\theta$) \rev{for a fixed capillary number ($\cac=10^{-4}$). The dimensionless time ($t_0$ to $t_4$) refers to end of each stage}.}
	\label{fig:2}
\end{figure}

%---------------------------------------------------------------
%
\noindent 
%\fig\ref{fig:2} depicts the instantaneous phase ($\phi$) flow profiles for the broad ranges of the flow rate ratio ($0.1 \leq \qr \leq 10$) and contact angle ($120^{\circ} \leq \theta \leq 180^{\circ}$) under the squeezing ($\cac < 10^{-2}$) flow regime. The pure continuous ($\phi=0$) and pure dispersed ($\phi=1$) phases are represented by the dark blue and dark red colours contours in \fig\ref{fig:2}.  
%
Based on the instantaneous evolution of phases  ($\phi$) and liquid-liquid interface (LLI), a detailed discussion on the periodic time cycle and droplet pinch-off mechanism at a fixed contact angle ($\theta = 135^{\circ}$) have been made in our recent study \citep{venkat2022a}.  
%Based on the instantaneous evolution of phases  ($\phi$) and liquid-liquid interface (LLI), a recent study \citep{venkat2022a} has presented a detailed discussion on the periodic time cycle and droplet pinch-off mechanism at fixed contact angle ($\theta = 135^{\circ}$).  
In continuation, \fig\ref{fig:2} depicts a combined influence of contact angle ($120^{\circ} \leq \theta \leq 180^{\circ}$) and flow rate ratio ($0.1 \leq \qr \leq 10$) on the periodic time cycle of the droplet formation\rev{, which} is classified into five stages (S0 - initial, S1 - filling, S2 - squeezing, S3 - pinch-off, and S4 - stable droplet) under the squeezing ($\cac < 10^{-2}$) flow regime.  
%
%\noindent The droplet formation based on the instantaneous phase contours is described as shown in \fig\ref{fig:2} and classified into the following stages: (a). S-0: initial, (b). S-1: filling, (c). S-2: squeezing, (d). S-3: pinch-off and (e). S-4: stable droplet. 
%

\noindent 
First, both phases are injected simultaneously through respective channel inlets. The initial stage (S0) ends at time $t_0=0$ when the vertical channel is completely filled with the dispersed phase (DP). Subsequently, DP starts intruding into the primary channel at the starting of the filling stage (S1) at time $t_0$. The dispersed phase evolves as a convex shape and expands until it reaches the opposite wall of the main channel at the end of S1 at time $t_1$.
\noindent 
At $\qr=1$, the dispersed phase touches the top wall for \rev{both} hydrophobic ($120^{\circ} \leq \theta \leq 150^{\circ}$) \rev{and} superhydrophobic ($150^{\circ} < \theta \leq 180^{\circ}$) \rev{conditions}. \rev{In this case,} both CP and DP\rev{, having equal viscosity,}  are flowing with the same flow rate. Therefore, the wall effects have become more potent \rev{with increasing} contact angle. \rev{The filling time is thus smoothly increasing with increasing contact angle.}
\rev{Similarly,} there is a smooth increase in the \rev{filling} time for extreme flow rate conditions, $\qr=10$ and $1/10$ under both the hydrophobic and superhydrophobic conditions. The filling or growing stage S1 ends \rev{at time $t_1$ when} the continuous phase flow is blocked \rev{by the dispersed phase due to the equilibrium between the forces arising from interfacial tension and squeezing pressure}. However, a narrow gap exists between the wall and interface that allows the continuous phase to flow. Nonetheless, the gap between the interface and channel wall is reduced, leading to the hydrodynamic force exerted by the surrounding fluid on the droplet causes to deform \citep{VanSteijn2010}. 

\noindent 
The filling time ($t_{\text f}=t_1-t_0$) is correlated as a function of $\qr$, $\cac$ and $\theta$ ($0.1\le\qr\le 10$, $120^{\circ} \leq \theta \leq 180^{\circ}$, and $10^{-4}\le \cac \le 10^{-3}$) as follows:
\begin{equation}
	t_{\text{f}}={A} \cac^{B} \qr^{C}
	 \label{eq:tf}
\end{equation}
where 
\begin{gather*}
A=\alpha + \beta \theta^{2.5} +\gamma \theta^3,\qquad 
B=\alpha +\beta \theta^{2.5}+\gamma \theta^{-2},\qquad 
C=\alpha + \beta \theta^{2.5}+\gamma \theta^3  
\end{gather*}
Statistical non-linear regression analysis is performed to obtain the correlation coefficients \rev{of $t_{\text{f}}$} with DataFit (trial version) and MATLAB tools and their values are presented in \tab\ref{tab:1}.
\begin{table}[!t]
	\begin{center}
		\caption{Statistical correlation coefficients and parameters.}
		\label{tab:1}
		\scalebox{0.8}
		{
			\begin{tabular}[t]{|l|c|c|c|c|c|c|c|c|} \hline
				\multicolumn{2}{|l|}{Relation}&	$\alpha$  &$\beta$   &$\gamma$ &$\zeta$ & $\delta_{\text{min}} (\%)$ & $\delta_{\text{max}} (\%)$ & $R^2$\\ \hline
%---------------------
		\eqn(\ref{eq:tf})&$A$  &$1.3067$ &$1.7163 \times 10^{-5}$  & $-1.2358 \times 10^{-6}$ &-	&0.0063  &5.8633&0.9992\\
				& $B$    &$0.0421$ &$-1.1329 \times 10^{-7}$ & $-131.5744$  &-	&  && 	\\
				& $C$    &$-1.0260$  &$4.9780 \times 10^{-7}$  & $-4.2857 \times 10^{-8}$  &-	&  && \\ \hline 
%---------------------
			\eqn(\ref{eq:ts})	&$A$    &$7.8576$  &$-0.6008$  & $-30.048$  &-	&0.2204	&6.6781 &0.9929\\
				&$B$    &$0.0019$  &$0.0010$   & $-0.0011$  &-	&  	&&\\
				&$C$    &$-0.0766$  &$-0.0170$  & $6.5148 \times 10^{-5}$  &-	&  &&\\ 
				&$D$    &$5199.1216$  &$9615.2693$  & $1298.1974$  &-	& &&\\ 
				&$E$    &$0.0026$  &$-0.1570$  & $0.0056$  &-	& &&\\ 
				&$F$    &$-7.8343 \times 10^{5}$  &$1.2337 \times 10^4$  & $44.3553$  &-&	&	& \\ 	\hline 
%---------------------				
			\eqn(\ref{eq:tb})		&$A$    &$0.0037$  &$-1.6227 \times 10^{-4}$  & $4.4983 \times 10^{-5}$ &-  &0.2397  &8.1216&0.9940	\\
				&$B$    &$0.0317$  &$-3.1369 \times 10^{-3}$  & $9.0569 \times 10^{-4}$ &- &  && 	\\
				&$C$    &$3.9705$  &$-1.8292$  & $0.2755$ & $-0.0135$ &  &&\\ 
				&$D$    &$-230.58$  &$101.08$  & $-14.533$  & $0.718$ &  &&\\ 
				&$E$    &$3.6861 \times 10^{-5}$  &$-0.0022$  & $-2.7062 \times 10^{-6}$ &-&  && \\ 
				&$F$    &$26.015$  &$-13.193$  & $2.2006$  & $-0.1219$&  && \\ 	\hline	
%---------------------				
				\eqn(\ref{eq:pcpm})&$A$    &$5597.2289$  &$-9.18 \times 10^{-3}$  & $2.83\times 10^{-5}$&-&0.0337  &7.7721& 0.9976 	\\
				&$B$    &$-2.8920 \times 10^{5}$ &$-5.8981 \times 10^{3}$ & $1.9482 \times 10^{-3}$&-&  && 	\\
				&$C$    &$1340.7634$  &$-2.90\times 10^{-2}$  & $2.0 \times 10^{-3}$&-&  &&   \\ 
				&$D$    &$-1840.9596$  &$1572.3580$  & $-20476.5479$&-&  &&   \\ 
				&$E$    &$-1409.3022$  &$9.2296\times 10^{6}$  & $-8.0480 \times 10^{7}$&-&  &&   \\ \hline 
%---------------------				
			\eqn(\ref{eq:t3p})&${A}$  	&$-0.3352$  &$20.4687$  & $-104.5415$ &-& 0.0846 &6.0235&0.9955	\\
			&${B}$    &$-2926.2562$ &$-145.4043$  & $943.3530$&-&  && 	\\
			&${C}$    &$-2.0346$  &$0.3018$ & $2.0033$&-&  && \\ 
			&${D}$    &$6.2976$  &$-0.5135$ & $-27.9788$&-&  &&  \\ 
			&${E}$    &$-2.5380$  &$13.9446$  & $-5390.5463$&-&  && \\
			&${F}$    &$6.5432 \times 10^{-5}$  &$0.3142$  	  &$-0.7944$ &-&  && \\ \hline	
%---------------------				
			\eqn(\ref{eq:2rm})&${A}$    &$1.3166$  &$-19.3372$  & $10786.2346$ &-&0.0031  &4.9505&0.9516	\\
				&${B}$    &$-134.1631$ &$-7.8858$  	& $69.9675$ &-&  &&	\\
				&${C}$    &$-183.6308$  &$-46.2430$  & $135.6212$ &-&  &&  \\ 
				&${D}$    &$0.1515$  &$-3.8498$  	& $2.5792$ &-&  &&  \\ 
				&${E}$    &$-0.1871$  &$-1.8985 \times 10^{-4}$  & $0.0273$ &-&  &&\\
				&${F}$    &$-1.5830 \times 10^{-5}$  &$-0.0136$  	&$4.9666 \times 10^{-5}$ &-&  && \\	\hline 
			\end{tabular}
		}
	\end{center}
\end{table}

\noindent
The growing dispersed phase moves downstream as time progresses, and this stage is called squeezing (S\rev{2}) as the surrounding fluid starts exerting viscous shear on the interface due to the pressure developed upstream of the main channel. The dispersed phase now has two sides: front and rear. The front side shape does not change with time and position; however, the interface curvature changes from convex to concave on the rear side. For a smaller contact angle, the interaction between the channel wall surface and droplet becomes stronger. Hence, the squeezing time ($t_{\text s}=t_2-t_1$) increases with the contact angle as the surface offers higher resistance. 

\noindent
The time required by the squeezing stage $t_{\text{s}}=(t_2-t_1)$ is best correlated as a function of $\qr$, $\cac$ and $\theta$ ($0.125\le\qr\le 10$, $120^{\circ} \leq \theta \leq 180^{\circ}$, and $10^{-4}\le \cac \le 10^{-3}$) as follows. 
\begin{gather}
	t_{\text{s}} = X_{\text{s}}t_{\text{f}} \label{eq:ts}
\end{gather} 
where  
\begin{gather*} 
	X_{\text{s}} = A+B/\cac+C\qr+D\qr^2+E\qr^3+F\qr^4, \\
A=\alpha+\beta m/\log(m) +\gamma m^{-2}\log (m),\qquad 
B=\alpha+\beta m^{2.5}+\gamma e^{m}, \qquad
C=\alpha/(1+\beta \theta+\gamma \theta^2), \\
D=\alpha+\beta m\log (m)+\gamma \log (m), \qquad
E=\alpha/(1+\beta q +\gamma q^2),\qquad
F=\theta/(\alpha+\beta \theta-\gamma \theta^2)\\
m=10^{-3}\theta,\qquad q=10^{-1}\theta 
\end{gather*}
Statistical non-linear regression analysis is performed to obtain the correlation coefficients \rev{of $X_{\text{s}}$} with DataFit (trial version) and MATLAB tools and their values are presented in \tab\ref{tab:1}.

\noindent 
On moving further, the interface on the rear side of the dispersed phase collapses as time progresses and forms a droplet. Hence, this stage is called pinch-off (S\rev{3}). The degree of confinement of the droplets reduces when the shear forces dominate the breakup process \citep{Glawdel2012a,Bashir2014}. For larger $\theta$, the degree of confinement promotes the breakup, whereas for smaller $\theta$, the degree of confinement suppresses the breakup. The wall provides less resistance to the droplet in the superhydrophobic regime and accelerates the droplet formation. It can be observed that the droplet shape is also changing when the surface wettability is different. Therefore, each stage of the droplet formation is directly influenced by the contact angle. 

\noindent
The time required by the spontaneous pinch-off (S3) stage is $t_{\text{b}}=(t_3 - t_2)\lll 1$, and best correlated as a function $\qr$, $\cac$ and $\theta$ ($0.125\le\qr\le 10$, $120^{\circ} \leq \theta \leq 180^{\circ}$, and $10^{-4}\le \cac \le 10^{-3}$) as follows:

\begin{gather}
	t_{\text{b}}= X_{\text{b}}t_{\text{s}} \label{eq:tb}
\end{gather}
where,  
\begin{gather*}
X_{\text{b}} =A+B/x_1+C/\qr+D/x_1^2+E/\qr^2+F/(x_1 \qr),\\
A=\alpha+\beta/m^{1.5}+\gamma/m^2, \qquad
B=\alpha+\beta/m^{1.5}+\gamma/m^2,  \qquad
C=\alpha m^3+\beta m^2+\gamma m +\zeta, \\
D=\alpha m^3+\beta m^2+\gamma m +\zeta,  \qquad
E=\alpha+\beta s^3+\gamma/s^{1.5},  \qquad
F=\alpha m^3+\beta m^2+\gamma m +\zeta, \\
x_1=\log\cac,\qquad m=10^{-3}\theta,\qquad s=\log\theta
\end{gather*}
Statistical non-linear regression analysis is performed to obtain the correlation coefficients \rev{of $X_{\text{b}}$} with DataFit (trial version) and MATLAB tools and their values are presented in \tab\ref{tab:1}.

\noindent 
After the pinch-off, the droplet attains a stable shape and does not change anymore when it is allowed to move downstream further as the development of the droplet size is completely achieved, and this stage is called the stable droplet \rev{(S4)}. The stable droplet time is defined as, $t_{\text {sd}}=t_4-t_3$. 

\noindent
It is, therefore, concluded that the total time needed for one complete cycle \rev{(S0 to S3)} of the droplet breakup or pinch-off as $t_{\text{p}}=(t_{\text{f}} + t_{\text{s}}  + t_{\text{b}})$ and that of the stable droplet, $t_{\text{d}}=(t_{\text{p}} +t_{\text{sd}})$. The flow of CP and DP continues during the process of droplet formation. However, the time needed for each droplet formation stage depends upon the capillary number ($\cac$) and flow rate ratio ($\qr$), and contact angle ($\theta$). Further insights of droplet formation and its dynamics are provided in the following section in terms of the evolution of interface.
\begin{figure}[hpbt]
	\centering
	\subfloat[$\cac=10^{-4}$]{\includegraphics[width=0.49\linewidth]{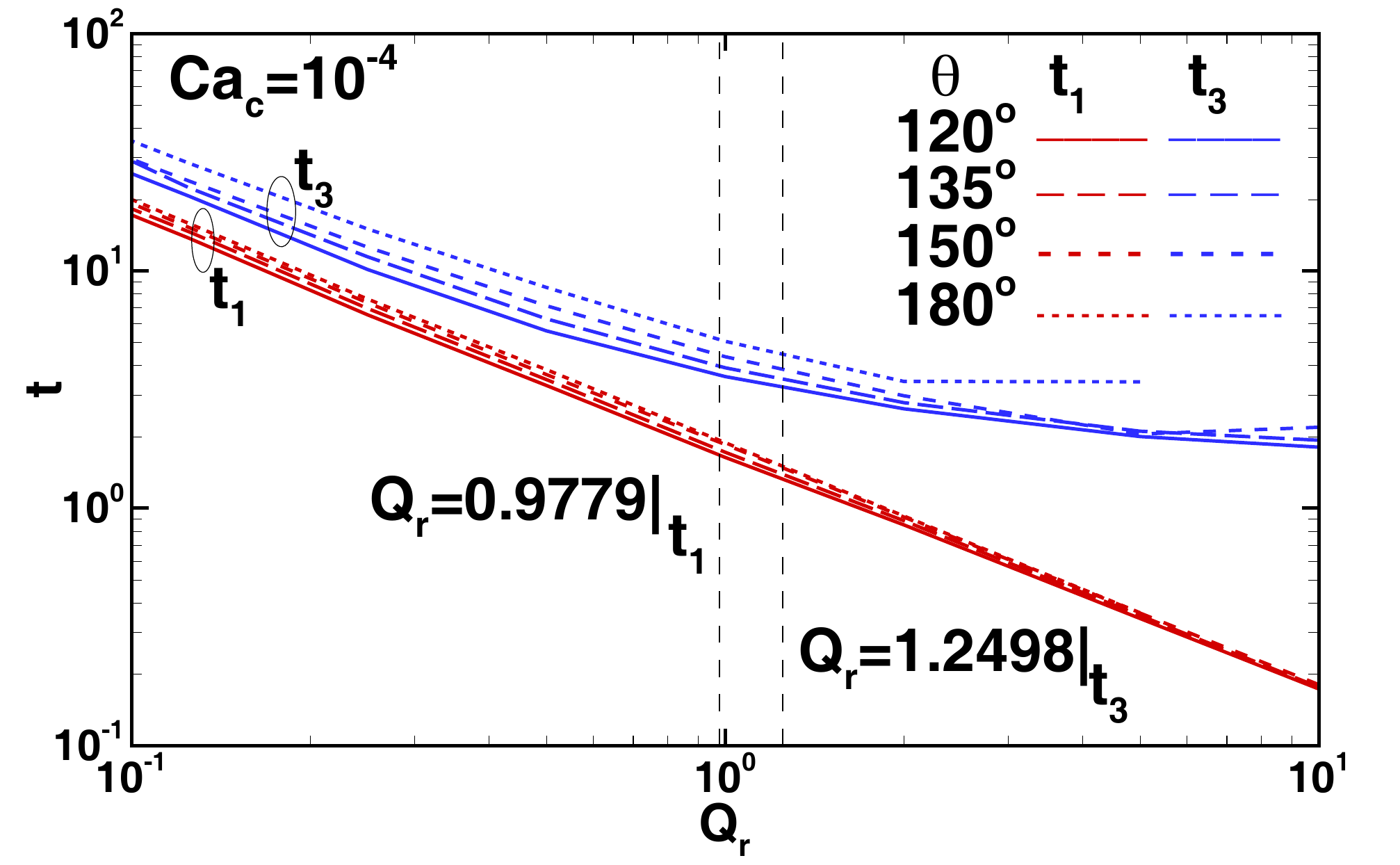}\label{fig:2aa}}
	\subfloat[$\cac=10^{-3}$]{\includegraphics[width=0.49\linewidth]{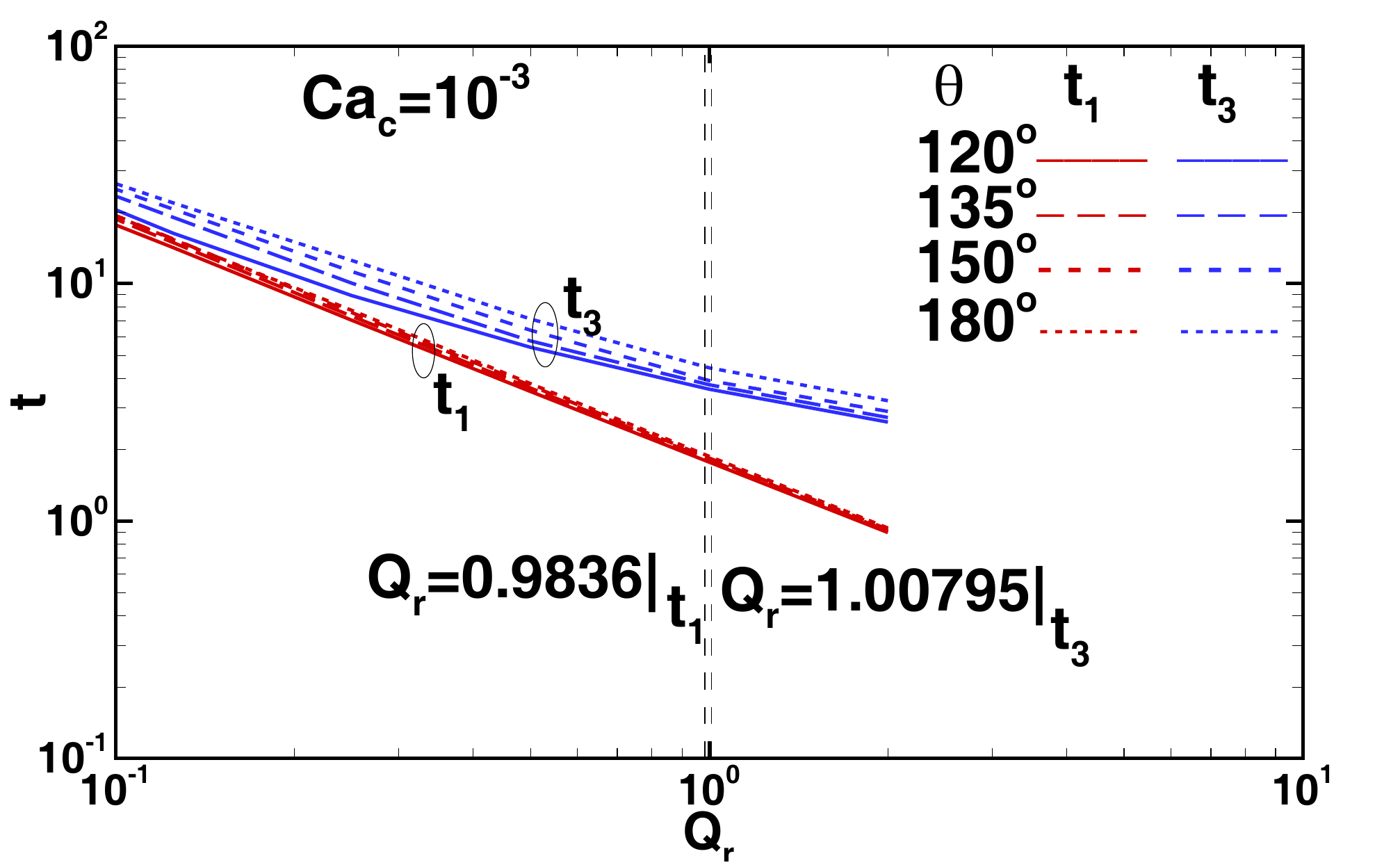}\label{fig:2ab}}\\
	\subfloat[\rev{$\cac=10^{-4}$}]{\includegraphics[width=0.49\linewidth]{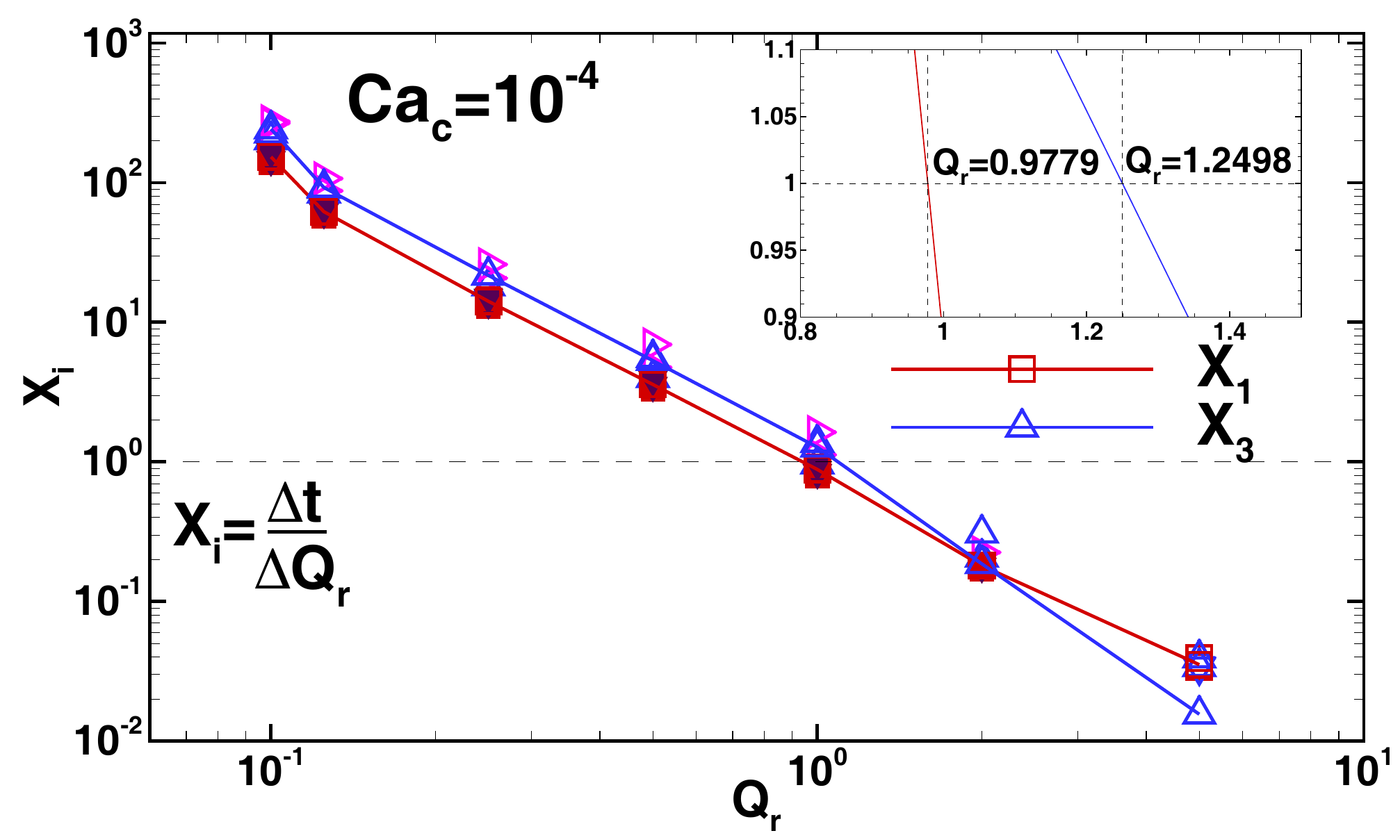}\label{fig:2ac}}
	\subfloat[\rev{$\cac=10^{-3}$}]{\includegraphics[width=0.49\linewidth]{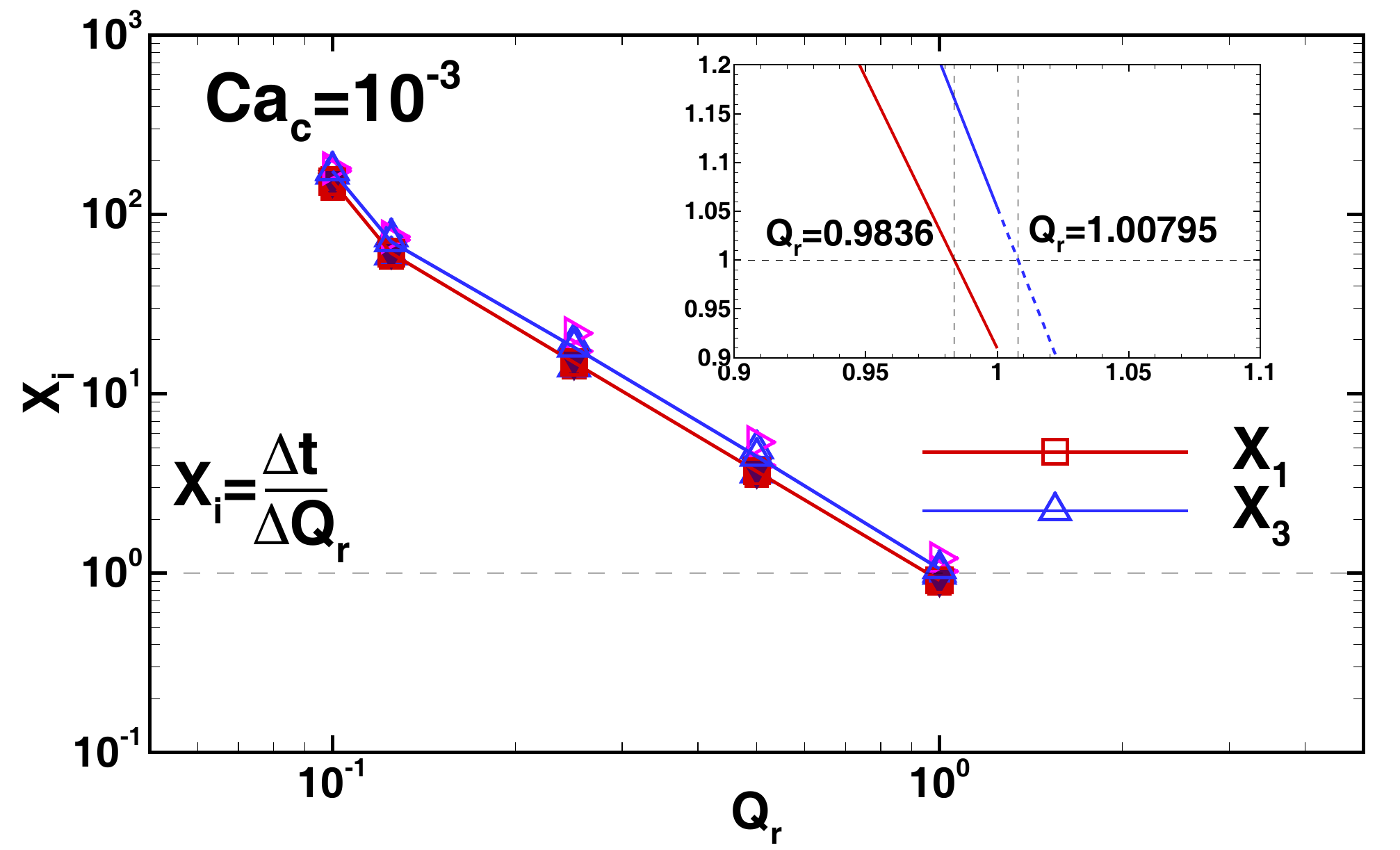}\label{fig:2ad}}\\
	\subfloat[\rev{$\cac=10^{-4}$}]{\includegraphics[width=0.5\linewidth]{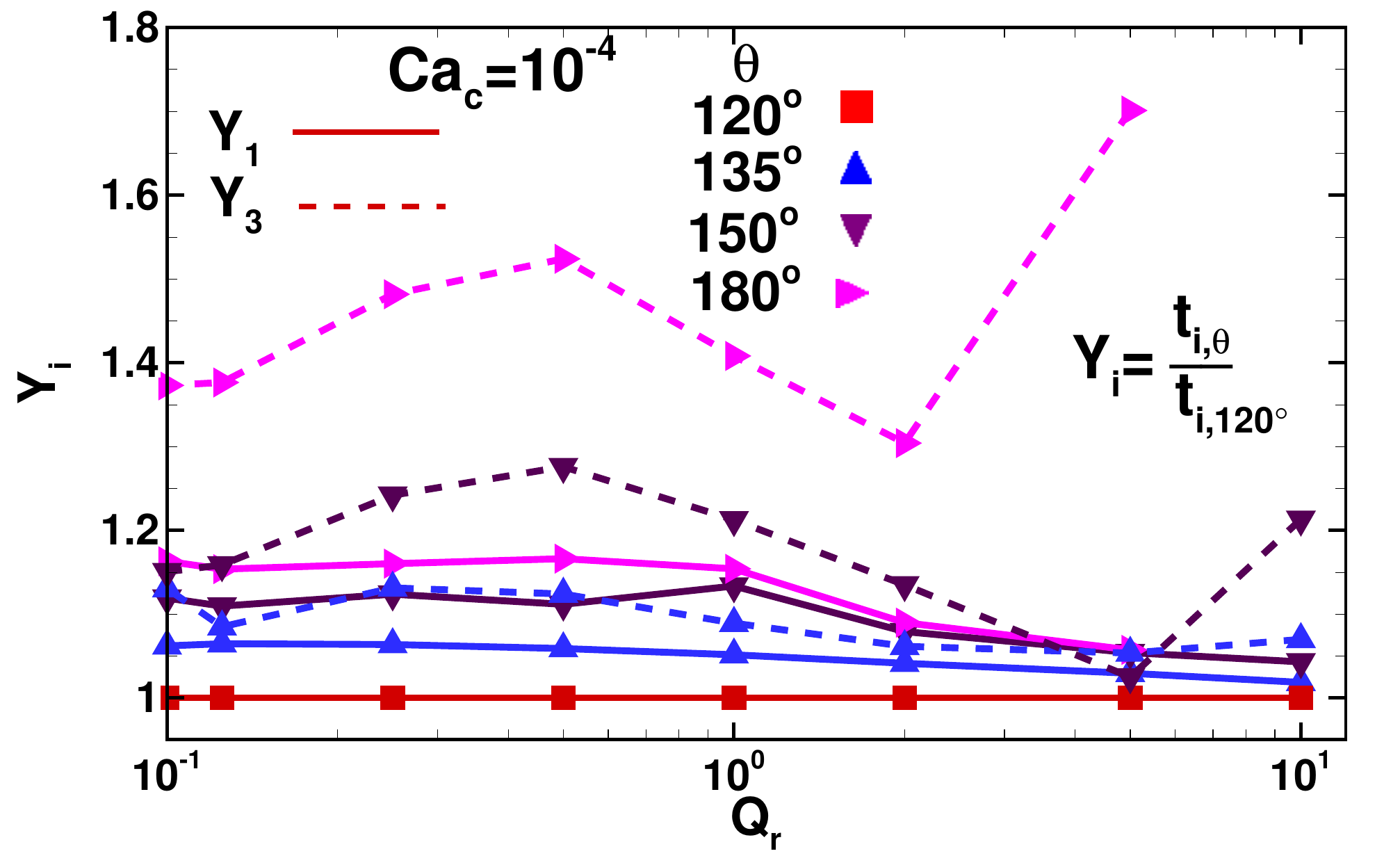}\label{fig:2ae}}
	\subfloat[\rev{$\cac=10^{-3}$}]{\includegraphics[width=0.5\linewidth]{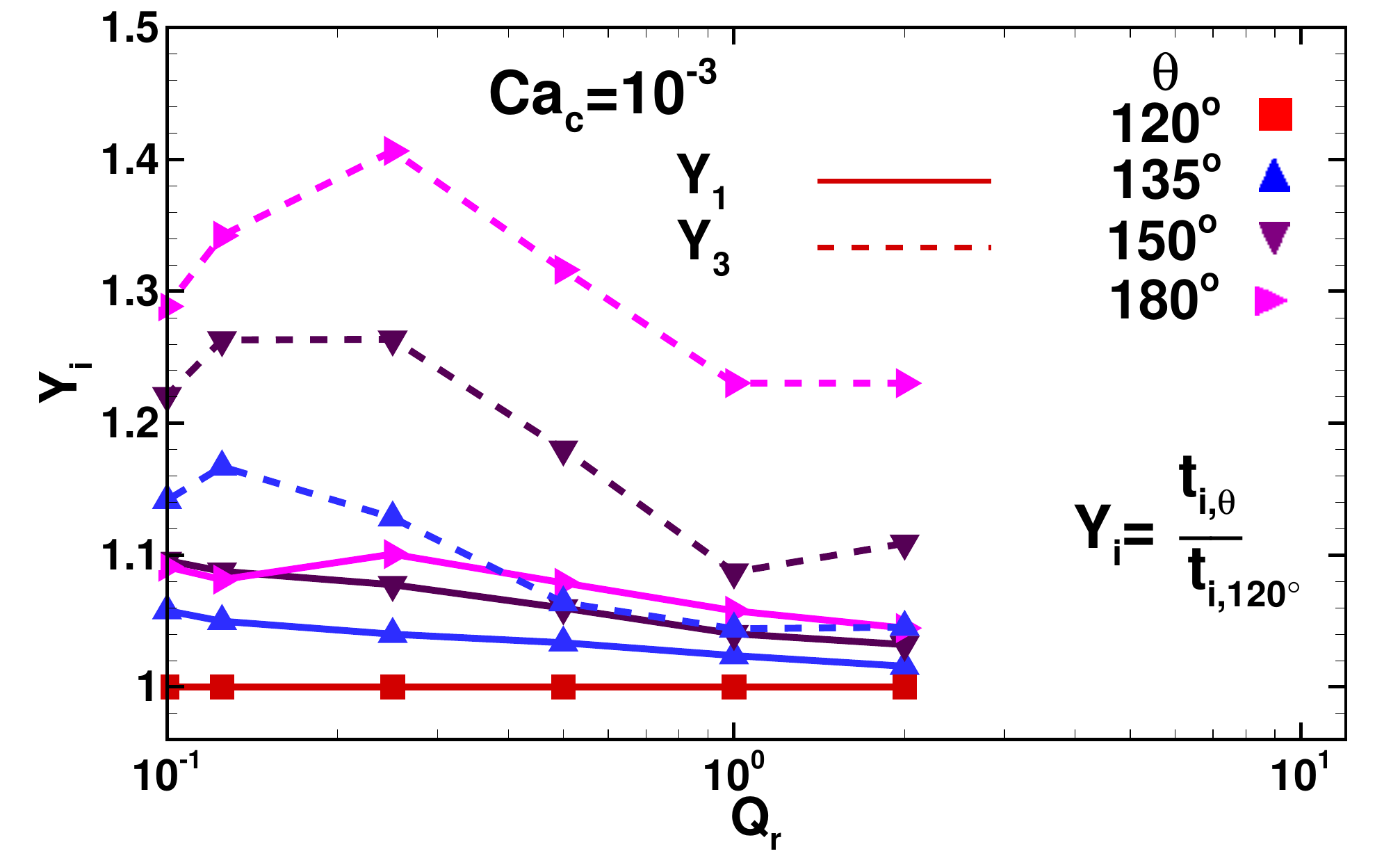}\label{fig:2af}}\\
	\caption{Flow map, time ratio ($Y_{\text i}$), gradient ($X_{\text i}$) of filling and pinch-off stages of the droplet as a function of contact angle ($\theta$) and flow rate ratio ($\qr$).}
	\label{fig:2a}
\end{figure}

\noindent
Overall, the flow map presents the droplet formation stages as a function of the contact angle and capillary number; both the filling ($\text t_1$) and pinch-off ($\text t_3$) time decrease with an increase in the flow rate ratio \rev{for a given contact angle}, as shown in \figs\ref{fig:2aa} and \ref{fig:2ab}. 
It can be observed that the contact angle effect is \rev{marginal} in the filling stage because of the less dominance of surface forces; however, the contact angle effect is dominant in the pinch-off stage as the interaction between the three phases: CP, DP, and the wall, is substantial. 
\rev{Further, a gradient ($X_{\text i}=\Delta t_{\text i}/\Delta Q_r$) has been plotted against the flow rate ratio ($Q_r$) in \figs\ref{fig:2ac} and \ref{fig:2ad} to elaborate the influence of $Q_r$ and $\theta$ on $\text t_1$ and $\text t_3$. The $X_{\text i}$ has shown inverse dependence on $Q_r$ and proportional dependence on $\theta$.} 
Both $X_1$ and $X_3$ decrease \rev{non-linearly (i.e., exponential or power-law)} up to a threshold value  \rev{of $\qr=0.9779$ and $\qr=1.2498$} for $\cac=10^{-4}$, and \rev{similarly, $\qr=0.9836$ and $\qr=1.00795$} for $\cac=10^{-3}$ (refer dashed line in \figs\ref{fig:2aa} to \ref{fig:2ad}) \rev{where the gradient becomes unity ($X_{\text i}=1$) for all values of contact angle}.
\rev{Beyond these critical values of $\qr$, the filling time is slightly influenced by the contact angle. However, the pinch-off time is strongly influenced by the superhydrophobic ($\theta\ge 150^{\circ}$) nature of the surface.}

\noindent
\rev{Furthermore,  time ratio ($Y_{\text i}=t_{\text i, \theta}/t_{\text i, 120^{\circ}}$) parameter is introduced and plotted in \figs\ref{fig:2ae} and \ref{fig:2af} to elucidate the effect of the contact angle on filling ($\text t_1$) and pinch-off ($\text t_3$) time.  It can clearly be observed that both filling and pinch-off time strongly depend on the contact angle, as reflected by $Y_{\text i}>1$, in contrast to non-noticeable effects seen in \figs\ref{fig:2aa} to \ref{fig:2ad}. A difference of about 5-10 \% between the estimates of $\text t_1$ and $\text t_3$ can be noticed vividly. It may thus be concluded that both flow rate ratio and contact angle strongly influence the droplet formation cycle. The evolution of interface is further explored in the subsequent section.}
\begin{figure}[!t]
	\centering
	\subfloat[Filling, $\theta=120^{\circ}$]{\includegraphics[width=0.45\linewidth]{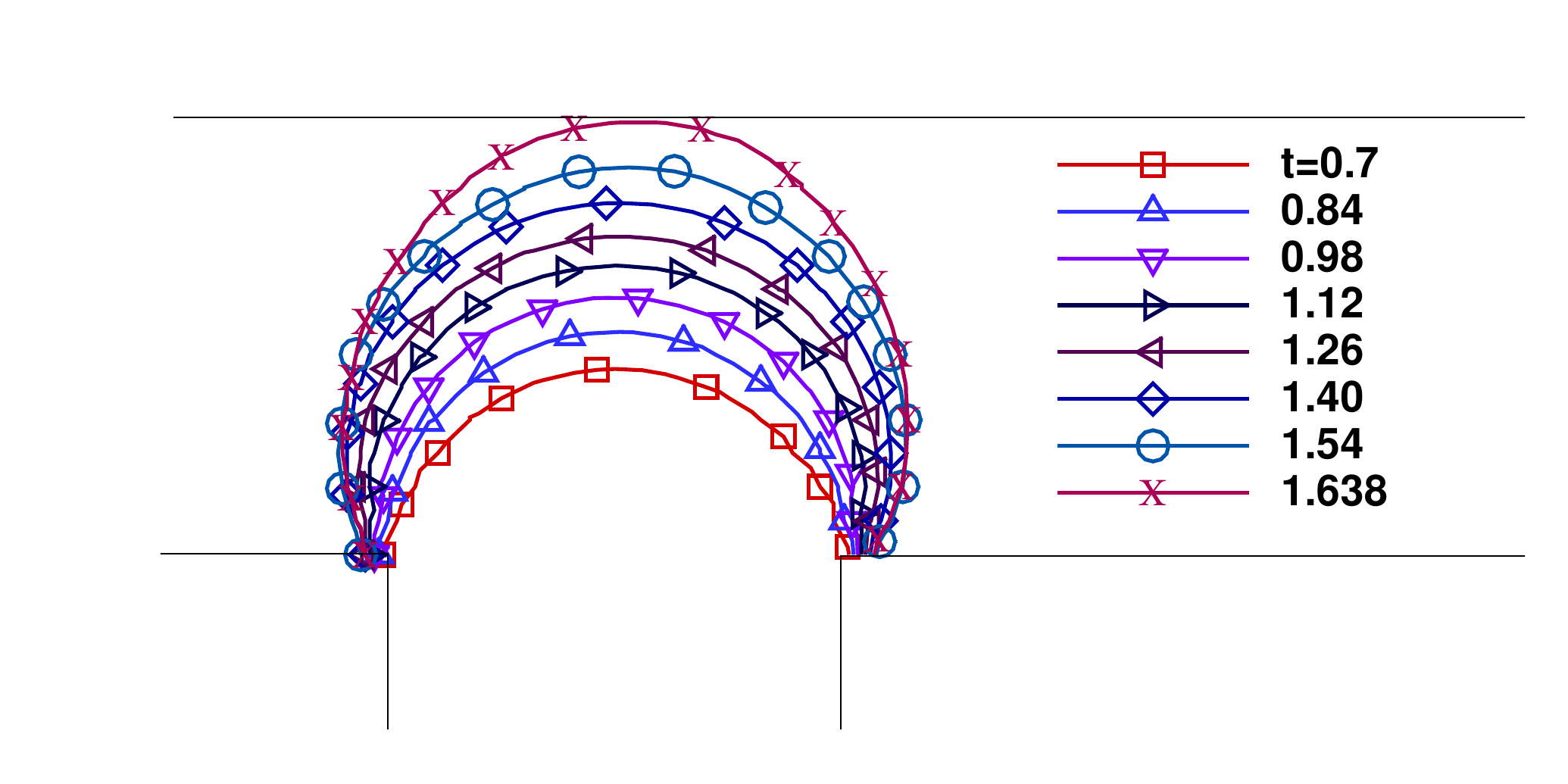}}
	\subfloat[Squeezing, $\theta=120^{\circ}$]{\includegraphics[width=0.45\linewidth]{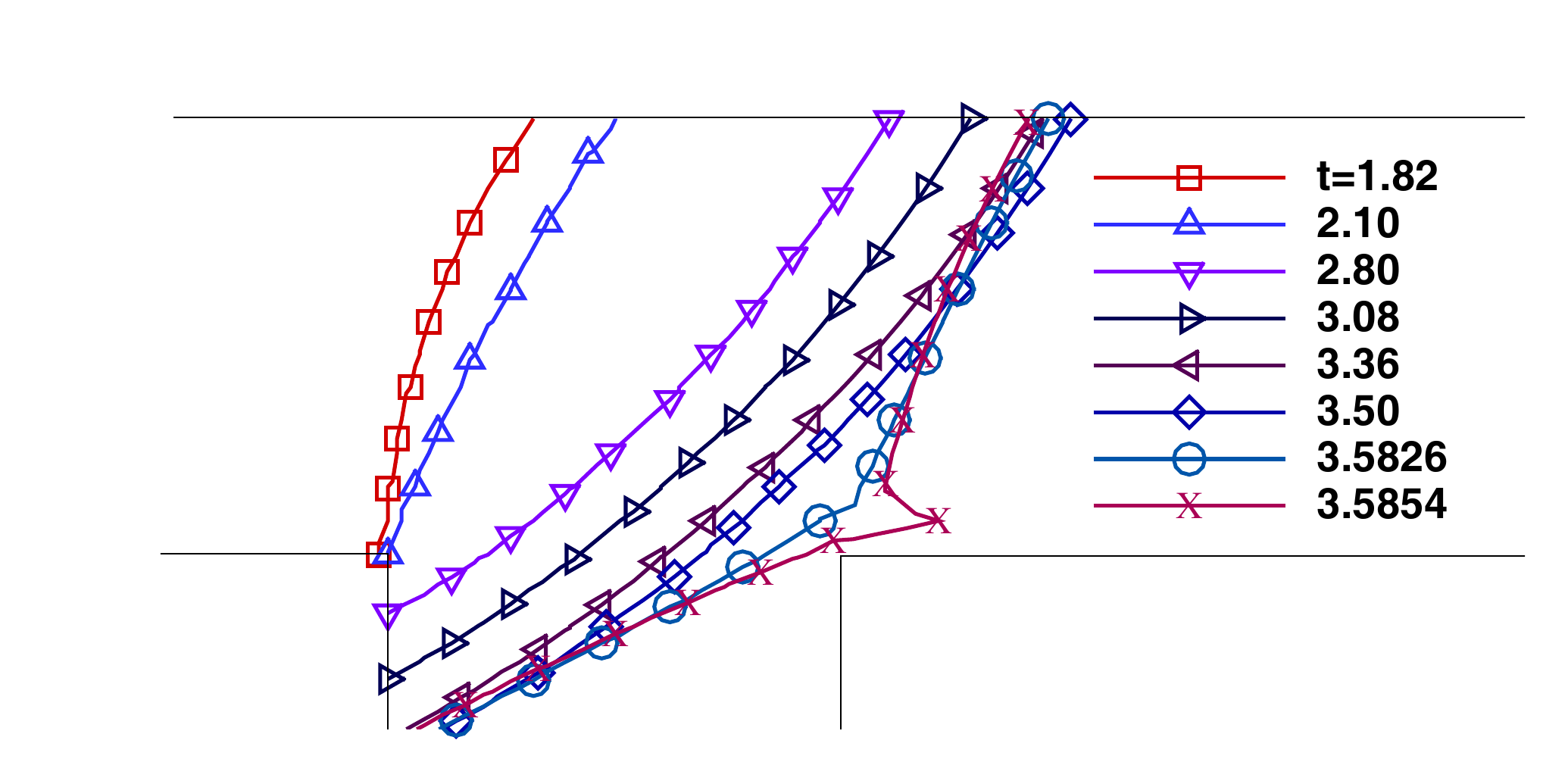}}\\
	\subfloat[Filling, $\theta=135^{\circ}$]{\includegraphics[width=0.45\linewidth]{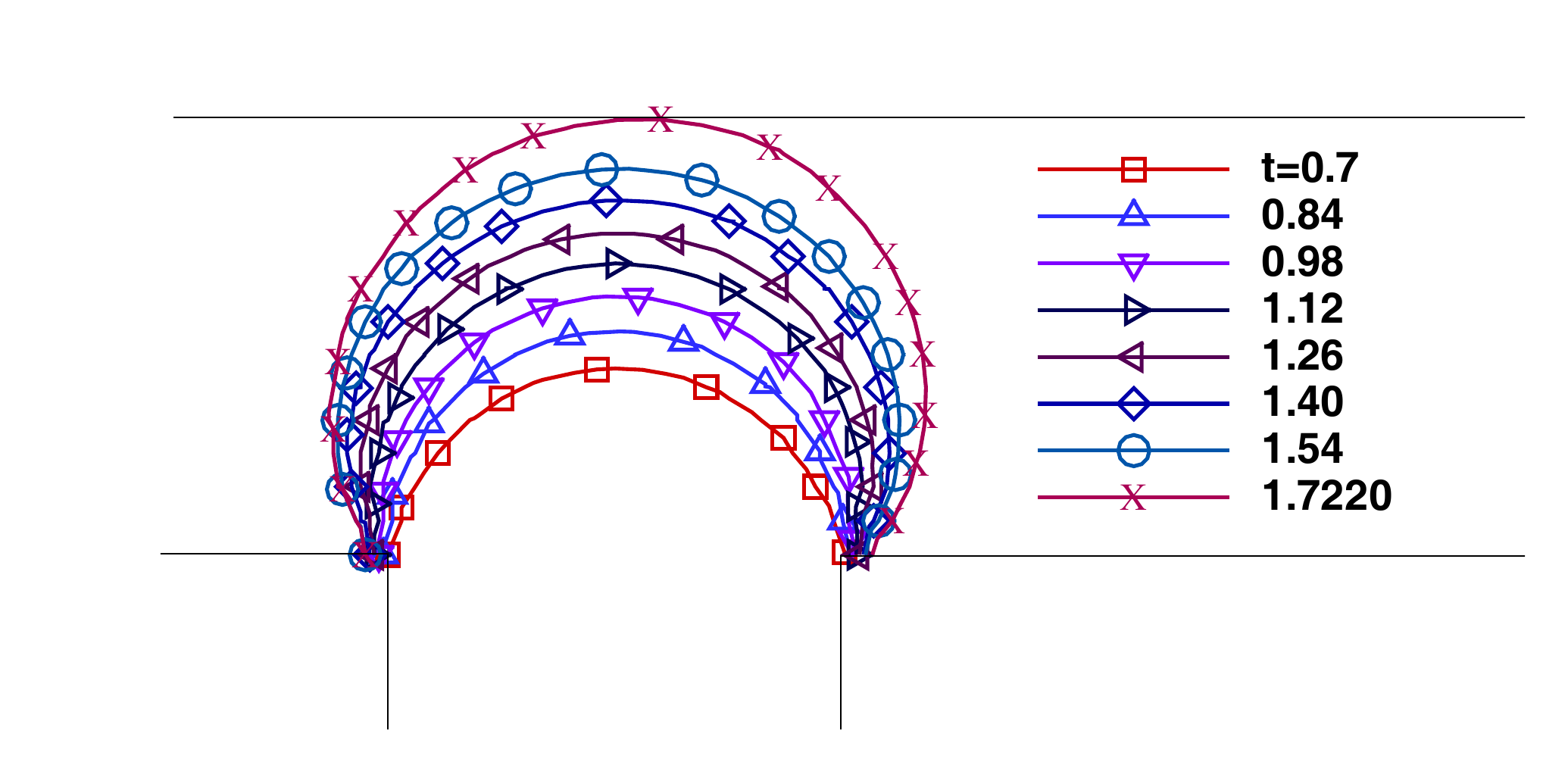}}
	\subfloat[Squeezing, $\theta=135^{\circ}$]{\includegraphics[width=0.45\linewidth]{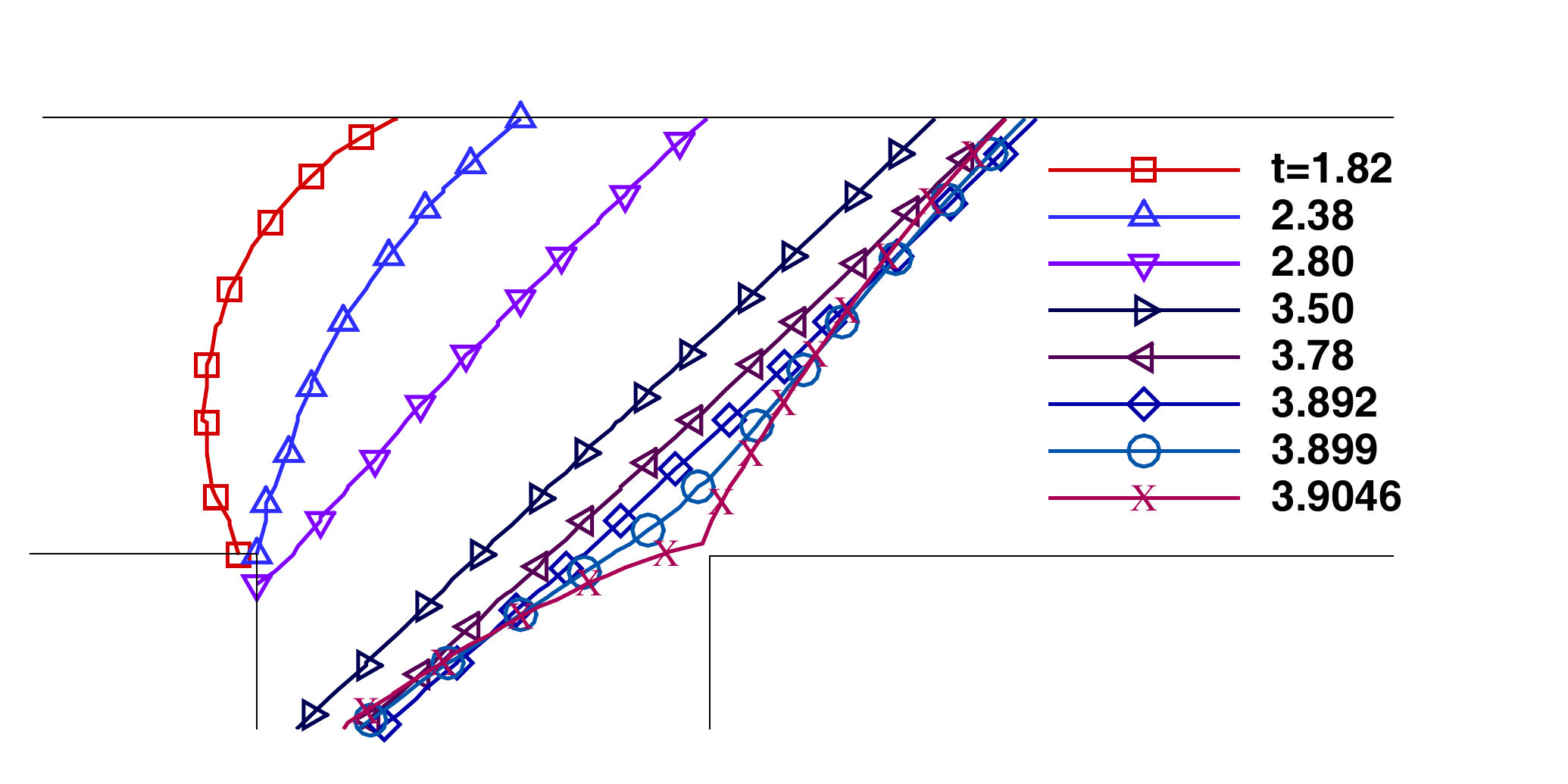}}\\
	\subfloat[\rev{Filling, $\theta=150^{\circ}$}]{\includegraphics[width=0.45\linewidth]{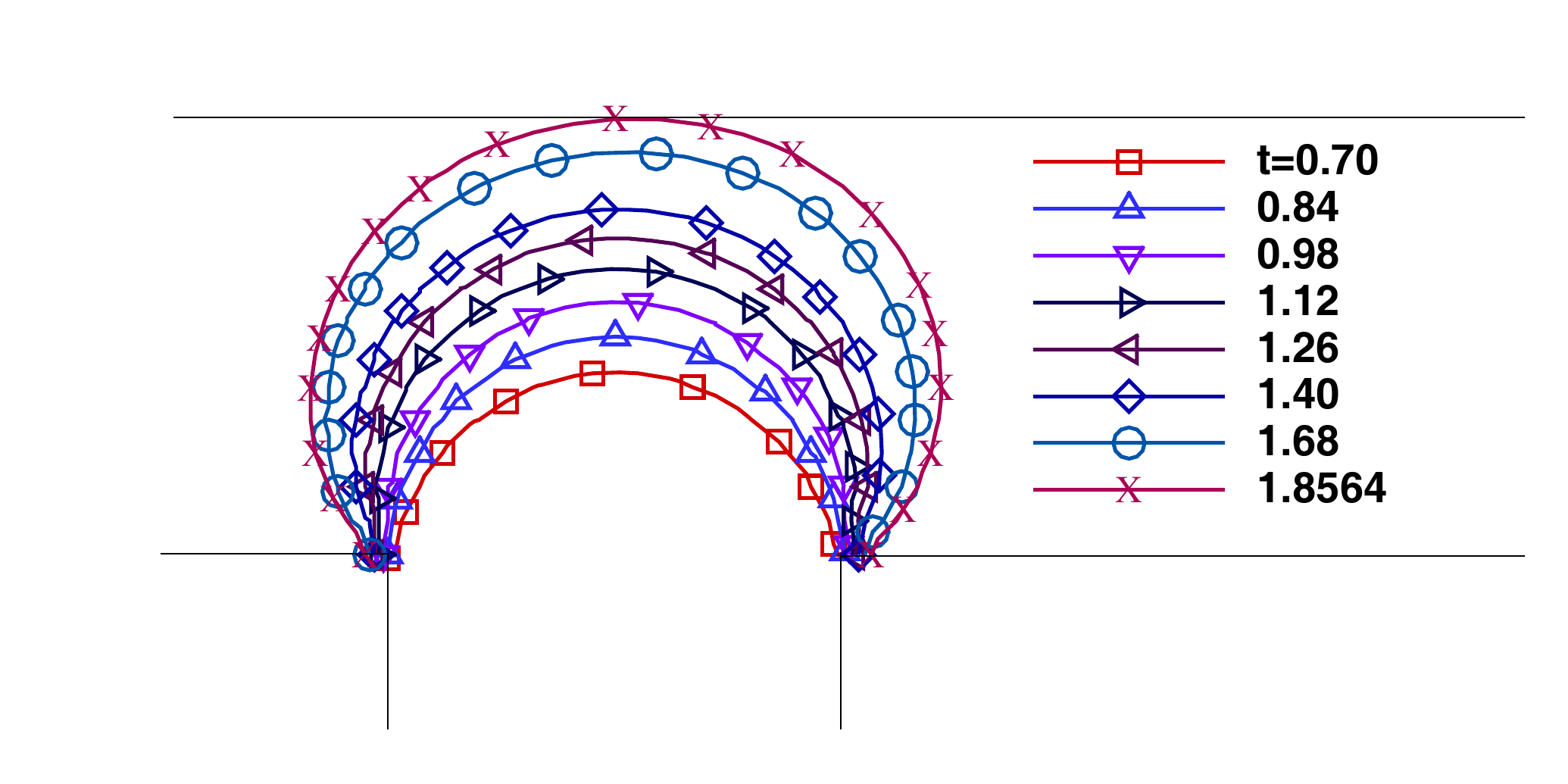}}
	\subfloat[\rev{Squeezing, $\theta=150^{\circ}$}]{\includegraphics[width=0.45\linewidth]{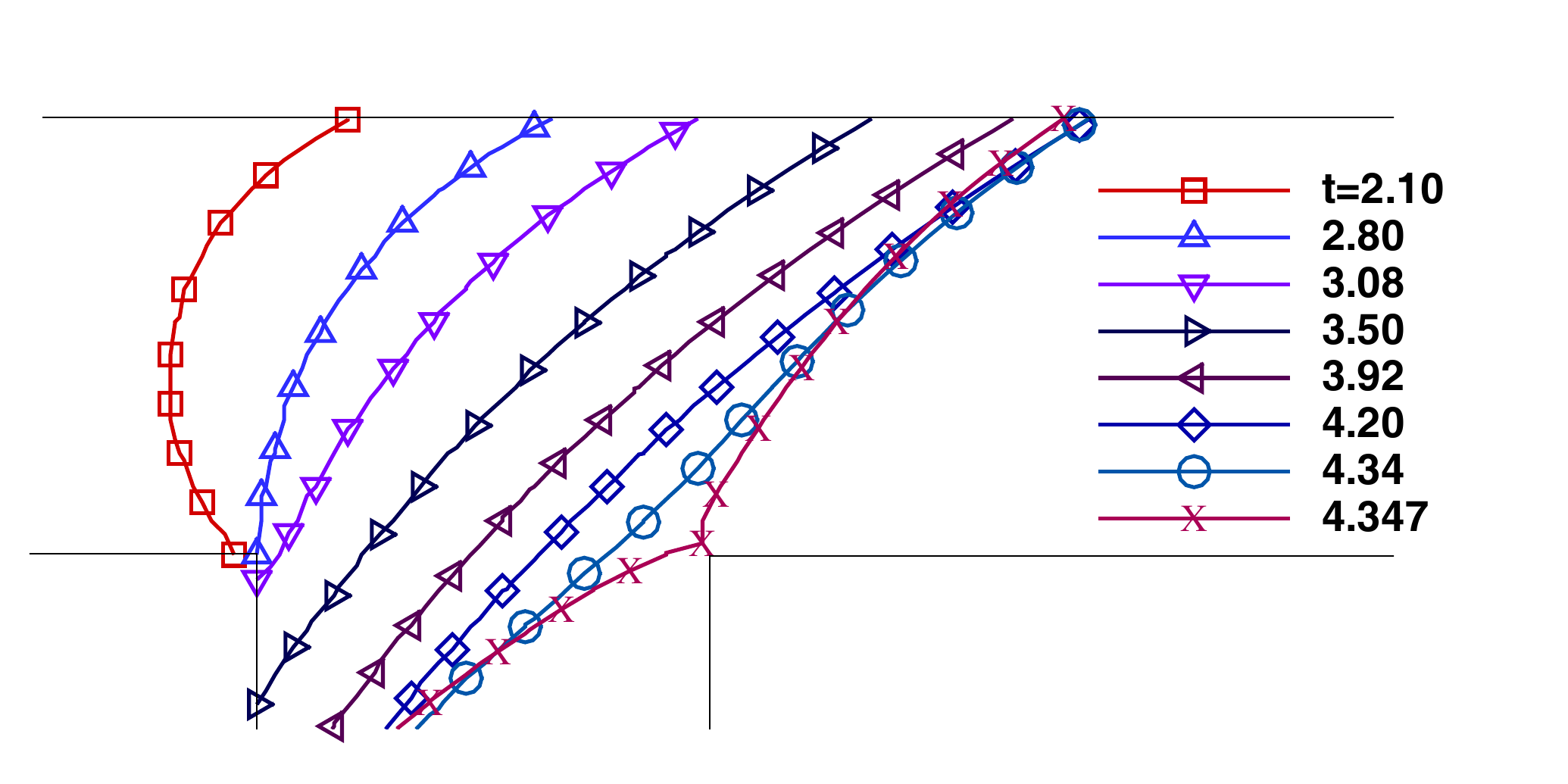}}\\
	\subfloat[\rev{Filling, $\theta=180^{\circ}$}]{\includegraphics[width=0.45\linewidth]{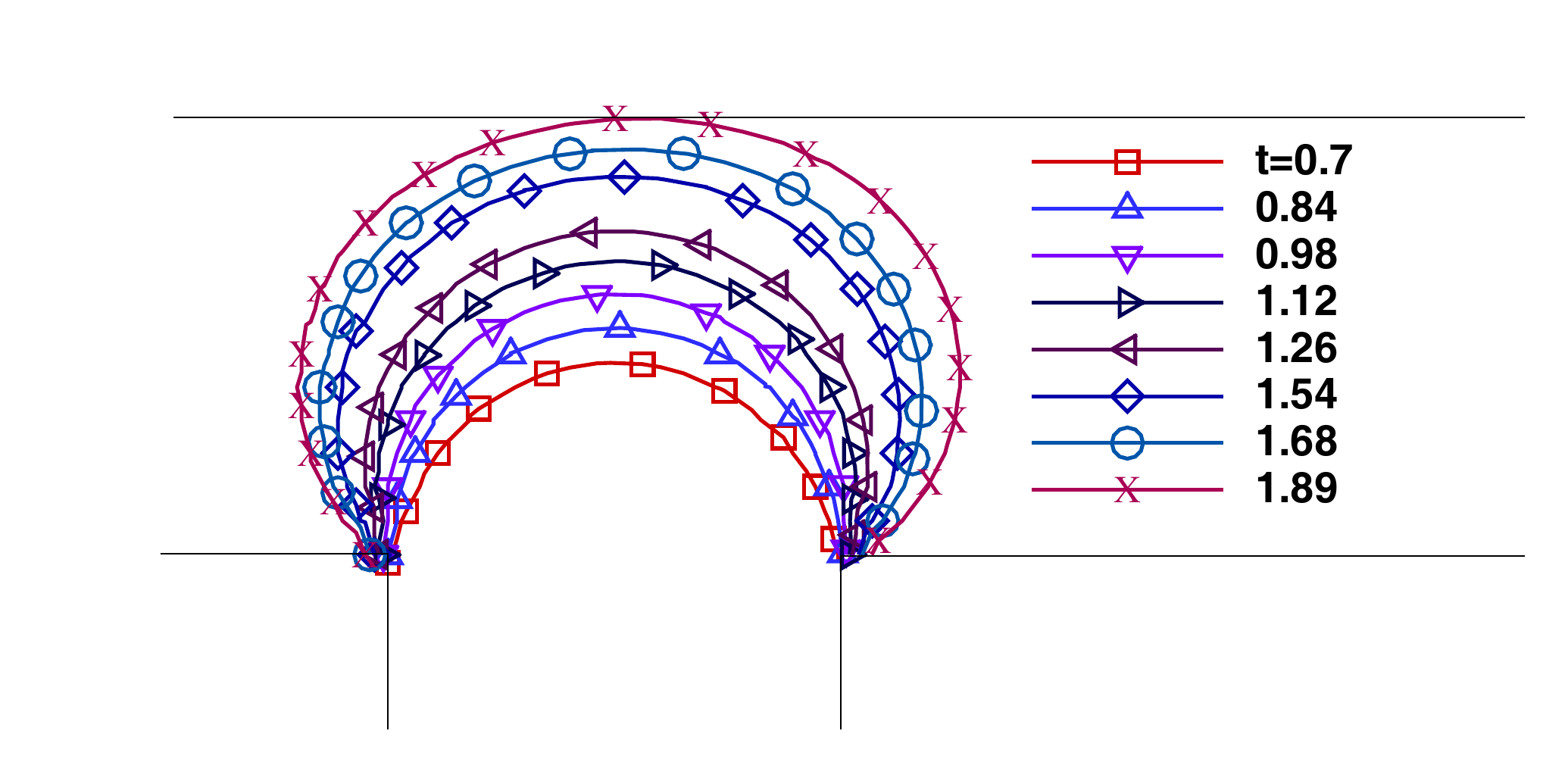}}
	\subfloat[\rev{Squeezing, $\theta=180^{\circ}$}]{\includegraphics[width=0.45\linewidth]{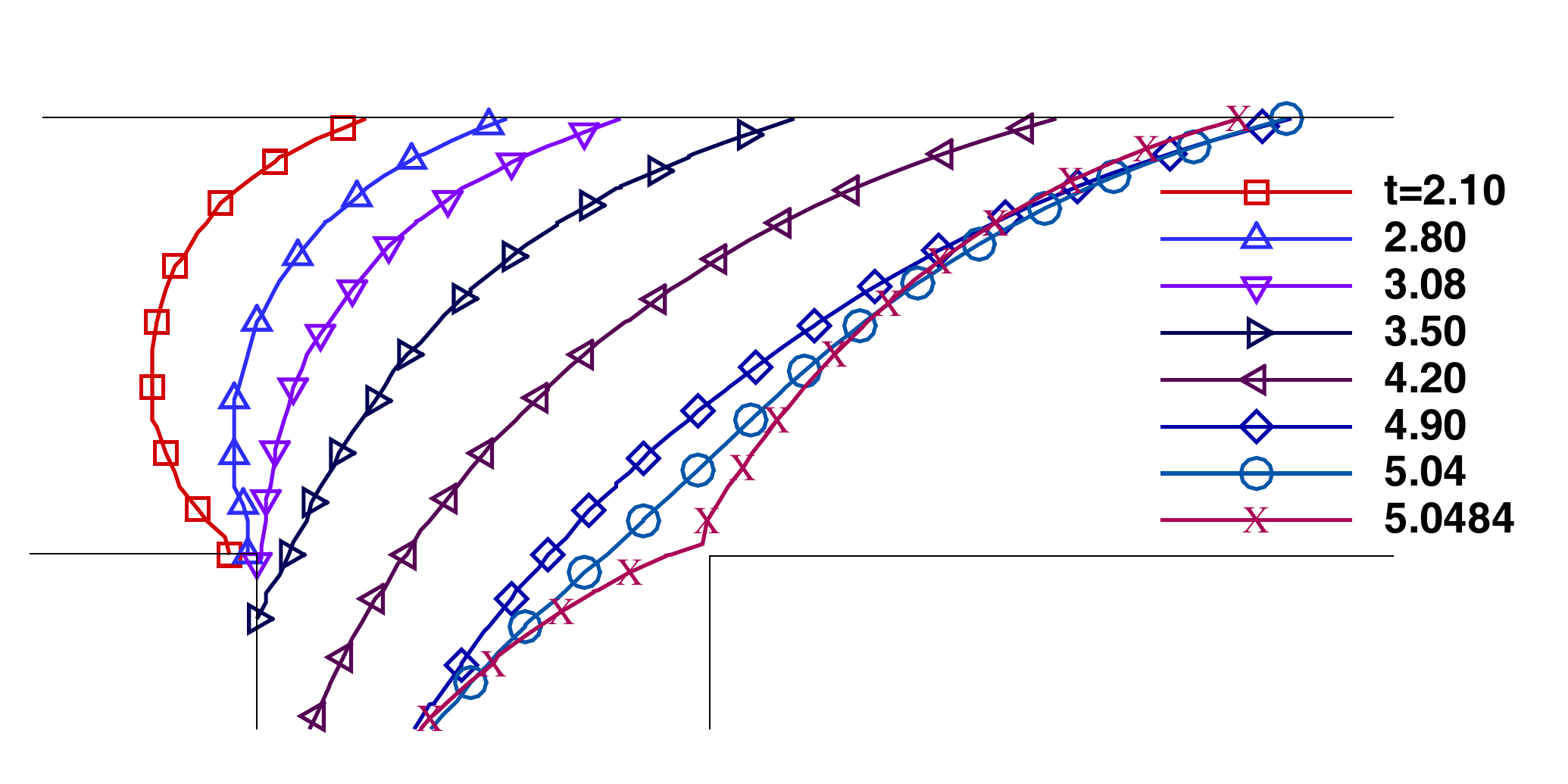}}
	\caption{Evolution of the interface during the filling and squeezing stage as a function contact angle ($\theta$) for $\cac=10^{-4}$ at $\qr=1$}
	\label{fig:3}
\end{figure}
%
%%%%%%%%......................%%%
\subsection{Evolution of the instantaneous interface}
%%........................%%
\noindent The evolution of the interface \rev{($x$ vs $y$ at $\phi=0.5$)} profile in the filling and squeezing stages with the time are depicted for $\qr=1$ at $\cac=10^{-4}$ in \fig\ref{fig:3}. The numerical data of curvature of the interface has been extracted and plotted as a function of time and contact angle. Depending upon the contact angle, the interface attains a specific shape and shows \rev{distinct} variation accordingly. In the first stage (filling), the droplet is expanding without showing much movement towards the downstream direction. It can be observed that at $\theta=120^{\circ}$, the filling time is equal to $1.638$ (refer to \fig\ref{fig:3}a), and for $\theta=180^{\circ}$, it has increased to \rev{1.890} (refer to \fig\ref{fig:3}e).  The interface is evolving slowly at $120^{\circ}$ and becoming smooth with an increase in the contact angle. The continuous phase is wetting channel walls more than the dispersed phase. The surface energy of the solid walls \rev{in contact with the fluid} \rev{de}creases with an increase in the contact angle from $\rev{120}^{\circ}$ to $180^{\circ}$.  \rev{Because the dispersed phase entering the horizontal channel would take up more fluid in the droplet formation owing to the force due to the contact angle that further leads to attachment to the bottom surface rather than to the top surface.} Hence, the filling time is increasing from \rev{$120^{\circ}-180^{\circ}$} due to the high resistance \rev{encountered by the dispersed phase} {from the solid surface} and changing the evolution of the interface. Thus, the total time for the interface to touch the opposite wall of the channel is also increasing.

\noindent 
Further, the dispersed phase starts moving downstream in the squeezing stage as the surrounding fluid pushes it, as shown in \figs\ref{fig:3}b, d, f and h. The evolution of the rear side of the interface is plotted as a function of time. At lower contact angles, $\theta=120^{\circ}$, the interface is seen in a convex shape early, which quickly transforms into concave (refer to \fig\ref{fig:3}b). The interface attaining a concave shape lately at higher contact angles, $\theta =135^{\circ}-180^{\circ}$ (refer to \fig\ref{fig:3}d, f and h). Therefore, it can be concluded that the interactive force between the channel wall and dispersed phase is getting reduced. The interface shape with larger curvature is attributed mainly due to the dominance of the capillary forces at the higher contact angles ($\theta =135^{\circ}-180^{\circ}$). 
\begin{figure}[!t]
	\centering
	\subfloat[Filling, $\qr=10$]{\includegraphics[width=0.45\linewidth]{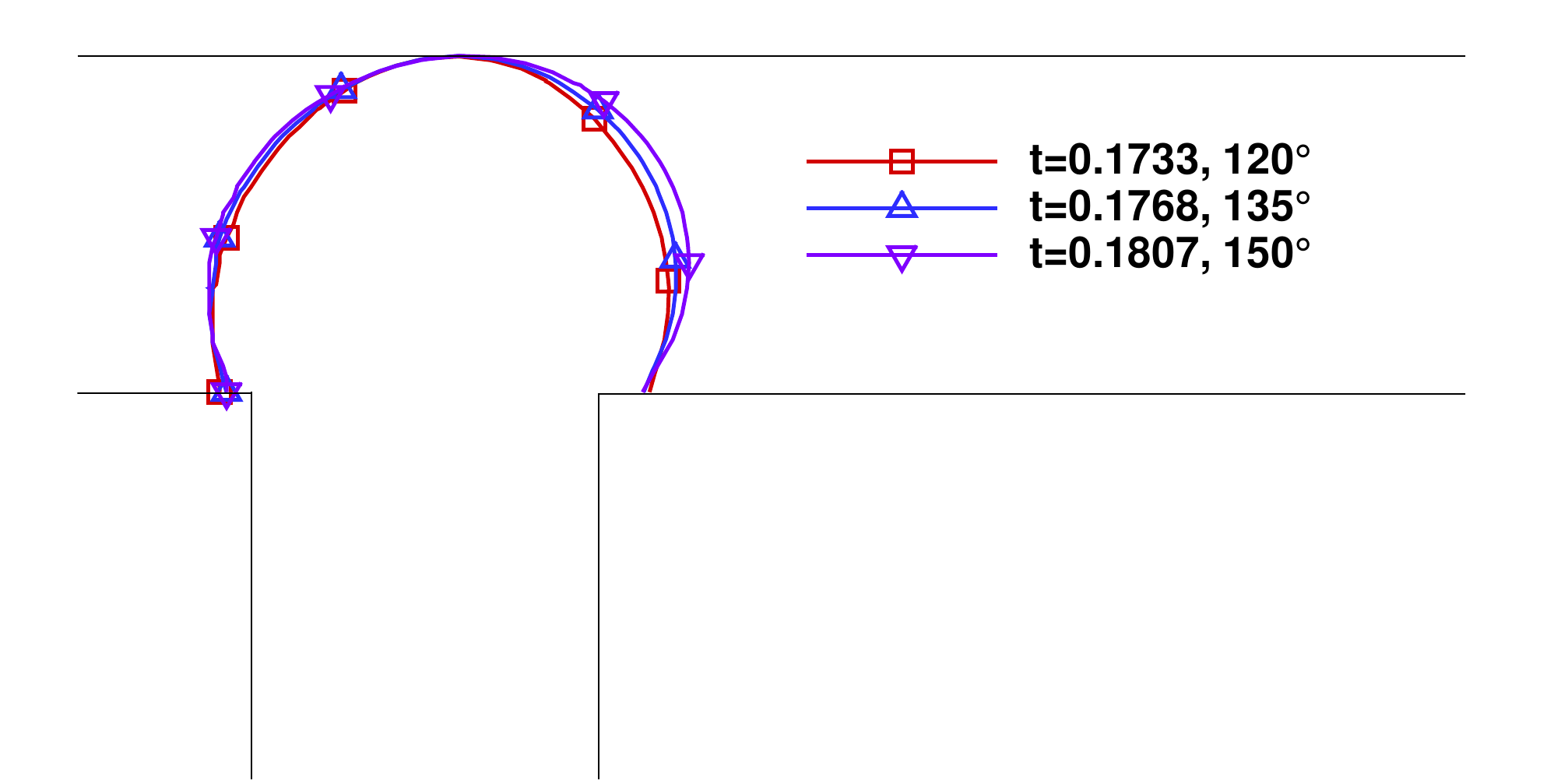}}
	\subfloat[Squeezing, $\qr=10$]{\includegraphics[width=0.45\linewidth]{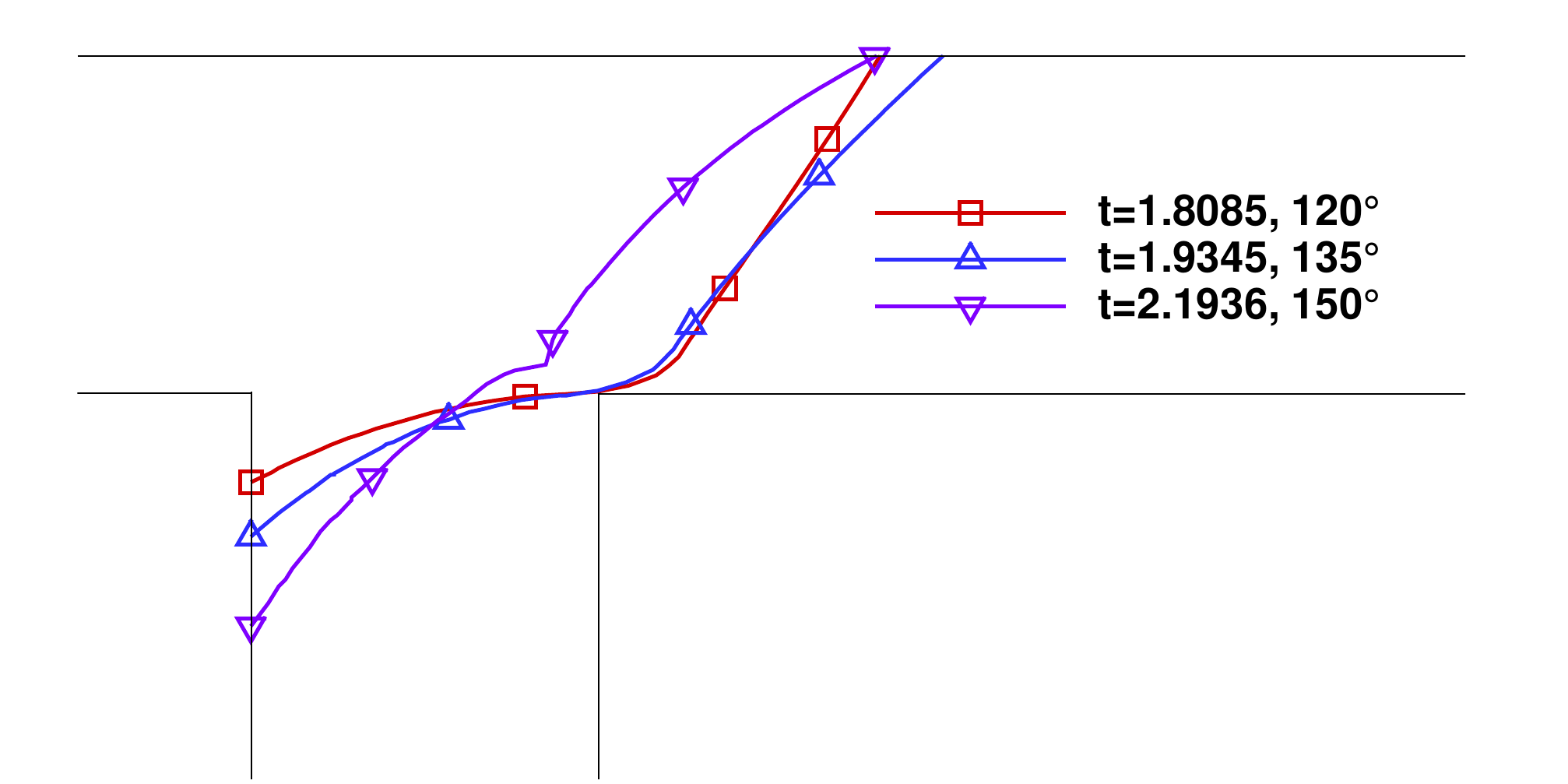}}\\
	\subfloat[\rev{Filling, $\qr=1$}]{\includegraphics[width=0.45\linewidth]{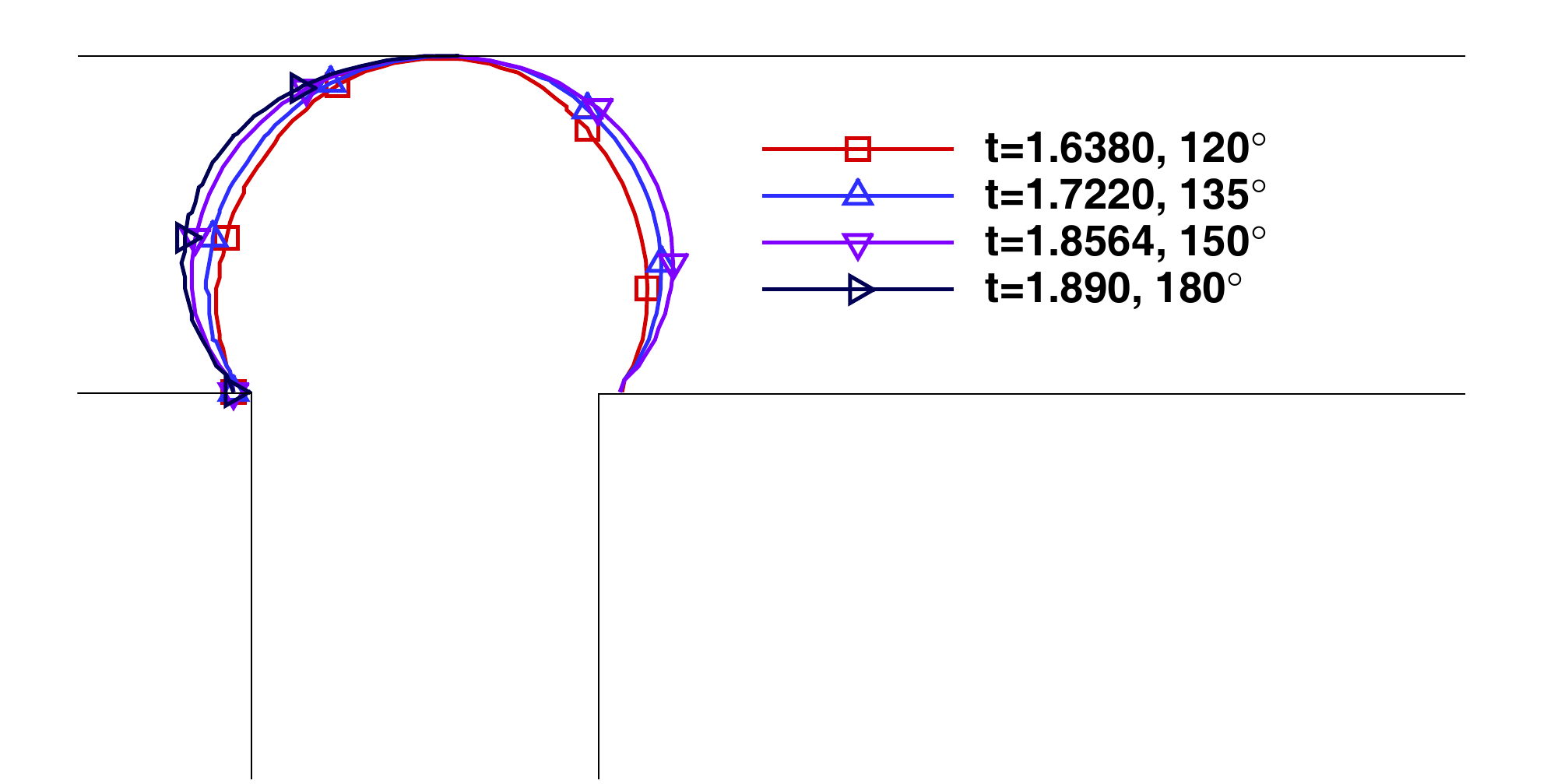}}
	\subfloat[\rev{Squeezing, $\qr=1$}]{\includegraphics[width=0.45\linewidth]{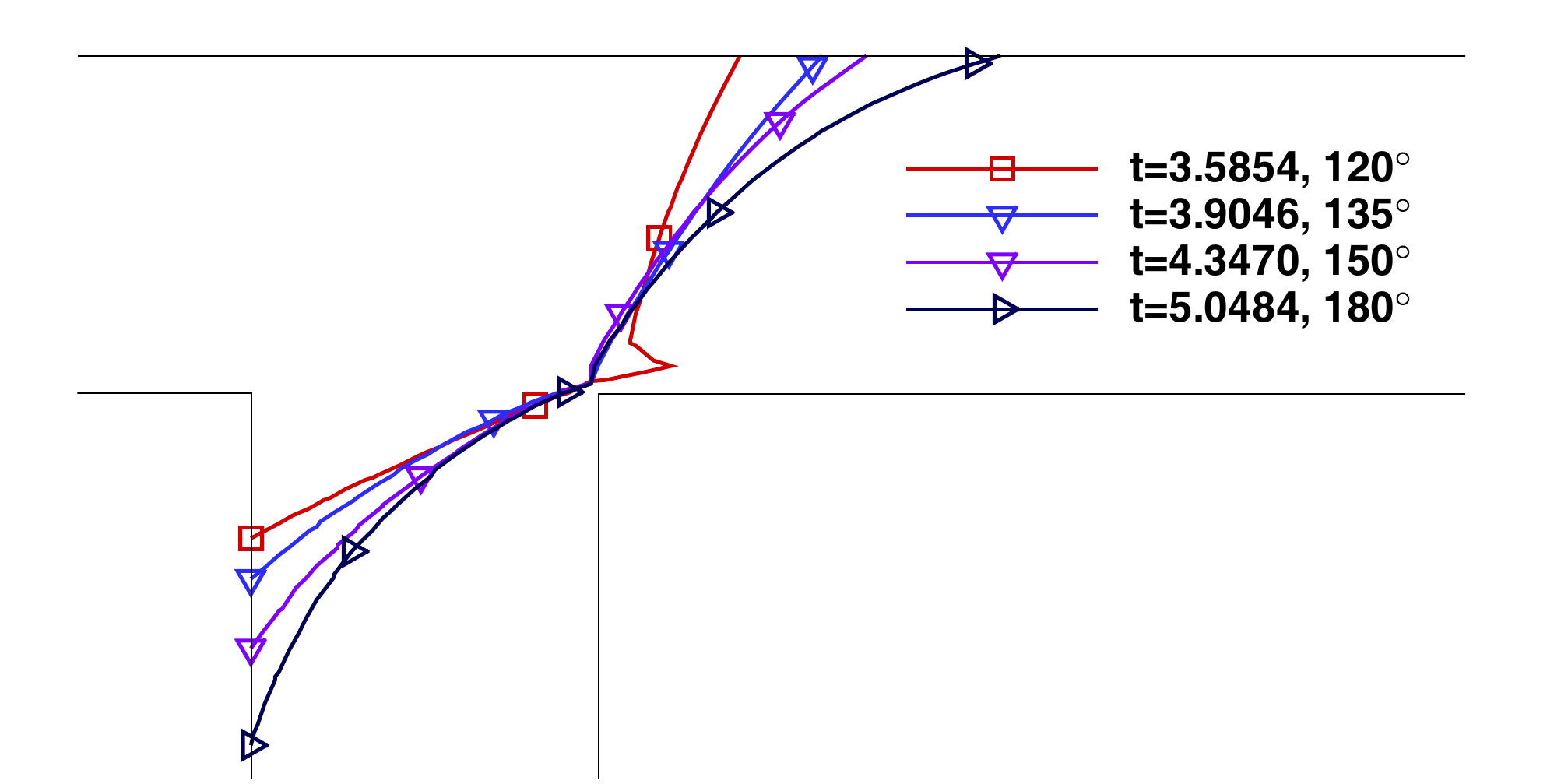}}\\
	\subfloat[Filling, $\qr=1/10$]{\includegraphics[width=0.45\linewidth]{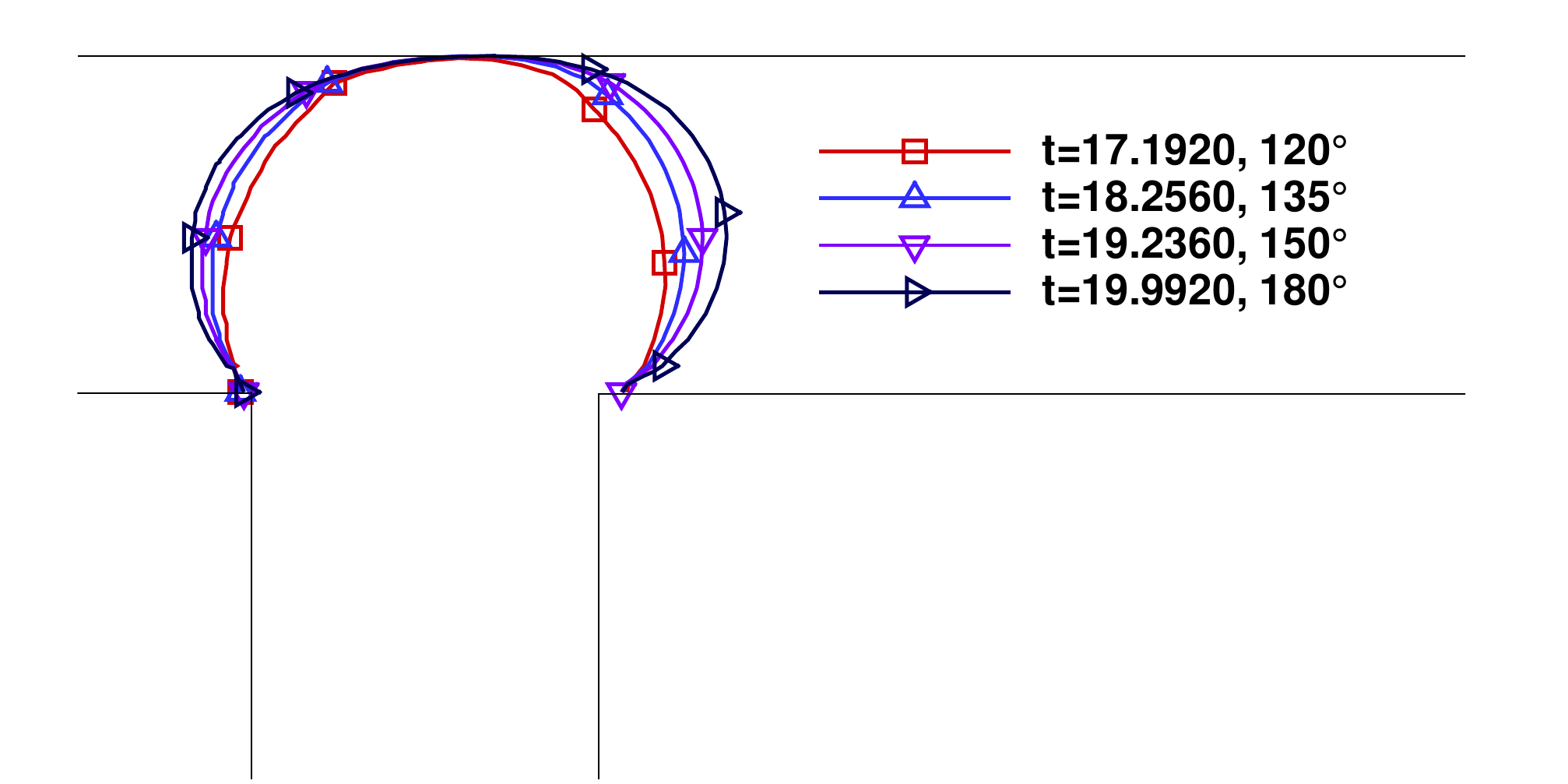}}
	\subfloat[Squeezing, $\qr=1/10$]{\includegraphics[width=0.45\linewidth]{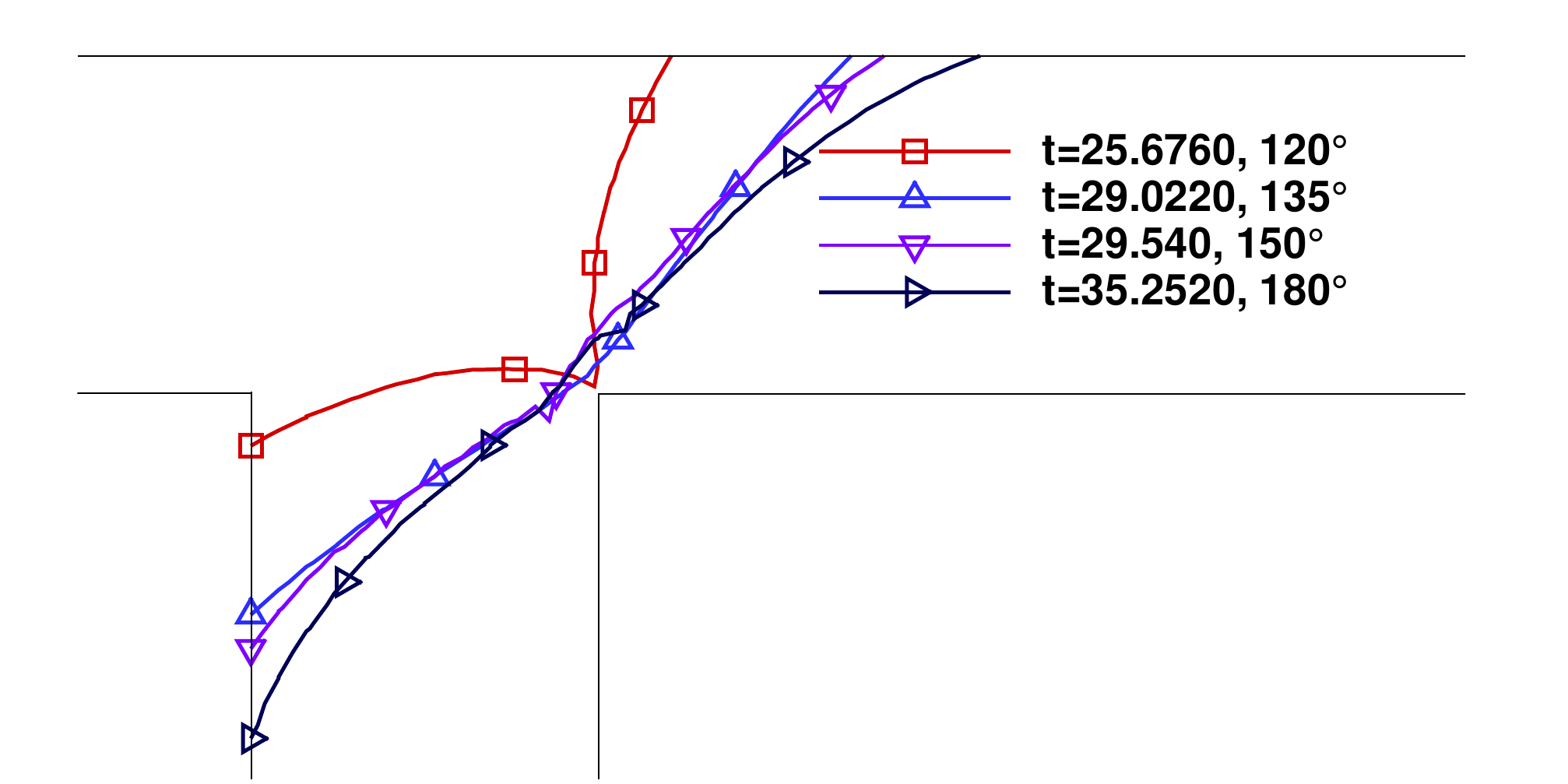}}\\
	\caption{Evolution of the interface during the filling and squeezing stages as a function of flow rate ratio ($\qr$) and contact angle ($\theta$) for $\cac=10^{-4}$}
	\label{fig:4}
\end{figure}

\noindent The dispersed phase enters into the primary channel until the interface collapses and forms a droplet, as shown in \fig\ref{fig:4}. 
In the filling stage, DP forces are dominant over CP. Hence, the droplet is expanding without showing much movement towards the downstream direction. It can be observed that the filling time is equal to $0.1733$ at $\theta=120^{\circ}$, which has increased to $0.1807$ at $\theta=150^{\circ}$ \rev{for $\qr=10$} (refer to \fig\ref{fig:4}a). Hence, the filling time is increasing from $120^{\circ}-150^{\circ}$ due to the high resistance from the wall. The continuous phase will flow through the small gap between the wall and interface in the squeezing stage \citep{VanSteijn2010}. The interaction between the solid wall and droplet is more vital in the case of $\theta=120^{\circ}$ and becoming weaker at $\theta=150^{\circ}$ as shown in \fig\ref{fig:4}b. Notably, the interface shape is stretching and trying to attain a circular shape for superhydrophobic conditions. Hence, the time of the squeezing stage is increasing from $t=1.8085$ to $2.1936$ (refer to \fig\ref{fig:4}b). After that, the interface becomes unstable and collapses from the mainstream. Moreover, the pinch-off point moves away from the T-junction towards the downstream when $\theta=120^{\circ}$ and breakup is happening exactly at the junction point for $135^{\circ} \leq \theta \leq 180^{\circ}$. Therefore, it can be concluded that the breakup location is the point where the forces responsible for the pinch-off are balanced. The filling time for $120^{\circ}-180^{\circ}$ is increasing from $t=1.638$ to \rev{1.890} for $\qr=1$ (refer to \fig\ref{fig:4}c). This is happening due to less contact of the dispersed phase with the solid. In the squeezing stage, the continuous phase flows around the dispersed phase until it pinches-off. The larger curvature obtains the interface shape due to the dominance of the capillary forces. 

\noindent 
The interface exhibited a convex shape for superhydrophobic surfaces; however, more planar-like behavior is observed for the hydrophobic surfaces. At the pinch-off point, the interface is seen as a sharp bending for $120^{\circ} \leq \theta \leq 150^{\circ}$ and a smooth for $\theta > 150^{\circ}$. At $\theta = 180^{\circ}$, the interface is evolving smoothly in the squeezing stage. Moreover, at the same time, the interaction between the wall and the interface is becoming weaker. Hence, the filling and squeezing timings are decreasing.
At lower $\qr=1/10$, the interface expands horizontally for the filling stage. The nose of the droplet is touching the opposite wall and moves towards the downstream by increasing the contact angle ($\theta$) from $120^{\circ}-180^{\circ}$, as shown in \fig\ref{fig:4}e. Hence, the filling time is also increasing. The interface curvature is changing into a concave shape faster for $120^{\circ} \leq \theta \leq 135^{\circ}$. However, the interface is sustaining to be in convex until the pinch-off time for $\theta \geq 150^{\circ}$. It is interesting to see that the pinch-off point at the \rev{right corner of} T-junction point for all values of the contact angles. The following section presents the effect of contact angle on the interface evolution in terms of the instantaneous evolution of the pressure in the continuous and dispersed phases.  
%%%%%%%%%%%%%%
\subsection{Pressure evolution}
\begin{figure}[!t]
	\centering
	\subfloat[$p_{\text{cp}}$, $\qr=10$]{\includegraphics[width=0.5\linewidth]{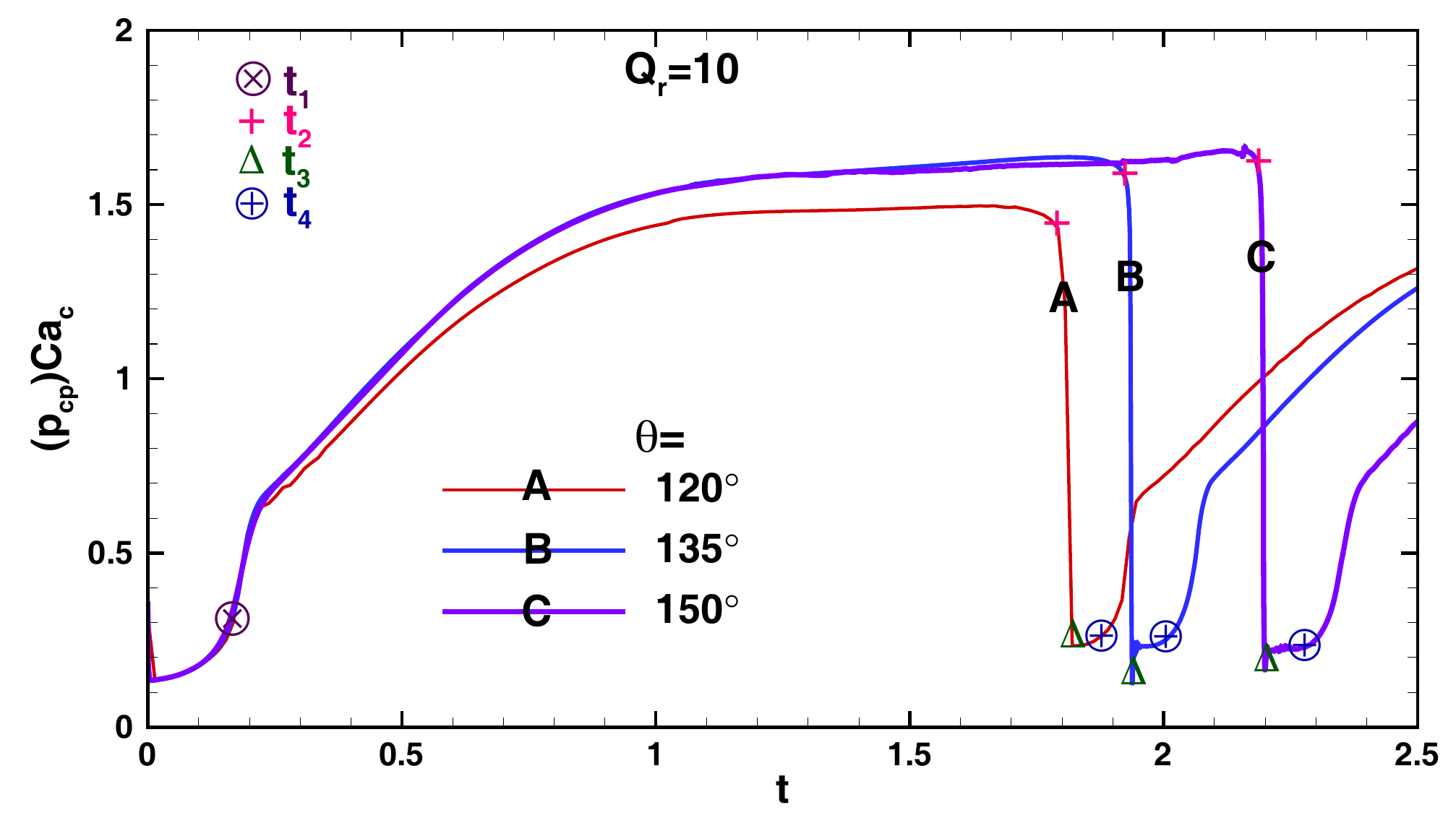}}
	\subfloat[$p_{\text{dp}}$, $\qr=10$]{\includegraphics[width=0.5\linewidth]{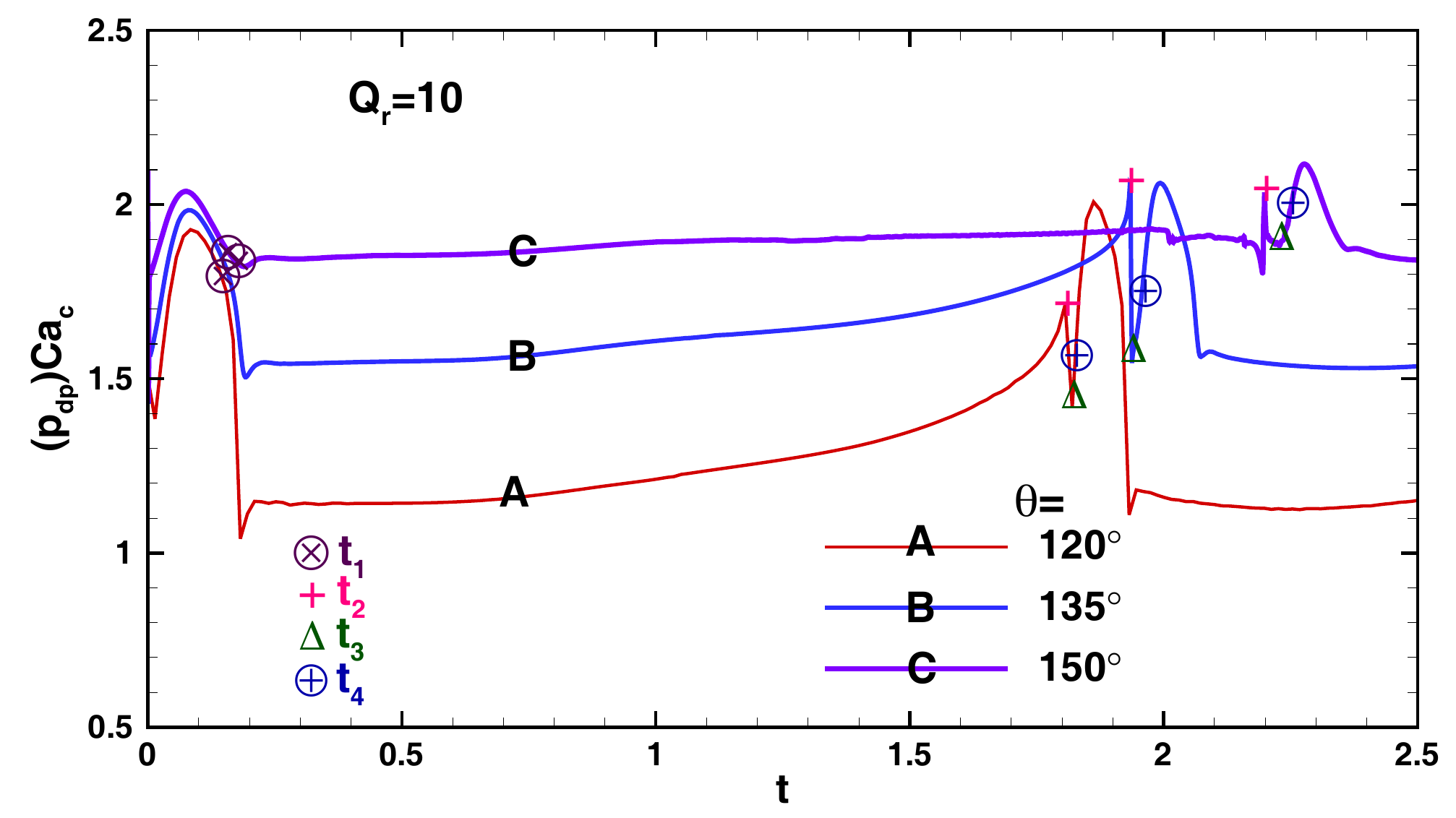}}\\
	\subfloat[\rev{$p_{\text{cp}}$, $\qr=1$}]{\includegraphics[width=0.5\linewidth]{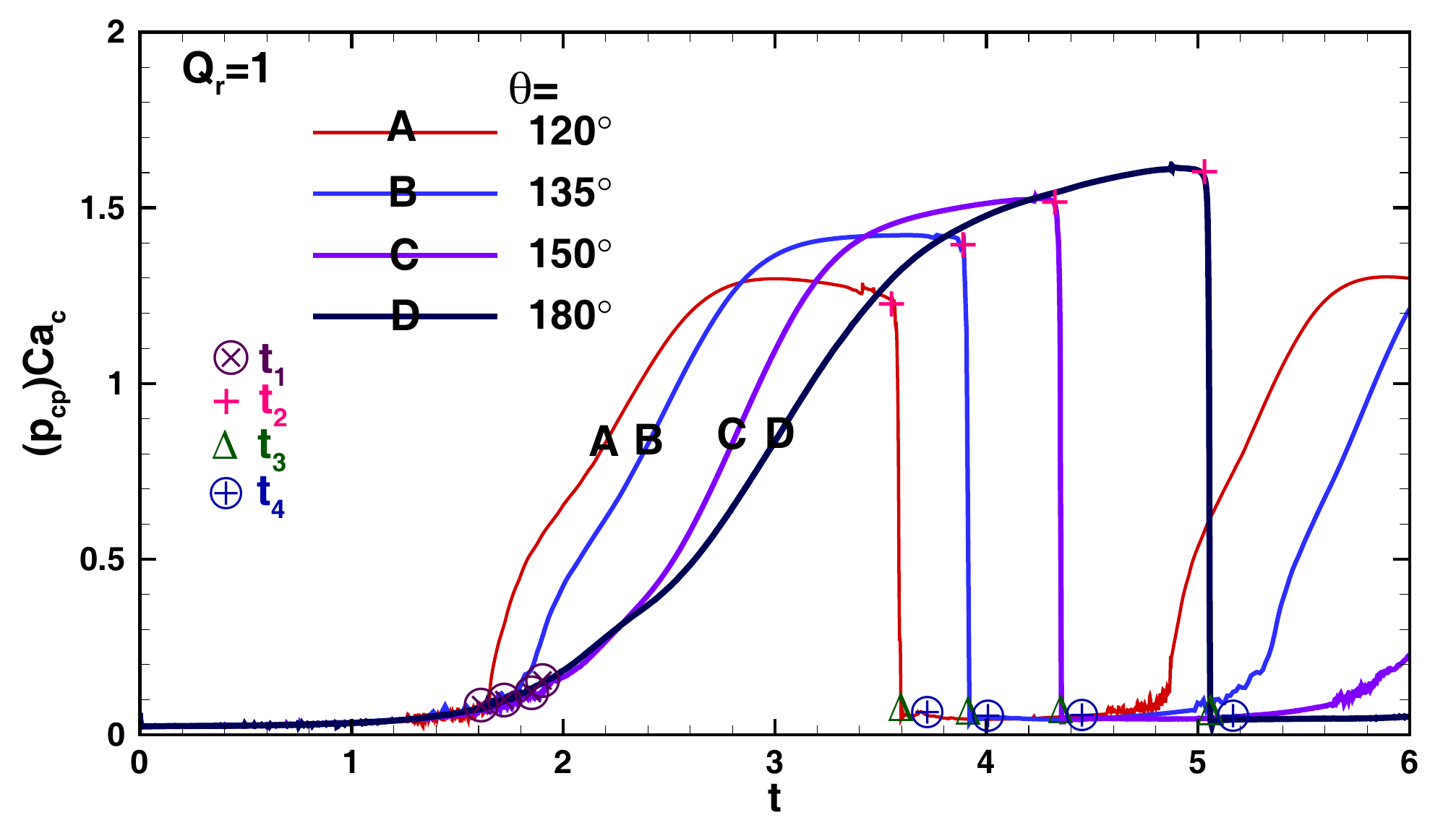}}
	\subfloat[\rev{$p_{\text{dp}}$, $\qr=1$}]{\includegraphics[width=0.5\linewidth]{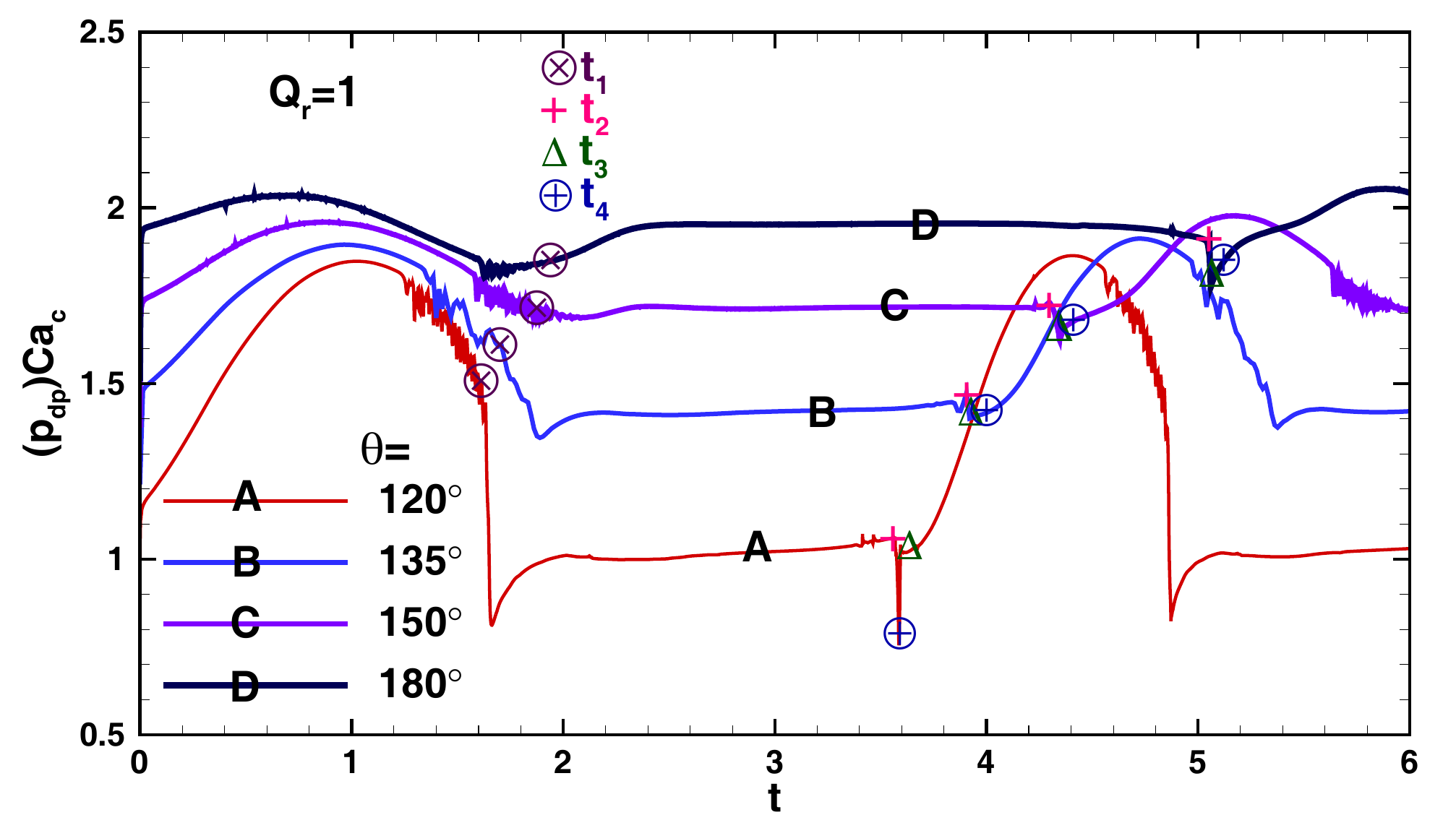}}\\
	\subfloat[$p_{\text{cp}}$, $\qr=1/10$]{\includegraphics[width=0.5\linewidth]{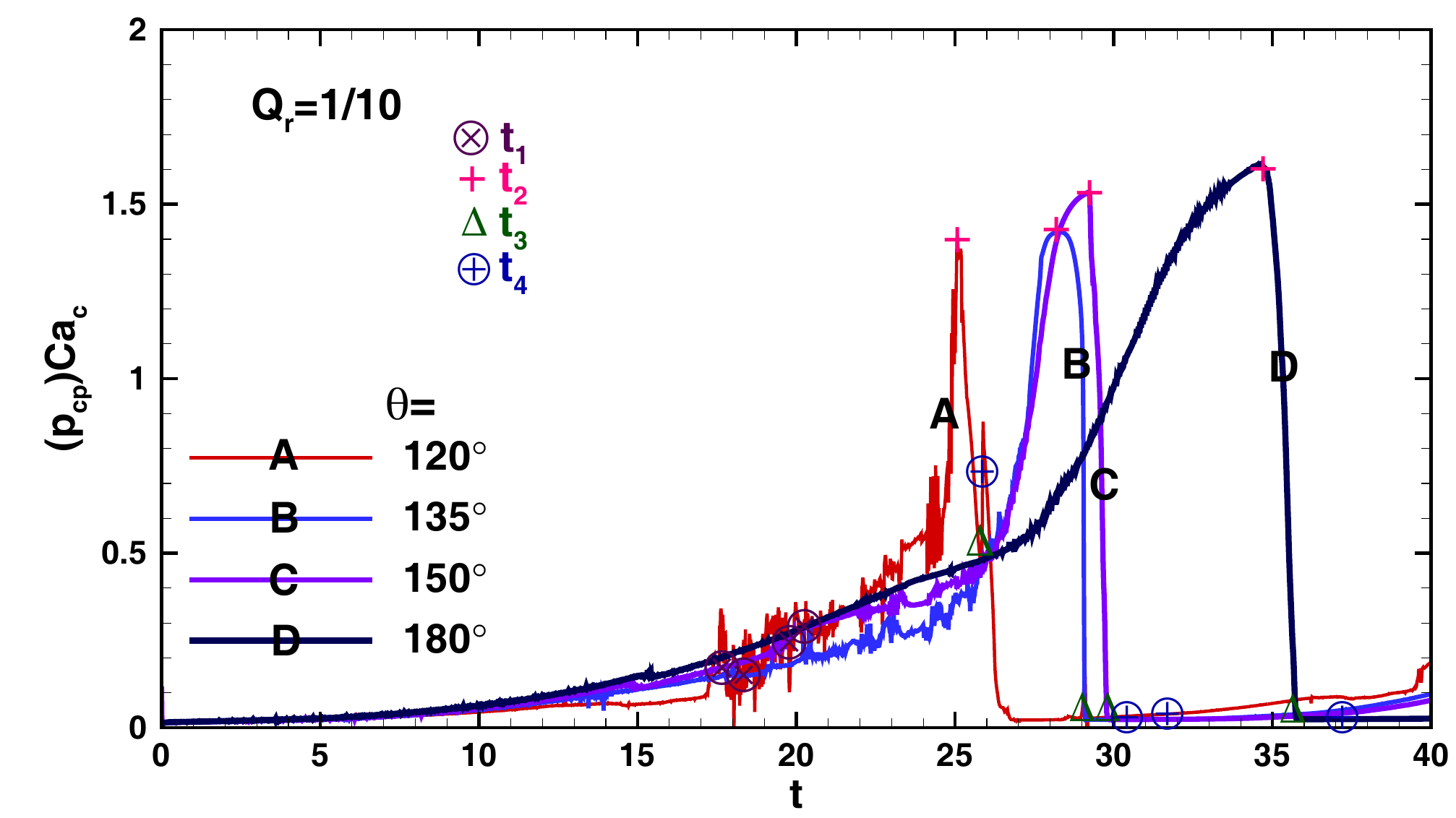}}
	\subfloat[$p_{\text{dp}}$, $\qr=1/10$]{\includegraphics[width=0.5\linewidth]{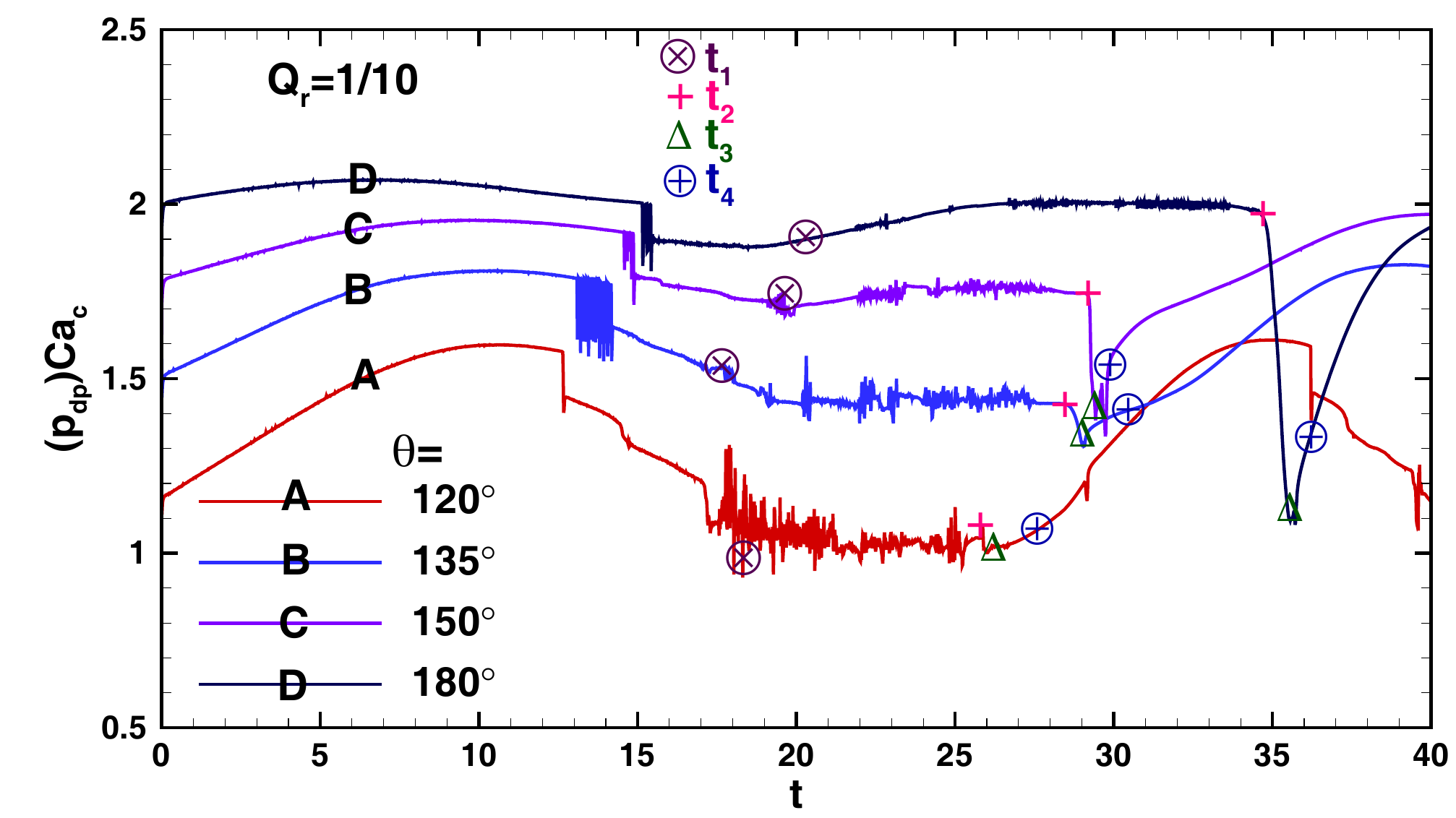}}
	\caption{Evolution of the pressure in the continuous ($p_{\text {cp}}$) and dispersed ($p_{\text {dp}}$) phases as a function the flow rate ratio ($\qr$), and contact angle ($\theta$) for $\cac=10^{-4}$. \rev{The filling, squeezing, pinch-off and stable droplet time are marked with $\bigotimes$, $+$, $\Delta$, and $\bigoplus$, respectively.}}
	\label{fig:5}
\end{figure}

\noindent  The surface wettability effects are more significant in the evolutions of pressure in the continuous ($p_{\text {cp}}$) and dispersed phases ($p_{\text {dp}}$). The pressure evolutions have been measured during the droplet formation by choosing two different locations, one in the continuous phase ($p_{\text {cp}}$) and the second in the dispersed phase ($p_{\text {dp}}$), as shown in \fig\ref{fig:1}a. 
During the early stage of the droplet formation, the dispersed phase invades into the main channel and slowly restricts the flow of the continuous phase. Once the filling stage is completed, the continuous phase flow is obstructed leads to a rise in the upstream pressure ($p_{\text {cp}}$). The process continues until $p_{\text {cp}}$ reaches a maximum value and then a sudden fall at the pinch-off point wherein the droplet breaks off from the dispersed phase. 
Earlier studies reported that pressure ($p_{\text {dp}}$) in the dispersed phase is constant \citep{Garstecki2006,Bashir2014}. However, we observed that $p_{\text {dp}}$ is not constant during the filling stage, but it oscillates in an anti-phase with $p_{\text {cp}}$ \citep{Abate2012}. After crossing the filling stage, $p_{\text {dp}}$ is showing gradual increase up to the pinch-off point. It is noticed that at the pinch-off, both $p_{\text {dp}}$ and $p_{\text {cp}}$ are showing a sudden falling trend, as shown in \fig\ref{fig:5}.

\noindent 
At higher values of $\qr (=10)$, $p_{\text {cp}}$ and $p_{\text {dp}}$ are not showing much variation with the contact angle ($\theta$) during the filling stage and it can be observed in \fig\ref{fig:5}a-b. However, in the squeezing stage, $p_{\text {cp}}$ and $p_{\text {dp}}$ increase with increase in $\theta$. This is attributed mainly due to the confinement of the flow. The variations in $p_{\text {dp}}$ with respect to $\theta$ are increasing at $\qr=1$ and $10$.

\noindent Maximum pressure in CP ($p_{\text {cp,max}}$) as a function of $\qr$, $\cac$ and $\theta$ ($0.1\le\qr\le 10$, $120^{\circ} \leq \theta \leq 180^{\circ}$, and $10^{-4}\le \cac \le 10^{-3}$) is given by the following correlation: 
\begin{gather}
	p_{\text{cp,max}}=A + B\cac^{0.5} + C x_2 + Dx_2^{2} + Ex_2^{3}
	\label{eq:pcpm}
	%\label{eqn:P_{cp,max}}
\end{gather}
where 
\begin{gather*}
x_2=\log\qr,\qquad A=\alpha/(1+\beta \theta+\gamma \theta^{2}),\qquad 
B=\alpha+\beta q^{1.5}+\gamma e^{q},\qquad
C=\alpha+\beta \theta^{2.5}+\gamma \theta^{3}, \\ 
D=\alpha+\beta w^{3}+\gamma (\log w)^2,\qquad 
E=\alpha+\beta/\theta^{1.5}+\gamma/\theta^{2},\qquad %\\
q=10^{-1} \theta,\qquad w=10^{-2} \theta,\qquad
\end{gather*} 
Statistical non-linear regression analysis is performed to obtain the correlation coefficients \rev{for $p_{\text{cp,max}}$}   with DataFit (trial version) and MATLAB tools and their values are presented in \tab\ref{tab:1}.

\begin{figure}[hbtp]
	\centering
	\subfloat[$\qr=10$]{\includegraphics[width=0.7\linewidth]{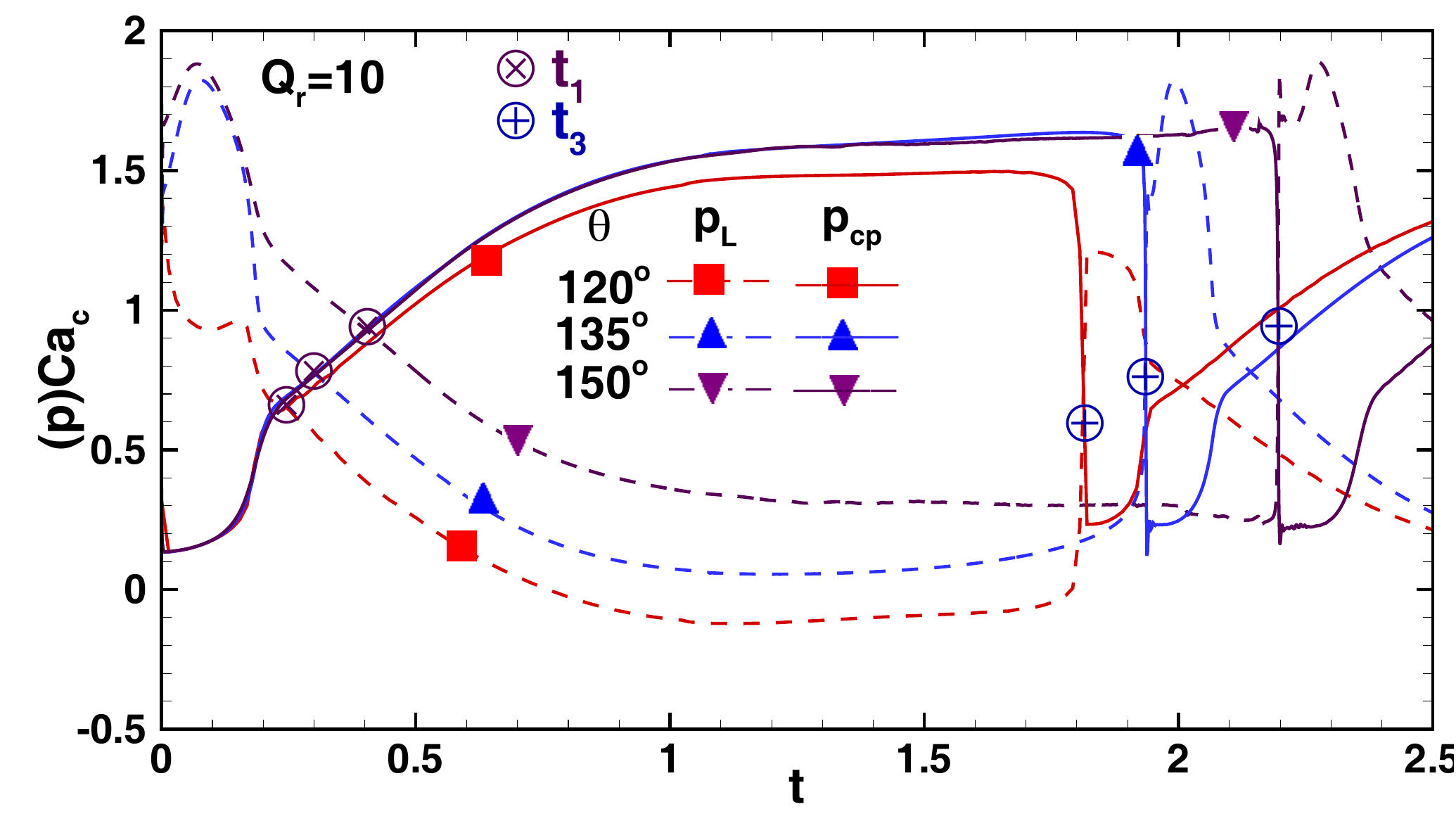}}\\
	\subfloat[\rev{$\qr=1$}]{\includegraphics[width=0.7\linewidth]{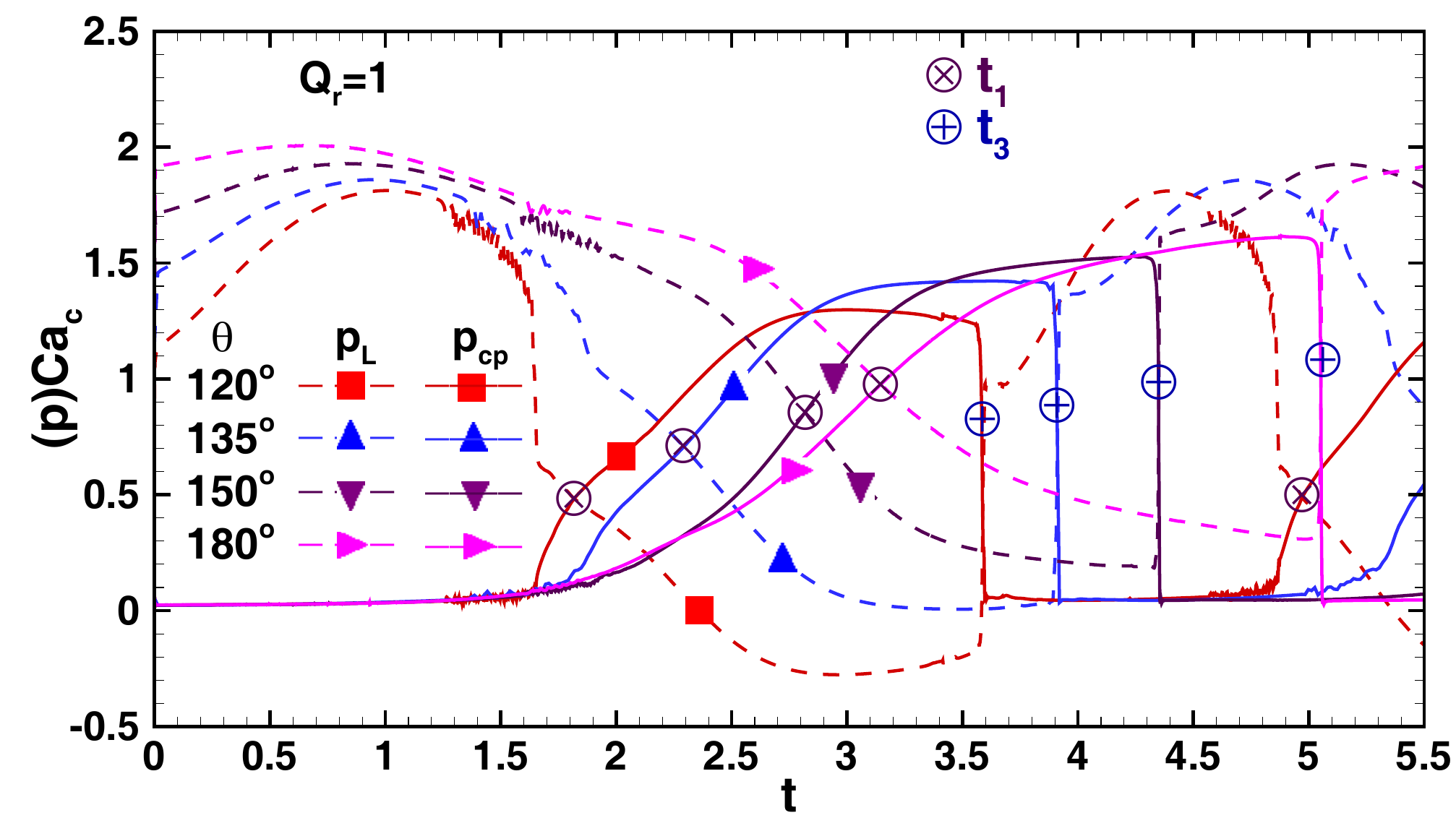}}\\
	\subfloat[$\qr=1/10$]{\includegraphics[width=0.7\linewidth]{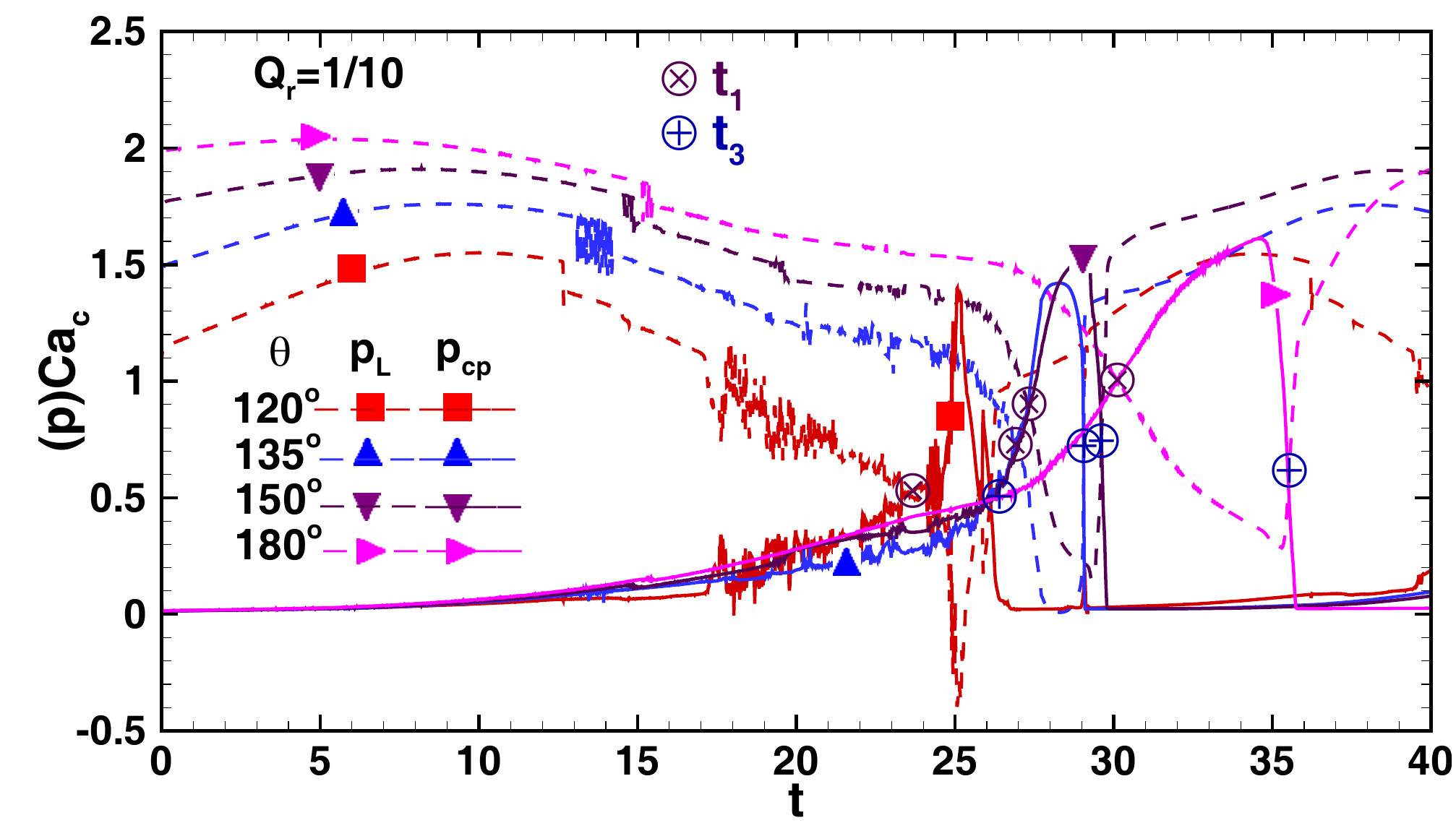}}
	\caption{Evolution of the instantaneous pressure for as a function of the flow rate ratio ($\qr$) and contact angle ($\theta$) for $\cac=10^{-4}$. \rev{The filling and pinch-off time are marked with $\bigotimes$ and $\bigoplus$, respectively.}}
	\label{fig:6}
\end{figure}

\noindent The Laplace pressure plays a vital role in the interface shape evolution and it is essentially the difference between $p_{\text {dp}}$ and $p_{\text {cp}}$. For instance, if $p_{\text{L}}$ is small, the radius of the curvature of the interface is large leads to adopting a flattened shape. Nevertheless, the interface adopts a more curved shape with a smaller radius of curvature when $p_{\text{L}}$ is large. The effect of the contact angle ($\theta$) on the Laplace pressure ($p_{\text{L}}$) is seen in \fig\ref{fig:6}. At higher flow rate ratios ($\qr=10$), $p_{\text{L}}$ is increasing in the starting period then slowly decreases (refer to \figs\ref{fig:6}a-c). Therefore, it indicates that the interface is transiting from a more curved into flattened shape. When $\qr=10$, $p_{\text{L}}$ is continuously decreasing for $\theta=120^{\circ}$. Whereas there is a rise and then decrease for $\theta=135^{\circ}$ and $150^{\circ}$. At $\qr=1$ and $1/10$, $p_{\text{L}}$ is not shown much variation for superhydrophobic conditions ($\theta>150^{\circ}$), but there is an increase in $p_{\text{L}}$ in the initial period for $\theta=120^{\circ}$ and $135^{\circ}$\rev{, as shown in  \figs\ref{fig:6}a-c}. 

\noindent 
It can also be observed from the point where both $p_{\text {cp}}$ and $p_{\text{L}}$ are intersecting indicates the end of the filling stage, marked with symbol \rev{$\bigotimes$}, ${t}_1$. ${t}_1$ is increasing as $\theta$ increases. For instance, ${t}_1=0.196$ for $\theta=120^{\circ}$ and $0.404$ for $\theta=150^{\circ}$ when $\qr=10$ (refer to \fig\ref{fig:6}a).

\noindent 
Further, the pinch-off time (${t}_3$) can be predicted from the point where $p_{\text {cp}}$ and $p_{\text{L}}$ are intersecting for the second time, marked with symbol \rev{$\bigoplus$}, ${t}_3$. It is increasing with increasing contact angle, as shown in \fig\ref{fig:6}. For instance, ${t}_3=1.810$ for $\theta=120^{\circ}$ and $2.196$ for $\theta=150^{\circ}$ when $\qr=10$ (refer to \fig\ref{fig:6}a).
A correlation is proposed to predict the pinch-off time (${t}_3$) based on the pressure evolution profiles as a function of contact angle (refer to \fig\ref{fig:6}) as follows:
\begin{gather}
	t_{3,p}=  A + B\cac^{0.5} + C/\qr + D/\qr^{2} + E/\qr^{3} + F/\qr^{4}
	\label{eq:t3p}
%	\label{eqn:t_{sq,p}}
\end{gather}
where 
\begin{gather*}
A=\alpha+\beta w^{2} \log w+\gamma s^{2},\qquad 
B=\alpha+\beta \theta^{0.5}+\gamma s,\qquad 
C=\alpha+\beta e^{p}+\gamma/p^{2},\\ 
D=\alpha+\beta/\log z+\gamma (\log z)/z^{2}, \qquad
E=\alpha+\beta/s+\gamma/\theta^{2},\qquad 
F=\alpha/(1+\beta p+\gamma p^{2}), \\
w= 10^{-2}\theta,\qquad s=\log\theta,\qquad p=\sec\theta,\qquad z=\cot\theta
\end{gather*}
Statistical non-linear regression analysis is performed to obtain the correlation coefficients \rev{for $t_{3,p}$} with DataFit (trial version) and MATLAB tools and their values are presented in \tab\ref{tab:1}.
\begin{figure}[!b]
	\centering
	{\includegraphics[width=0.8\linewidth]{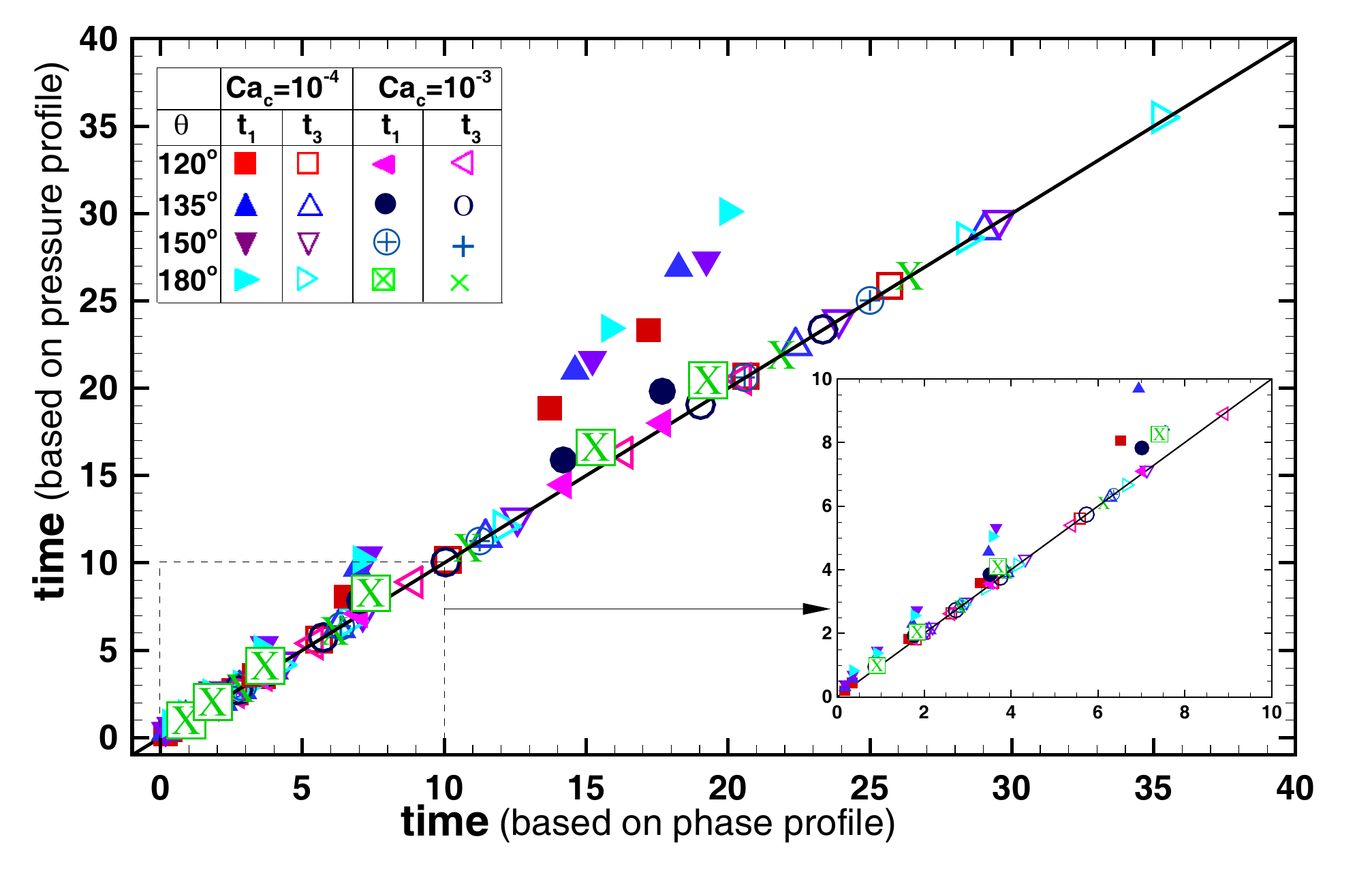}}
	\caption{\rev{Comparison of the filling (${t}_1$) and pinch-off times (${t}_3$) obtained from the phase and pressure profiles as a function of the contact angle and $\qr$.}}
	\label{fig:6a}
\end{figure}

\noindent
\fig\ref{fig:6a} shows the comparison of the filling ($t_1$, filled symbols) and pinch-off ($t_3$, unfilled symbols) time based on the phase and pressure profiles. The pressure in the continuous phase starts to rise after the dispersed phase reaches the opposite wall. Hence, the filling time ($t_1$) predicted based on the pressure profile, is higher than the phase profile. Thus, the filling time ($t_1$) points show more deviation. However, the pinch-off time ($t_3$) matches the phase and pressure profiles for all the contact angles. In culmination, the droplet pinch-off mechanism can also be elucidated in depth by installing pressure sensors even without the flow visualization and image processing of the phase profiles. 
\begin{figure}[htbp]
	\centering
	\subfloat[$\theta=120^{\circ}$]{\includegraphics[width=0.88\linewidth]{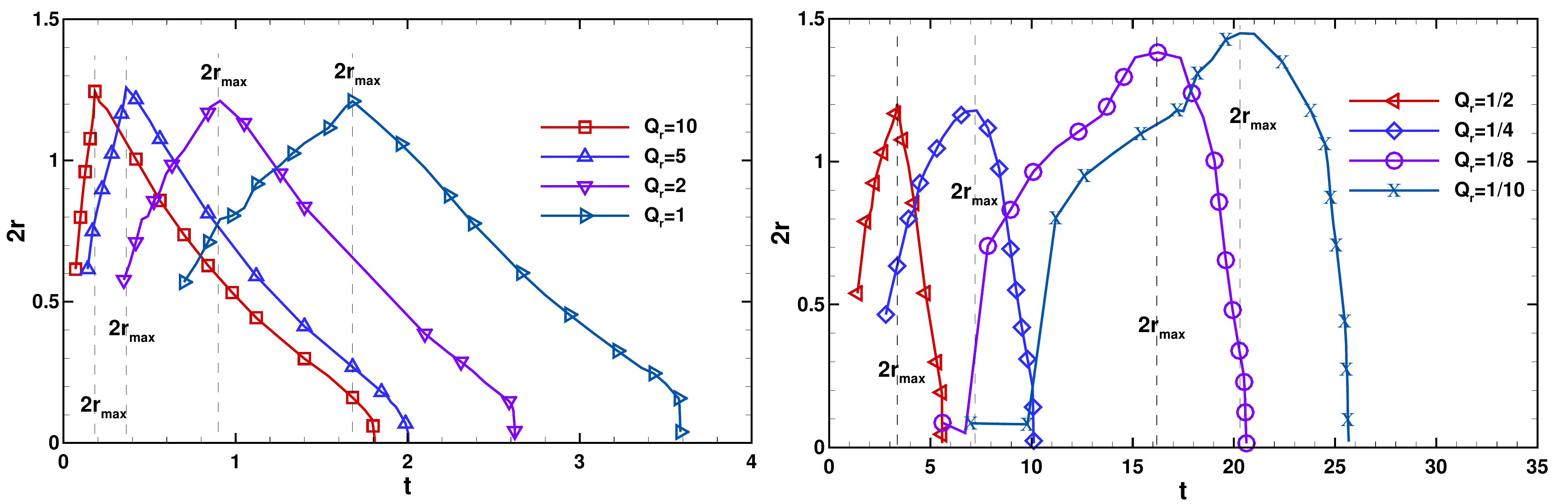}}\\
	\subfloat[$\theta=135^{\circ}$]{\includegraphics[width=0.88\linewidth]{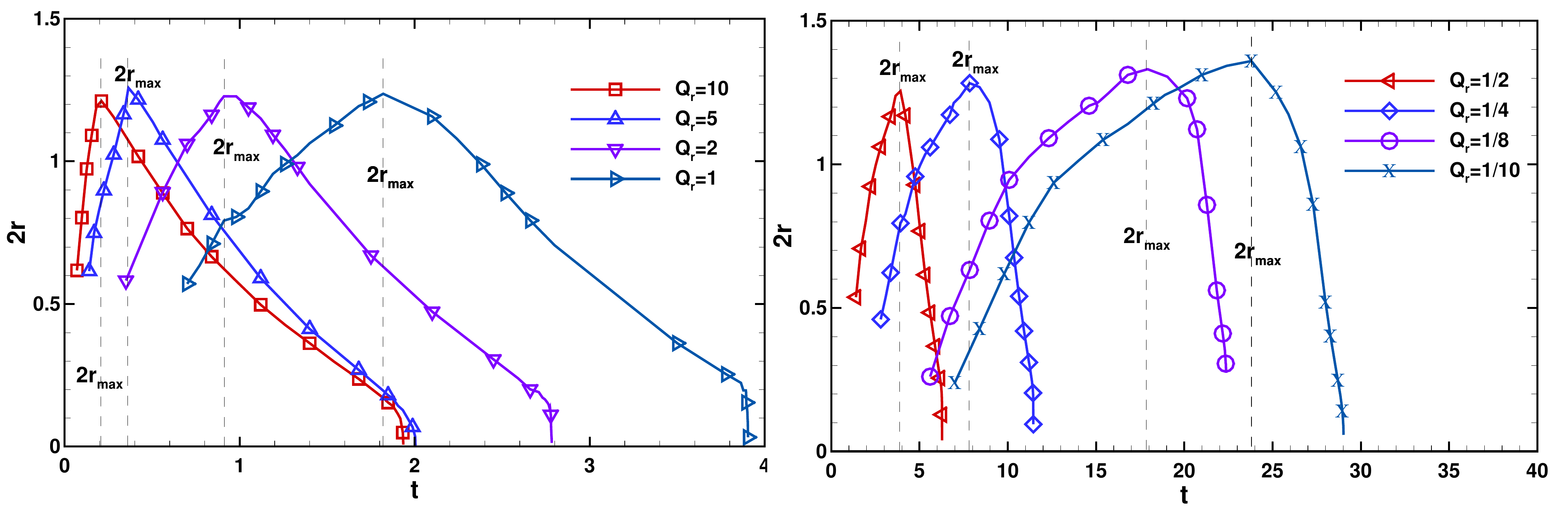}}\\
	\subfloat[$\theta=150^{\circ}$]{\includegraphics[width=0.88\linewidth]{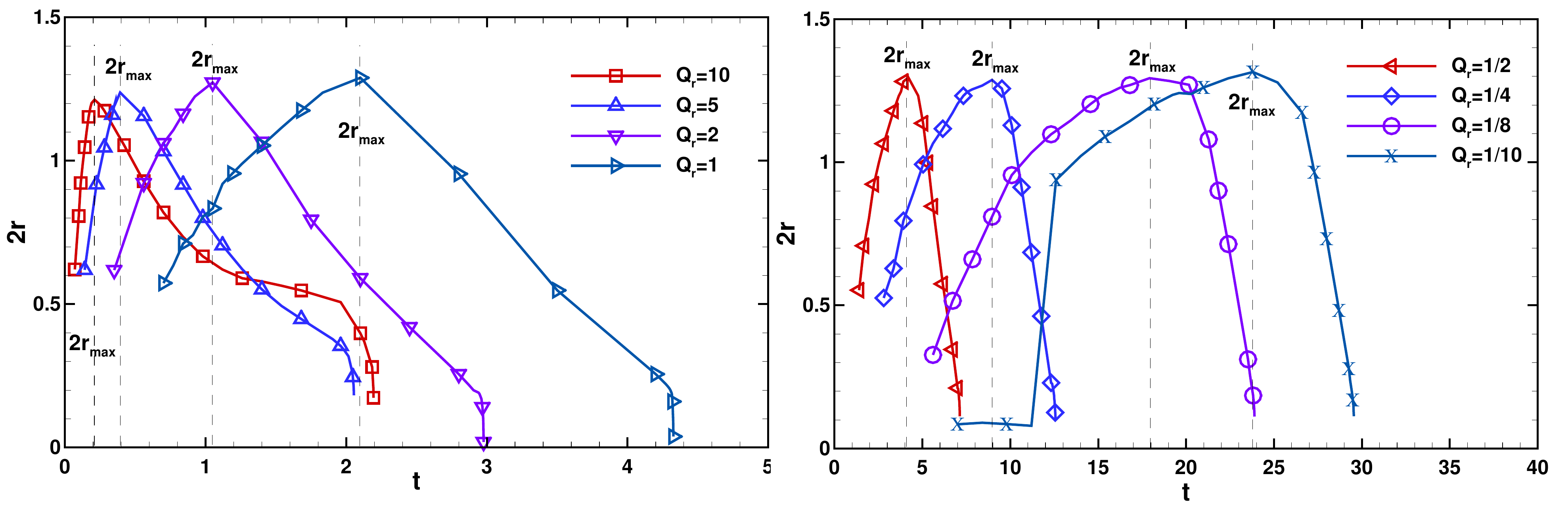}}\\
	\subfloat[$\theta=180^{\circ}$]{\includegraphics[width=0.88\linewidth]{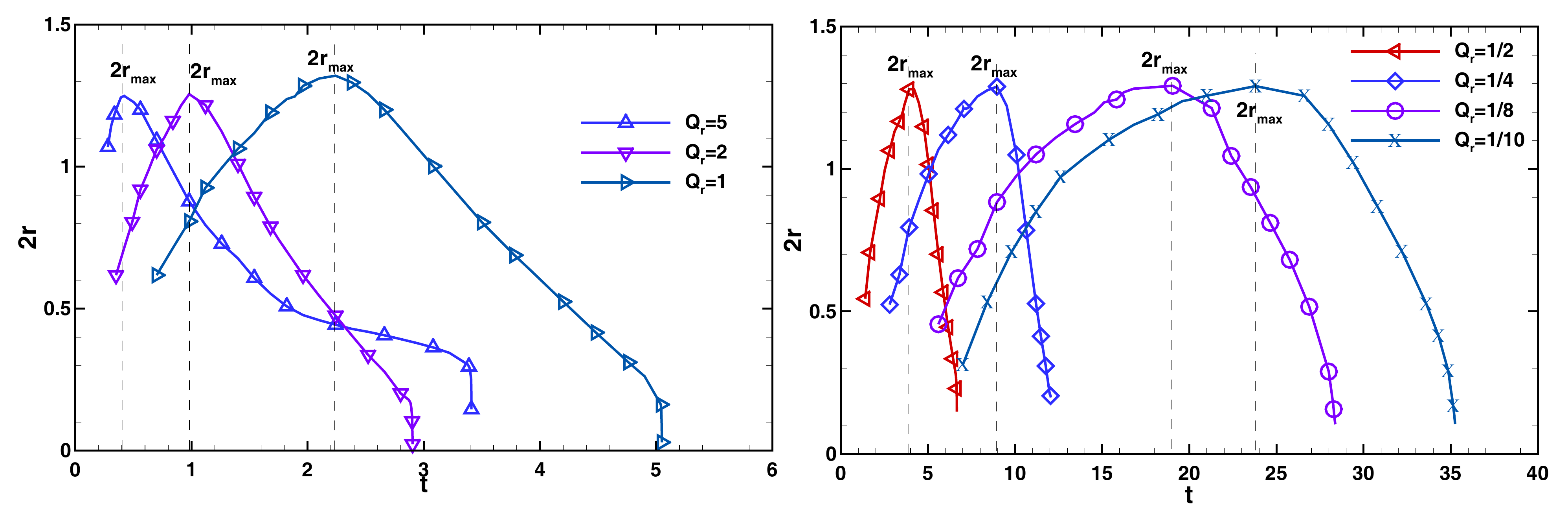}}	
	\caption{The interface neck thickness as a function of dimensionless time and contact angle ($\theta$) at $\cac=10^{-4}$.}
	\label{fig:7}
\end{figure}
%---------------------------------
\subsection{The droplet pinch-off mechanism}
%---------------------------------
\noindent In this section, the droplet pinch-off mechanism is elucidated by analyzing the evolution of the interface and the Laplace pressure as a function of contact angle.
In the filling stage, the dispersed phase starts penetrating the primary channel. The interface expands and progresses towards the opposite wall. The interface deforms as it experiences the hydrodynamic forces and attaining shape with a larger radius at the rear side and smaller at the front side.  The corresponding Laplace pressure is sharply decreasing with time. Hence, the interface adopts a curvy shape with a smaller radius of curvature in the initial time and slowly transits into a flattened shape as time progresses.

\noindent
\fig\ref{fig:7} describes the neck thickness ($2r$) as a function of the flow rate ratio ($\qr$). For hydrophobic conditions ($120^{\circ}\leq\theta<150^{\circ}$), $2r$ increases in the filling stage until it reaches a threshold value ($2r_{\text {max}}$,  marked with dashed lines) for lower $\qr$ values range from $1/2$ to $1/10$ (refer to \fig\ref{fig:7}a,b). 
\rev{While $2r_{\text {max}}$ is increasing with decreasing $\qr$ ($1/2\ge\qr\ge1/10$), it} is achieving the same value for higher $\qr$ ($10\leq\qr\leq1$).
\rev{However, the neck thickness ($2r$) has shown contrasting nature (refer to \fig\ref{fig:7}c,d) for superhydrophobic conditions ($150^{\circ}<\theta\leq180^{\circ}$).}
Further, it can also be observed that the trend of the droplet pinch-off follows the same for both hydrophobic and superhydrophobic conditions.
\noindent A correlation is developed to predict the value of $2r_{\text{max}}$ as a function of the contact angle as follows:
\begin{gather}
	2r_{\text{max}}= A + B \cac + C/\qr + D/\qr^{2} + E/\qr^{3} + F/\qr^{4}
	\label{eq:2rm}
%	\label{eqn:2r_{max}}
\end{gather}
where 
\begin{gather*}
{A}=\alpha+\beta/q^{2}+\gamma e^{-q}, \qquad
{B}=\alpha+\beta \cos\theta+\gamma \sin\theta, \qquad
{C}=\alpha+\beta (\log m)^{2}+\gamma/m^{0.5}, \\
{D}=\alpha+\beta \theta^{2.5}+\gamma \theta^{3},\qquad 
{E}=\alpha+\beta \theta \log \theta+\gamma \theta^{0.5},\qquad 
{F}=\alpha/(1+\beta \theta+\gamma \theta^{2}),\\
q=10^{-1}\theta,\qquad m=10^{-3}\theta 
\end{gather*} 
Statistical non-linear regression analysis is performed to obtain the correlation coefficients with DataFit (trial version) and MATLAB tools and their values are presented in \tab\ref{tab:1}.

\begin{figure}[hbtp]
	\centering
	\subfloat[$\qr=10$]{\includegraphics[width=0.5\linewidth]{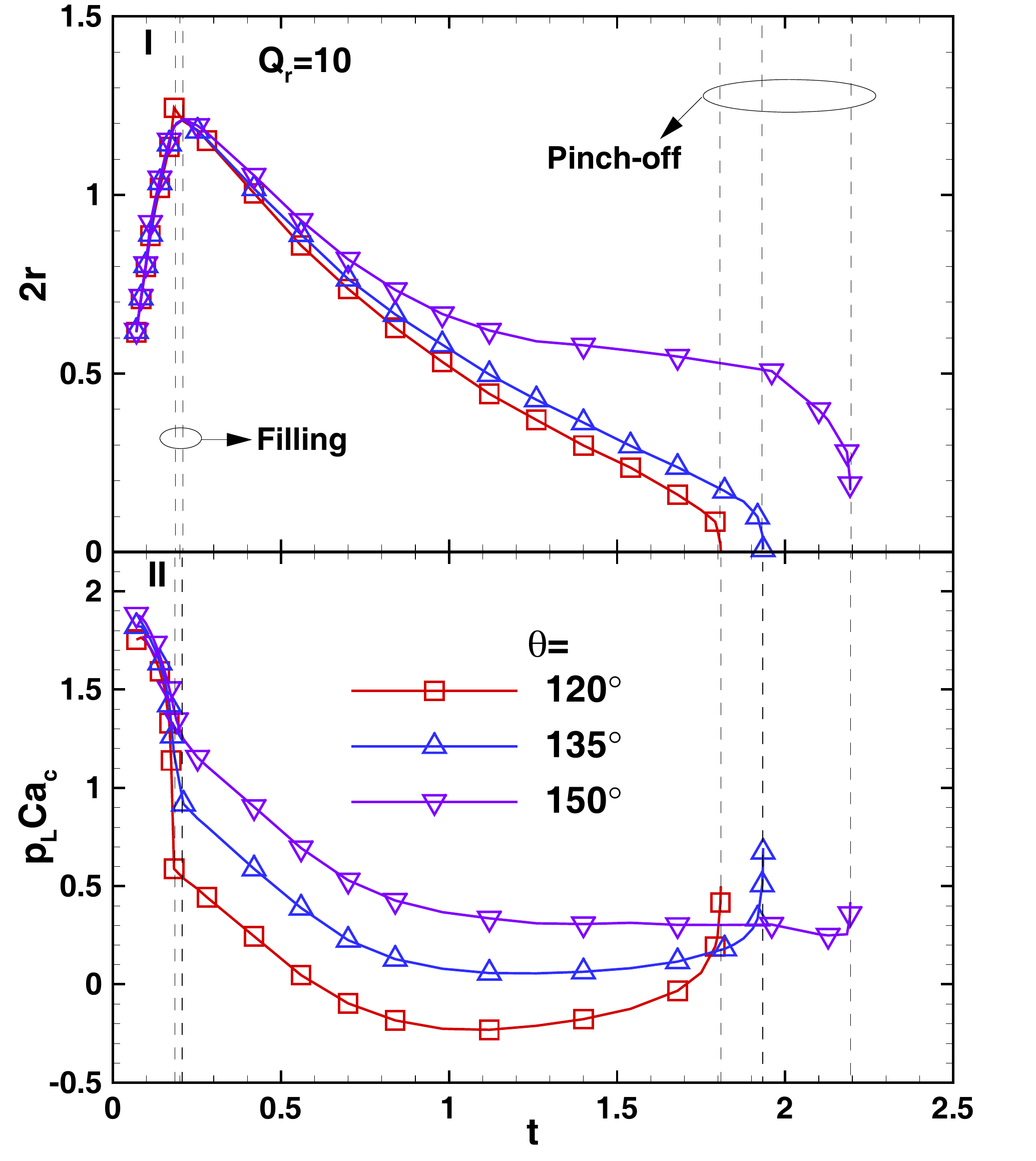}}
	\subfloat[\rev{$\qr=1$}]{\includegraphics[width=0.5\linewidth]{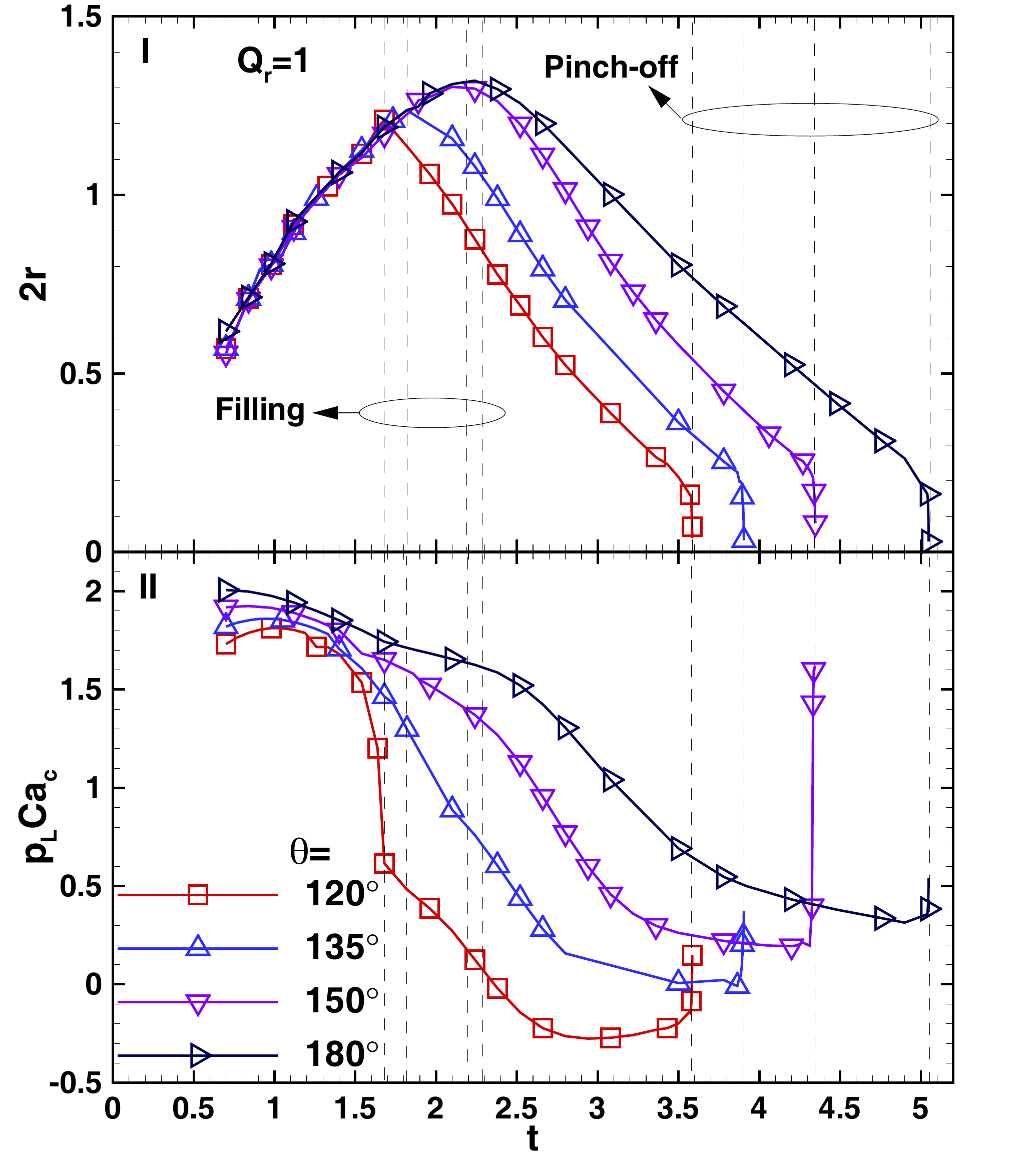}} \\
	\subfloat[$\qr=1/10$]{\includegraphics[width=0.5\linewidth]{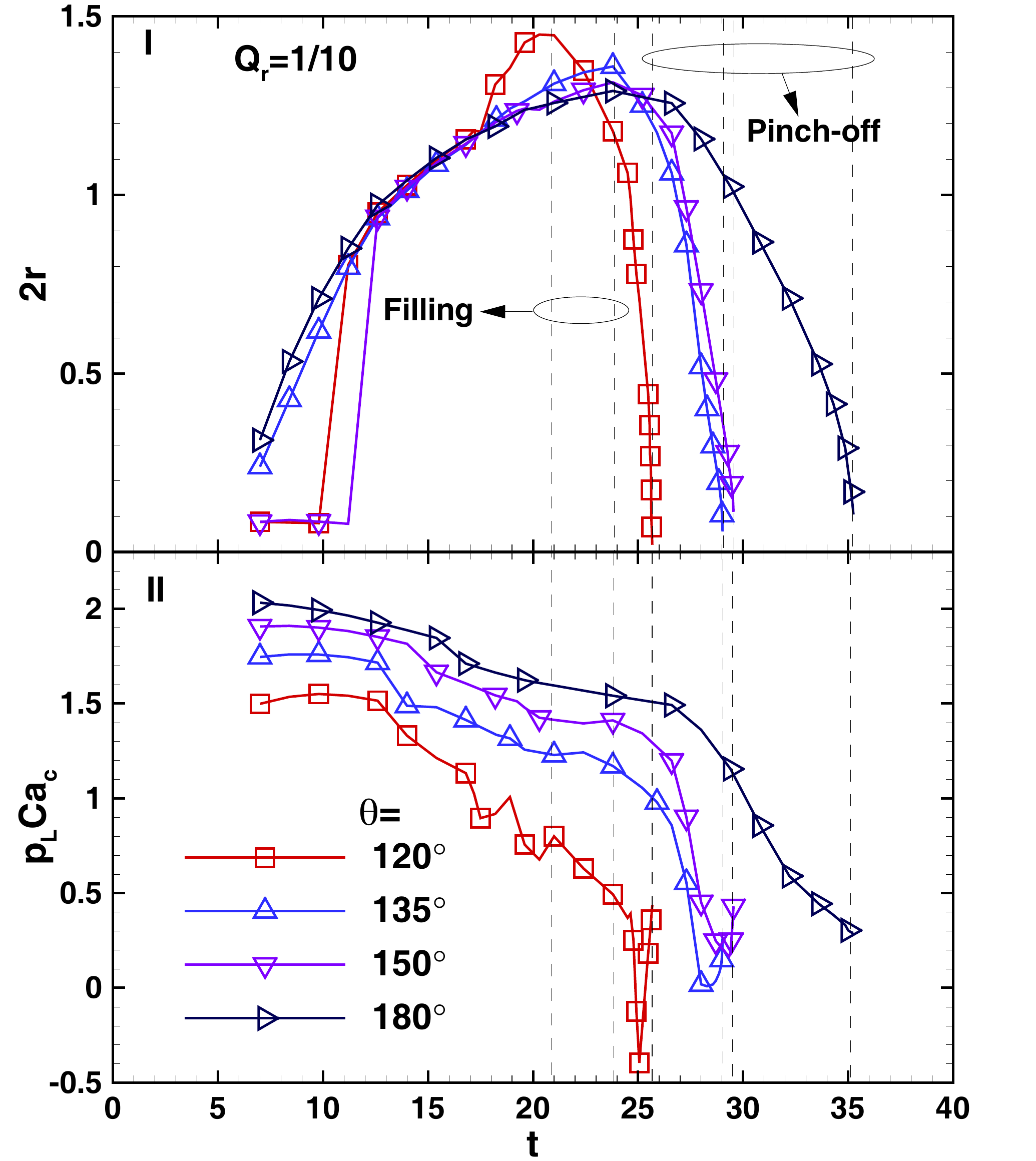}}
	\caption{The droplet dynamics for $\cac=10^{-4}$ at $\theta=120^{\circ}-180^{\circ}$ (a) $\qr=10$, (b) $\qr=1$, and (c) $\qr=1/10$.}
	\label{fig:8}
\end{figure}

%%%
\noindent
At the initial filling stage, $2r$ increases smoothly up to a threshold value (marked with dashed lines). After that, the \rev{growth} rate \rev{of} radius at the backside decreases, as shown in \fig\ref{fig:8}. The filling stage (dashed lines) is the point where both interfacial and shear forces are balanced. It is known that in the squeezing regime, the flow is essentially the pressure-driven. 
\noindent 
In the squeezing stage, the radius at the backside decreases with time smoothly. The droplet moves toward the downstream direction as its length increases further. At the pinch-off point, the radius approaches the minimum value, where the forces are balanced on the droplet, which is marked with dashed lines, as shown in \fig\ref{fig:8}(I). 
When $\qr=10$, $2r$ increases at the early filling stage for the contact angle ($\theta$) ranging from $120^{\circ} \leq \theta \leq 150^{\circ}$. Nevertheless, in the squeezing stage, the radius at the backside is comparatively high, and the droplet is expanding for $\theta = 150^{\circ}$, as shown in \fig\ref{fig:8}a(I). The corresponding Laplace pressure ($p_{\text{L}}$) across the interface is shown \fig\ref{fig:8}a(II). The Laplace pressure ($p_{\text{L}}$) is dropping smoothly in the filling stage for all the values of $\theta$. On moving further, $p_{\text{L}}$ in the squeezing or necking decreases sharply for $\theta =120^{\circ}$ and smoothly for $150^{\circ}$. It can be observed that there is a sudden shoot-up in the $p_{\text{L}}$ at the pinch-off point. Nonetheless, the pinch-off is taking place early in the case of $\theta=120^{\circ}$ and extending for $150^{\circ}$. 

\noindent When $\qr=1$, $2r$ increases smoothly up to the threshold value in the early filling stage for all the values of $\theta$. Subsequently, $2r$ decreases smoothly in the squeezing stage until the pinch-off point is marked with dashed lines, as shown in \fig\ref{fig:8}b(I). However, there is a noticeable change in $p_{\text{L}}$ acting on the interface for $\theta=120^{\circ}$. $p_{\text{L}}$ suddenly falls at the end of the filling stage, and subsequently, it decreases smoothly, as shown in \fig\ref{fig:8}b(II). At the pinch-off point, there is a sudden shoot-up in $p_{\text{L}}$.  

\noindent When $\qr=1/10$, $2r$ follows a sharp increase in the early filling stage. It can also be observed that for $\theta=120^{\circ}$, $2r$ is shown smooth increasing variation in the filling stage and then sudden shoot-up and again drop. However, $2r$ shows a smooth variation for $\theta\ge135^{\circ}$, as shown in \fig\ref{fig:8}c(I). The corresponding $p_{\text{L}}$ is shown in \fig\ref{fig:8}c(II). When $\theta=120^{\circ}$, $\Delta p$ shows an oscillatory behavior in the filling stage. On moving further, $p_{\text{L}}$ is falling suddenly in the squeezing and then again showing a shoot-up at the pinch-off point. The Laplace pressure for $\theta\ge135^{\circ}$ shows a smooth decrease and sudden rise at the pinch-off point. Therefore, it can be concluded that the solid surface is hydrophobic in nature, i.e., $\theta=120^{\circ}$, the force due to the surface tension is strongly resisting the shear exerted by the surrounding fluid.
%--------------------------------------
\section{Conclusions}
%%--------------------------------------------
\noindent The effect of surface wettability on two-phase flow and dynamics of droplet pinch-off in a cross-flow microfluidic system has been studied systematically and modeled using the Navier-Stokes equation coupled with the conservative level set method. The critical physicochemical determinants varied in a wide range: capillary number, $\cac <10^{-2}$;  the contact angle, $120^{\circ} \leq \theta \leq 180^{\circ}$; flow rate ratio, $1/10 \leq \qr \leq 10$. The droplet formation stages are based on the phase contours characterized as initial, filling, squeezing, pinch-off, and stable droplet. The time required for each stage of the droplet formation is a complex function of the flow rate ratio, contact angle, and capillary number. Comparison of the filling and pinch-off time based on the phase and pressure profiles as a function of the contact angle provided a new insight that the droplet pinch-off mechanism can be explained by installing the pressure sensors in the microchannels. Instantaneous interface profiles have been captured microscopically and analyzed in-depth to elucidate the droplet dynamics further. Under the hydrophobic conditions the ($120^{\circ} \leq \theta \leq 150^{\circ}$), the interface shape profile transiting from convex into concave immediately and slowly for the superhydrophobic conditions ($150^{\circ} < \theta \leq 180^{\circ}$). The pressure profiles in the continuous and dispersed phases have been compared. The pressure in the dispersed phase shows an oscillating behavior and evolution as anti phase with the pressure in the continuous phase. Maximum pressure in the continuous phase is a complex function of the dimensionless parameters. The Laplace pressure acting on the interface is quantitatively higher for the hydrophobic conditions, and hence the interface adopts a flattened shape. However, the interface adopts a more curved shape due to the higher Laplace pressure for superhydrophobic conditions. The interface neck width increases in the filling stage up to a threshold value and then decreases until the pinch-off point. A similar trend is observed for all contact angles qualitatively; however, it differed quantitatively. The maximum value of the neck width is a complex function of the flow rate ratio, contact angle, and capillary number. The insights obtained from the present study would guide the designing of the microfluidic devices dispensing droplets. 
%
%%%%%%%%%%%%%%%%%%%%%%%%%%%%%%%%%%%%%%%%%%
\section*{Declaration of Competing Interest}
\noindent 
The authors declare that they have no known competing financial interests or personal relationships that could have appeared to influence the work reported in this paper.
%All authors declare that they have no conflict of interest. The  authors  certify  that  they  have  NO  affiliations  with  or  involvement  in  any  organization or entity with any financial interest (such as honoraria; educational grants; participation in speakers’ bureaus; membership, employment, consultancies, stock ownership, or other equity interest; and expert testimony or patent-licensing arrangements), or non-financial interest (such as personal or professional relationships, affiliations, knowledge or beliefs) in the subject matter or materials discussed in this manuscript.
%%%%%%%%%%%%%%%%%%%%%%%%%%%%%%%%%%%%%%%%%%
\section*{Acknowledgments}
\noindent 
R.P. Bharti would like to acknowledge Science and Engineering Research Board (SERB), Department of Science and Technology (DST), Government of India (GoI) for providence of MATRICS grant (File No. MTR/2019/001598). 
%%%%%%%%%%%%%%%%%%%%%%%%%%%%%%%%%%%%%%%%%%
%
%
{%\fontsize{10}{10pt}\selectfont
%-------------------------------------
%\section*{Nomenclature}
\noindent 
\nomenclature[z0]{\textit{Abbreviations}}{}
\nomenclature[g0]{\textit{Greek letters}}{}
\nomenclature[d0]{\textit{Dimensionless groups}}{}
%\nomenclature[s0]{\textit{Superscripts}}{}
%
%-------- Abbreviations
%
 \nomenclature[zbdf]{BDF}{backward differentiation formula}
 \nomenclature[zcfd]{CFD}{computational fluid dynamics}
 \nomenclature[zcp]{CP}{continuous phase}
 \nomenclature[zcsp]{CSP}{continuum surface force}
 \nomenclature[zdp]{DP}{disperse phase}
% \nomenclature[zdae]{DAE}{differential algebraic equations}
% \nomenclature[zfdm]{FDM}{finite difference method}
 \nomenclature[zfem]{FEM}{finite element method}
% \nomenclature[zfvm]{FVM}{finite volume method}
 \nomenclature[zfsi]{FSI}{fluid-solid interaction}
% \nomenclature[zlbm]{LBM}{lattice Boltzmann method}
 \nomenclature[zlli]{LLI}{liquid-liquid interface}
 \nomenclature[zlsm]{LSM}{level set method}
% \nomenclature[zpardiso]{PARDISO}{parallel direct solver}
% \nomenclature[zpfm]{PFM}{phase field method}
% \nomenclature[zvof]{VOF}{volume of fluid}
%
%-------- List of Symbols
%
%\nomenclature[adeff]{$d_{\text{eff}}$}{effective droplet diameter (\eqn\ref{eq:deff}), m}
\nomenclature[aDt]{$\mathbf{D}$}{rate of strain tensor (\eqn\ref{eq:3}), s$^{-1}$}
%\nomenclature[aFsigma]{$\mathbf{F}_{\sigma}$}{interfacial tension force (\eqn\ref{eq:ift}), N}
%\nomenclature[aFi]{${F}_{\text{i}}$}{magnitude of inertial force, N} 
%\nomenclature[aFv]{${F}_{\text{v}}$}{magnitude of viscous force, N} 
%\nomenclature[aFs]{${F}_{\sigma}$}{magnitude of interfacial tension force, N} 
\nomenclature[aLu]{$L_{\text{u}}$}{upstream length of the main channel, $\mu$m} 
\nomenclature[aLd]{$L_{\text{d}}$}{downstream length of the main channel, $\mu$m}
\nomenclature[aLm]{$L_{\text{m}}$}{length of the main channel, $\mu$m}
\nomenclature[aLs]{$L_{\text{s}}$}{length of the side channel, $\mu$m}
\nomenclature[ahm]{$h_{\text{max}}$}{maximum size of mesh element, $\mu$m}
\nomenclature[aNe]{$N_{\text{e}}$}{total number of mesh elements, -}
\nomenclature[aP]{$p$}{pressure, Pa} 
\nomenclature[aPcp]{$p_{\text{cp}}$}{pressure in CP at point `cp', -} 
\nomenclature[aPcp]{$p_{\text{dp}}$}{pressure in DP at point `dp', -} 
\nomenclature[aPl]{$p_{\text{L}}$}{Laplace pressure, -} 
\nomenclature[aQc]{$Q_{\text{c}}$}{flow rate of CP, m$^{3}$/s}
\nomenclature[aQd]{$Q_{\text{d}}$}{flow rate of DP, m$^{3}$/s}
\nomenclature[aQr]{$Q_{\text{r}}$}{flow rate ratio (\eqn\ref{eq:7}), -}
%\nomenclature[afdd]{$f_{\text{dd}}$}{droplet detachment frequency, s$^{-1}$}
\nomenclature[at]{$t$}{time, s}
\nomenclature[atf]{$t_0$}{initial time, -}
\nomenclature[atf]{$t_1$}{filling time, -}
\nomenclature[atf]{$t_2$}{squeezing time, -}
\nomenclature[atf]{$t_3$}{pinch-off time, -}
\nomenclature[atf]{$t_4$}{stable droplet time, -}
\nomenclature[atf]{$t_{\text{f}}$}{filling (S-1) stage time, -}
\nomenclature[ats]{$t_{\text{s}}$}{squeezing (S-2) stage time, -}
\nomenclature[atb]{$t_{\text{b}}$}{breakup (S-3) stage time, -}
\nomenclature[atsd]{$t_{\text{sd}}$}{stable droplet (S-4) stage time, -}
\nomenclature[aRec]{$Re_{\text{c}}$}{Reynolds number for CP (\eqn\ref{eq:7}), -}
%\nomenclature[aRed]{$Re_{\text{d}}$}{Reynolds number for DP (\eqn\ref{eq:dimp1}), dimensionless}
%\nomenclature[aRer]{$Re_{\text{r}}$}{ratio of Reynolds numbers (\eqn\ref{eq:dimp2}), dimensionless}
\nomenclature[aCac]{$Ca_{\text{c}}$}{capillary number for CP (\eqn\ref{eq:7}), -}
%\nomenclature[aCad]{$Ca_{\text{d}}$}{capillary number for DP (\eqn\ref{eq:dimp1}), dimensionless}
%\nomenclature[aCar]{$Ca_{\text{r}}$}{ratio of capillary numbers (\eqn\ref{eq:dimp2}), dimensionless}
%\nomenclature[aCactrans]{$Ca_{\text{c,trans}}$}{transitional or threshold capillary number for CP (\eqn\ref{eq:10}), dimensionless}
%\nomenclature[aCadtrans]{$Ca_{\text{d,trans}}$}{transitional capillary number for DP (\eqn\ref{eq:10}), dimensionless}
%\nomenclature[aCartrans]{$Ca_{\text{r,trans}}$}{ratio of transitional capillary numbers (\eqn\ref{eq:10}), dimensionless}
\nomenclature[aU]{$\mathbf{u}$}{velocity vector, m/s}
\nomenclature[awc]{$w_{\text{c}}$}{width of the main channel, $\mu$m}
\nomenclature[awd]{$w_{\text{d}}$}{width of the side channel, $\mu$m}
\nomenclature[awr]{$w_{\text{r}}$}{channel width ratio (\eqn\ref{eq:7}), -}
\nomenclature[a2r]{$2r$}{neck width, -}
%\nomenclature[aRcmin]{$\rcmin$}{local minimum radius of curvature, -}
\nomenclature[ax]{$x$}{stream-wise coordinate}
\nomenclature[ay]{$y$}{transverse coordinate}
%\nomenclature[ayo]{$\text{Y}$}{critical $Re$ normalized w.r.t. corresponding unconfined Newtonian flow (\eqns\ref{nrec}, \ref{nrec1}), dimensionless}
%
%-------- Greek symbols
%
\nomenclature[gbeta]{$\beta$}{slip length, $\mu$m}
%\nomenclature[glambda]{$\lambda$}{wall confinement ratio ($=\beta^{-1}$), dimensionless}
%\nomenclature[gdPl]{$\Delta p_{\text{L}}$}{pressure drop across the interface in between points `dp' and 'cp', -} 
\nomenclature[ggamma]{$\gamma$}{re-initialization or stabilization parameter  (\eqn\ref{eq:6}), m/s}
\nomenclature[gepsilon]{$\epsilon_{\text{ls}}$}{interface thickness controlling parameter  (\eqn\ref{eq:6}), m}
\nomenclature[gmuc]{$\mu_{\text{c}}$}{viscosity of CP, Pa.s}
\nomenclature[gmud]{$\mu_{\text{d}}$}{viscosity of DP, Pa.s}
\nomenclature[gmur]{$\mu_{\text{r}}$}{viscosity ratio (\eqn\ref{eq:7}), -}
\nomenclature[grhoc]{$\rho_{\text{c}}$}{density of CP, kg/m$^3$}
\nomenclature[grhod]{$\rho_{\text{d}}$}{density of DP, kg/m$^3$}
\nomenclature[grhor]{$\rho_{\text{r}}$}{density ratio (\eqn\ref{eq:7}), -}
\nomenclature[gsigma]{$\sigma$}{liquid-liquid interfacial tension, N/m}
\nomenclature[gsigma]{$\sigma_{\text{sl}}$}{solid-liquid interfacial tension, N/m}
\nomenclature[gsigma]{$\sigma_{\text{sv}}$}{solid-vapor interfacial tension, N/m}
\nomenclature[gtau]{$\tau$}{extra stress tensor (\eqn\ref{eq:3}),  N/m$^2$}
\nomenclature[gphi]{$\phi$}{level set function, dimensionless}
\nomenclature[gkappa]{$\kappa$}{curvature of the interface, m}
\nomenclature[gtheta]{$\theta$}{contact angle, degrees}
%\nomenclature[gtaudd]{$\tau_{\text{dd}}$}{droplet detachment time, dimensionless}
\nomenclature[gdelma]{$\delta_{\text{max}}$}{maximum percent relative error, -}
\nomenclature[gdelmi]{$\delta_{\text{min}}$}{minimum percent relative error, -}
\nomenclature[gdelav]{$\delta_{\text{avg}}$}{average percent relative error, -}
%
%-------- dimensionless groups
%
\nomenclature[dRe]{$Re$}{Reynolds number (\eqn\ref{eq:7}), -}
\nomenclature[dCa]{$Ca$}{capillary number (\eqn\ref{eq:7}), -}
%\nomenclature[dStr]{$\mathit{St}$}{Strouhal number (\eqn\ref{eq17}), dimensionless}
% \nomenclature[dPr]{$Pr$}{Prandtl number, $\frac{\nu}{\alpha}$}
% \nomenclature[dRe]{$Re$}{Reynolds number, $\frac{H u_{lid} \rho}{\mu}$}
% \nomenclature[dGr]{$Gr$}{Grashof number, $\frac{g\beta\Delta{T} H^3}{\nu^2}$}
% \nomenclature[dRi]{$Ri$}{Richardson number, =$\frac{Gr}{Re^2}=\frac{g\beta \Delta{T} H}{u_{lid}^{2}}$}
%
%------ subscripts and superscripts
%
%\nomenclature[sc]{$\text{c}$}{continuous phase}
%\nomenclature[sd]{$\text{d}$}{dispersed phase}
%\nomenclature[sr]{$\text{r}$}{ratio}
	\renewcommand{\nompreamble}{\vspace{1em}\fontsize{10}{8pt}\selectfont}
	{\printnomenclature[5em]}}
%
%\begin{thebibliography}{0000}
%\bibliographystyle{plainnat}
\bibliography{references}
%
% Bibliographic references with the natbib package:
% Parenthetical: \citep{Bai92} produces (Bailyn 1992).
% Textual: \citet{Bai95} produces Bailyn et al. (1995).
% An affix and part of a reference:
%   \citep[e.g.][Ch. 2]{Bar76}
%   produces (e.g. Barnes et al. 1976, Ch. 2).
% 
% \bibitem[Names(Year)]{label} or \bibitem[Names(Year)Long names]{label}.
% (\harvarditem{Name}{Year}{label} is also supported.)
%\bibitem[]{}
%
%\input{references.tex}
%
%\end{thebibliography}
%
\addcontentsline{toc}{section}{References} 
\clearpage
%
%\listoftables
%
%\clearpage
%
%\listoffigures
%
%\clearpage
%
%\input{tables.tex}
%
%\clearpage
%
%\renewcommand{\thesubfigure}{(\roman{subfigure})}
%
%\input{figures.tex}
%
\end{document}